# Non-line-of-sight imaging with arbitrary illumination and detection pattern

**Authors**

Xintong Liu[1], Jianyu Wang[1], Leping Xiao[2,3], Zuoqiang Shi[1,4], Xing Fu[2,3,*], Lingyun Qiu[1,4,*]

**Affiliations**

[1] Yau Mathematical Sciences Center, Tsinghua University, Beijing, China, 100084

[2] State Key Laboratory of Precision Measurement Technology and Instruments, Department of Precision Instrument, Tsinghua University, Beijing, China, 100084

[3] Key Laboratory of Photonic Control Technology (Tsinghua University), Ministry of Education, Beijing, China, 100084

[4] Yanqi Lake Beijing Institute of Mathematical Sciences and Applications, Beijing, China, 101408

[*] Correspondence and requests for materials should be addressed to fuxing@tsinghua.edu.cn (Xing Fu) and lyqiu@tsinghua.edu.cn (Lingyun Qiu).

# Abstract

Non-line-of-sight (NLOS) imaging aims at reconstructing targets obscured from the direct line of sight. Existing NLOS imaging algorithms require dense measurements at rectangular grid points in a large area of the relay surface, which severely hinders their availability to variable relay scenarios in practical applications such as robotic vision, autonomous driving, rescue operations and remote sensing. In this work, we propose a Bayesian framework for NLOS imaging with no specific requirements on the spatial pattern of illumination and detection points. By introducing virtual confocal signals, we design a confocal complemented signal-object collaborative regularization (CC-SOCR) algorithm for high quality reconstructions. Our approach is capable of reconstructing both albedo and surface normal of the hidden objects with fine details under the most general relay setting. Moreover, with a regular relay surface, coarse rather than dense measurements are enough for our approach such that the acquisition time can be reduced significantly. As demonstrated in multiple experiments, the new framework substantially enhances the applicability of NLOS imaging.

# Introduction

The technique of imaging objects out of the direct line of sight has attracted increasing attention in recent years[1–26]. A typical non-line-of-sight (NLOS) imaging scenario is looking around the corner with a relay surface, where the target is obscured from the vision of the observer. NLOS imaging aims to recover the albedo and surface normal of the hidden targets with the measured photon information. Potential applications of NLOS imaging include but are not limited to robotic vision, autonomous driving, rescue operations, remote sensing and medical imaging.

To achieve NLOS reconstruction, laser pulses of high temporal resolution are used to illuminate several points on the relay surface, where the first diffuse reflection occurs. After that, photons enter the NLOS domain and are bounced back to the visible surface again by the unknown targets. The hidden targets can be reconstructed with the time-resolved photon intensity measured at several detection points on the visible surface. Commonly used time-resolved detectors are single-photon avalanche diodes (SPAD)[27]. The imaging system is confocal if the illumination point coincides with the detection point for each spatial measurement, otherwise being non-confocal. Besides, we call the measurements regular if the illumination and detection points are uniformly distributed in a rectangular region.

According to how the hidden surface is represented, existing imaging algorithms are divided into three categories: point-cloud-based[28], mesh-based[29] and voxel-based methods[1,8,9,30–35]. Among these categories, voxel-based algorithms yield to be the most efficient ones with low time complexity[32] and fine reconstruction results[34]. For voxel-based methods, the reconstruction domain is discretized with three-dimensional grid points and the albedo is represented as a grid function.

The first voxel-based NLOS reconstruction method is the back-projection algorithm proposed by Velten et al.[1]. The measured photon intensity is modeled as a linear operator applied to the albedo, and the targets are reconstructed by applying the adjoint operator to the measured data. Further improvements of the back-projection method include rendering approaches for fast implementations[2,16] and filtering techniques[33,36] for noise reduction. The light-cone-transform[30] (LCT) proposed by O'Toole et al. describes the physical process as a convolution of the light cone kernel and the hidden target. In this way, the reconstruction is formulated as a deblurring problem and can be realized efficiently using the fast Fourier transform. The directional light cone transform[31] (D-LCT) generalizes this method and simultaneously reconstructs the albedo and surface normal of the hidden target. The frequency-wavenumber migration[8] (F-K) method uses the wave equation to reconstruct the albedo and can also be implemented efficiently in the frequency domain. The LCT, D-LCT and F-K methods only work directly under confocal settings. Although it is possible to transfer the data collected in non-confocal setups to confocal ones, the approximation error cannot be neglected[34]. To reconstruct the hidden object under non-confocal settings, the phasor field[32] (PF) method formulates the NLOS detection process as one of diffractive wave propagation and provides a direct inversion solution with low time complexity. Its recent extension with SPAD arrays reconstructs live low-latency videos of NLOS scenes[37]. The signal-object collaborative regularization[34] (SOCR) method considers priors on both the reconstructed target and the measured signal, which leads to high-quality reconstruction with little background noise.

Despite these breakthroughs, two major obstacles of existing methods toward practical applications are the need for a large relay surface and dense measurement. When there are limitations on the shape and size of the relay surface, these algorithms may fail due to the lack of data. Besides, dense measurement results in a long acquisition time, which poses a significant challenge for applications such as auto-driving where the observer may move at high speed.

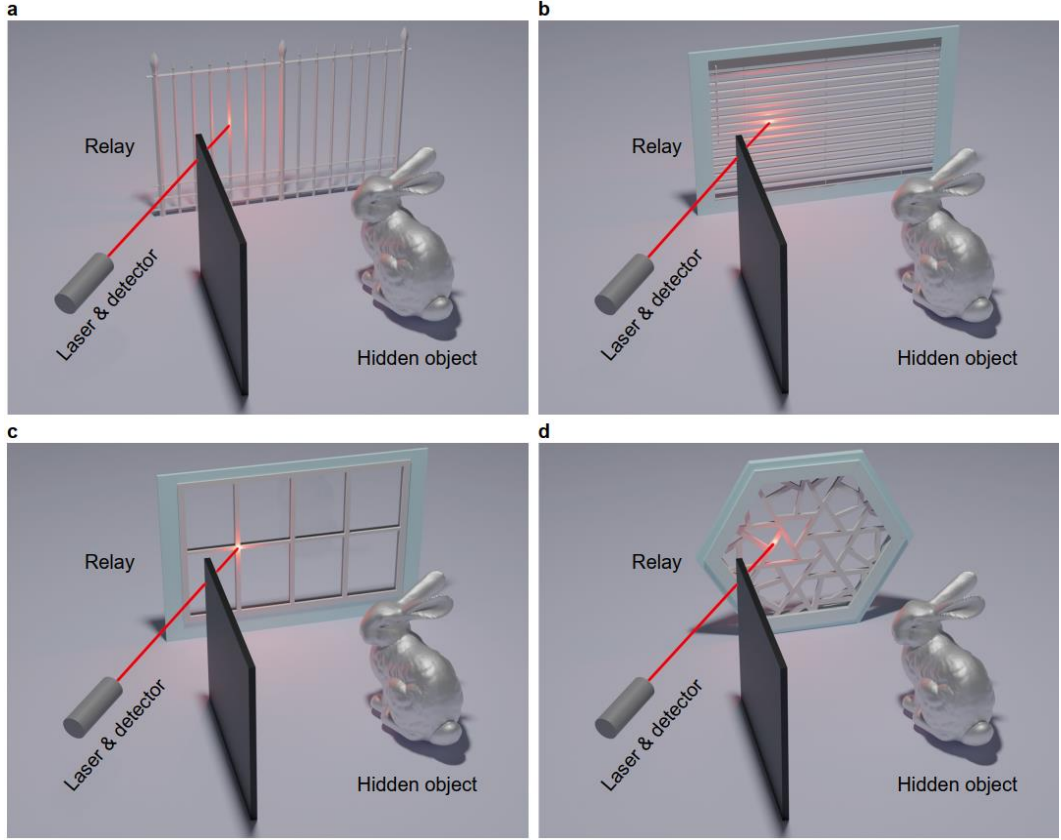

**Fig. 1 Irregular illumination and detection patterns for NLOS imaging. a** The relay is a fence. **b** The relay is a horizontal shutter. **c** The relay is an array of window edges. **d** The relay is a set of several sticks sparsely and randomly distributed.

In this work, we propose a Bayesian framework for NLOS reconstruction that is not limited by the spatial pattern of illumination and detection points. By introducing the virtual confocal signal at rectangular grid points, we design joint regularizations for the measured signal, virtual confocal signal and the hidden target. We put forward a confocal complemented signal-object collaborative regularization (CC-SOCR) framework, which reconstructs both the albedo and surface normal of the hidden target. The proposed method works quite well under the most general setting, allowing regular and irregular measurement patterns in both confocal and non-confocal scenarios. Besides, our approach provides sparse reconstructions of the targets with clear boundaries and negligible background noise, even in cases with very coarse and noisy measurements. Notably, the proposed method suggests a paradigm shift, liberating the research of NLOS imaging from relying heavily on the assumption of a large-size relay surface with regular shape and entire region (wall, ground) ever since the technique was first proposed. To the best of our knowledge, this work demonstrates high quality NLOS reconstruction for the first time, in the scenarios with the relay surfaces having discrete scattering regions, irregular shape, or very limited size, enabling the hidden object reconstruction with far more types of realistic relay surfaces such as window shutter, window frame, and fence, which significantly broadens the scope of NLOS imaging applications. As shown in Fig. 1, the illumination and detection patterns are irregular but manifest in ubiquitous scenes of daily lives. Reconstruction results of the bunny with synthetic confocal signals[38] detected at the entire relay surface and these four scenarios are provided in Supplementary Figures 1 – 5. Besides, our method can significantly reduce the acquisition time and accelerate the imaging process by using sparse measurements for the conventional scenario of a large relay surface.

## Results

**The NLOS physical model.** The goal of NLOS imaging is to take a collection of measured transient data and find the target that comes closest to fitting these signals. In this work, we adopt the physical model proposed in SOCR[34]. Let $x'_i$ and $x'_d$ be the illumination and detection points on the visible surface, and we call $(x'_i, x'_d)$ an active measurement pair, or simply a pair in the following. The photon intensity measured at time $t$ is given by

$$\tau(x'_i, x'_d, t) = \int_{\Omega} \frac{(x'_d - x) \cdot \mathbf{n}(x)}{\|x'_i - x\|^2 \|x'_d - x\|^3} f(x) \delta(\|x'_i - x\| + \|x'_d - x\| - ct) dx \quad (1)$$

in which $\Omega$ is the three-dimensional reconstruction domain, $f(x)$ denotes the albedo value of the point $x$, $\mathbf{n}(x)$ is the unit surface normal at $x$ that points toward the visible surface. The unit vector $\mathbf{n}(x)$ can be arbitrarily chosen for points with zero albedo value. By denoting $\mathbf{u} = f\mathbf{n}$, equation (1) is written equivalently as

$$\tau(x'_i, x'_d, t) = \int_{\Omega} \frac{(x'_d - x) \cdot \mathbf{u}(x)}{\|x'_i - x\|^2 \|x'_d - x\|^3} \delta(\|x'_i - x\| + \|x'_d - x\| - ct) dx \quad (2)$$

Noting that the intensity is linear with $\mathbf{u}$, the physical model can be written as $\boldsymbol{\tau} = A\mathbf{u}$ in the discrete form. The albedo and surface normal can be obtained directly from $\mathbf{u}$. Indeed, the albedo of a voxel $x$ is given by the norm of the vector $\mathbf{u}(x)$. The surface normal of a voxel $x$ is obtained by normalizing the vector $\mathbf{u}(x)$. The surface normal is not defined where the albedo is zero.

**Four types of signals.** To reconstruct the hidden target, signals measured at several active measurement pairs are needed. Consider a collection of $N$ pairs $B = \{(x_i^k, x_d^k)\}_{k=1}^{N}$, in which $x_i^k$ and $x_d^k$ are the $k^{th}$ illumination and detection points. For each pair, the photon intensity of the first $T$ time bins is collected. In order to obtain high quality reconstructions, we consider four types of signals for $B$: the ideal signal $\mathbf{b}_{ideal}$, the simulated signal $\mathbf{b}_{sim}$, the measured signal $\tilde{\mathbf{b}}$, and the approximated signal $\mathbf{b}$. In practice, the ideal physical model is nonlinear and depends on the reflection property of the hidden target. The signal $\mathbf{b}_{ideal}$ does not contain noise and is considered to be generated with the ideal nonlinear physical model. The simulated signal $\mathbf{b}_{sim} = A_{\mathbf{b}}\mathbf{u}$ is generated using equation (1) and the hidden target. The measured signal $\tilde{\mathbf{b}}$ is inevitably corrupted with noise, which is considered as certain deterioration of the ideal signal. The degradation is related to detection efficiency and background noise, whose distribution is hard to estimate and may vary from one scenario to another. To tackle this problem, we introduce the approximated signal $\mathbf{b}$, which serves as a better approximation of the ideal signal than the measured signal.

When the number of measurement pairs is small, the solution to the reconstruction problem may not be unique due to the lack of data. To overcome the rank deficiency of the measurement matrix, we introduce the virtual confocal signal at regular focal points. Suppose that the reconstruction domain $\Omega$ is discretized with $n_x \times n_y \times n_z$ voxels in the depth, horizontal and vertical directions. We denote by $D = \{(x^k, x^k)\}_{k=1}^{N_y \times N_z}$ the collection of confocal measurement pairs, which are the orthogonal projections of the voxels to a virtual planar surface perpendicular to the depth direction. The corresponding ideal, simulated and approximated signals for $D$ are denoted by $\mathbf{d}_{ideal}$, $\mathbf{d}_{sim} = A_{\mathbf{d}} \mathbf{u}$ and $\mathbf{d}$, respectively.

**A Bayesian framework.** We treat the reconstructed target $\mathbf{u}$, the measured signal $\tilde{\mathbf{b}}$ and the approximated signals $\mathbf{b}$, $\mathbf{d}$ as random vectors and formulate the imaging task as an optimization problem using Bayesian inference. The target and signals $\mathbf{u}$, $\mathbf{b}$ and $\mathbf{d}$ are obtained simultaneously by maximizing the joint posterior probability.

$$(\mathbf{u}^*, \mathbf{b}^*, \mathbf{d}^*) = \underset{\mathbf{u}, \mathbf{b}, \mathbf{d}}{\operatorname{argmax}} \ \mathbb{P}(\mathbf{u}, \mathbf{b}, \mathbf{d} | \tilde{\mathbf{b}}) \tag{3}$$

Three assumptions are made to formulate this as a concrete optimization problem.

Fig. 2 The proposed CC-SOCR method. a The CC-SOCR framework. For high quality reconstructions, the measured signal, approximated signal and virtual confocal signal are treated as random variables and solved simultaneously using Bayesian inference. The term $\Gamma(\mathbf{u},\mathbf{b})$ includes the sparseness of the approximated signal, as well as the sparseness and non-local self-similarity prior of the target. The term $\Upsilon(\mathbf{u},\mathbf{b},\tilde{\mathbf{b}})$ corresponds to an empirical Wiener filter, in which the simulated signal of the target serves as the pilot estimation. The term $\Xi(\mathbf{u},\mathbf{d})$ contains the sparseness of the virtual confocal signal, as well as the joint sparse representation of the local structures of the simulated signal and the virtual confocal signal. b The approximated signals and the reconstructed target of the instance of the statue (confocal, measured data). The measured data is provided in the Stanford dataset[8]. We assume the relay surface to be the region consisting of four letters 'N', 'L', 'O', and 'S'. The measured signal, approximated signal, virtual confocal signal and the reconstructed albedo are shown at the bottom.

Firstly, the conditional distribution of the measured signal $\tilde{\mathbf{b}}$ given the joint probability distribution of $\mathbf{u}$, $\mathbf{b}$ and $\mathbf{d}$ is
9

$$\mathbb{P}(\tilde{\mathbf{b}}|\mathbf{u},\mathbf{b},\mathbf{d}) = \mathbb{P}(\tilde{\mathbf{b}}|\mathbf{u},\mathbf{b}) = \exp\left(-\|\mathbf{b}-\tilde{\mathbf{b}}\|^2 - \Upsilon(\mathbf{u},\mathbf{b},\tilde{\mathbf{b}})\right) \qquad (4)$$

in which $\Upsilon$ is related to the joint prior distribution of $\mathbf{u}$, $\mathbf{b}$ and $\tilde{\mathbf{b}}$. With this assumption, $\mathbf{d}$ does not provide additional information to predict $\tilde{\mathbf{b}}$ when $\mathbf{b}$ is known.

Secondly, the joint prior distribution of $\mathbf{u}$ and $\mathbf{b}$ is

$$\mathbb{P}(\mathbf{u},\mathbf{b}) = \exp\left(-\|A_{\mathbf{b}}\mathbf{u}-\mathbf{b}\|^2 - \Gamma(\mathbf{u},\mathbf{b})\right) \qquad (5)$$

in which $\Gamma$ describes the prior distribution of $\mathbf{u}$ and $\mathbf{b}$. With this regularization term, we search for the target only in the set of real-world objects. Besides, $\mathbf{b}$ is less noisy than the measured data and is closer to the ideal signal of a certain real-world target, which helps to enhance the reconstruction quality.

Thirdly, the conditional distribution of $\mathbf{d}$ given $\mathbf{u}$ and $\mathbf{b}$ is

$$\mathbb{P}(\mathbf{d}|\mathbf{u},\mathbf{b}) = \exp\left(-\|R_{\mathbf{b}}(\mathbf{b},\mathbf{d}) - R_{\mathbf{d}}(\mathbf{b},\mathbf{d})\|^2 - \|A_{\mathbf{d}}\mathbf{u}-\mathbf{d}\|^2 - \Xi(\mathbf{u},\mathbf{d})\right) \qquad (6)$$

in which $R_{\mathbf{b}}(\mathbf{b},\mathbf{d})$ and $R_{\mathbf{d}}(\mathbf{b},\mathbf{d})$ are the subsets of the approximated signals $\mathbf{b}$ and $\mathbf{d}$ that share the same measurement pairs. $\Xi(\mathbf{u},\mathbf{d})$ is related to the joint prior distribution of the target $\mathbf{u}$ and the virtual signal $\mathbf{d}$.

With these assumptions, we derive a concrete optimization problem using the Bayesian formula.

Table 1: Comparisons of seven NLOS reconstruction algorithms

| Method | Scenario | | Reconstructed target | | Reconstruction quality | |
|---|---|---|---|---|---|---|
| | Confocal measurements | Non-confocal measurements | Albedo | Surface normal | Dense measurements | Coarse measurements |
| LOG-BP[33] | General | General | ✓ | ✗ | Medium | Very Low |
| LCT[30] | Regular | ✗ | ✓ | ✗ | High | Low |
| D-LCT[31] | Regular | ✗ | ✓ | ✓ | High | Low |
| F-K[8] | Regular | ✗ | ✓ | ✗ | High | Low |
| PF[32] | Regular | Regular | ✓ | ✗ | High | Low |
| SOCR[34] | Regular | Regular | ✓ | ✓ | Very High | Medium |
| CC-SOCR | General | General | ✓ | ✓ | Very High | High |

By 'regular' we mean illumination and detection points uniformly distributed in a rectangular region. By 'general' we mean arbitrary illumination and detection points.

$$\begin{aligned}(\mathbf{u}^*,\mathbf{b}^*,\mathbf{d}^*) &= \underset{\mathbf{u},\mathbf{b},\mathbf{d}}{\operatorname{argmax}}\ \mathbb{P}\left(\mathbf{u},\mathbf{b},\mathbf{d}|\tilde{\mathbf{b}}\right) \\ &= \underset{\mathbf{u},\mathbf{b},\mathbf{d}}{\operatorname{argmax}}\ \mathbb{P}\left(\tilde{\mathbf{b}}|\mathbf{u},\mathbf{b},\mathbf{d}\right)\mathbb{P}(\mathbf{u},\mathbf{b},\mathbf{d}) \\ &= \underset{\mathbf{u},\mathbf{b},\mathbf{d}}{\operatorname{argmax}}\ \mathbb{P}\left(\tilde{\mathbf{b}}|\mathbf{u},\mathbf{b}\right)\mathbb{P}(\mathbf{u},\mathbf{b},\mathbf{d}) \\ &= \underset{\mathbf{u},\mathbf{b},\mathbf{d}}{\operatorname{argmax}}\ \mathbb{P}\left(\tilde{\mathbf{b}}|\mathbf{u},\mathbf{b}\right)\mathbb{P}(\mathbf{d}|\mathbf{u},\mathbf{b})\mathbb{P}(\mathbf{u},\mathbf{b}) \\ &= \underset{\mathbf{u},\mathbf{b},\mathbf{d}}{\operatorname{argmin}}\ \|\mathbf{b}-\tilde{\mathbf{b}}\|^2 + \|R_\mathbf{b}(\mathbf{b},\mathbf{d})-R_\mathbf{d}(\mathbf{b},\mathbf{d})\|^2 + \|A_\mathbf{d}\mathbf{u}-\mathbf{d}\|^2 \\ &\quad + \|A_\mathbf{b}\mathbf{u}-\mathbf{b}\|^2 + \Upsilon\left(\mathbf{u},\mathbf{b},\tilde{\mathbf{b}}\right) + \Xi(\mathbf{u},\mathbf{d}) + \Gamma(\mathbf{u},\mathbf{b})\end{aligned} \quad (7)$$

in which the third equality follows from equation (4) and the last equality holds with equations (4), (5) and (6). By designing appropriate regularization terms $\Upsilon$, $\Xi$ and $\Gamma$, we obtain high quality reconstructions of the targets even in scenarios with highly incomplete measurements. The proposed framework and collaborative regularizations designed are illustrated in Fig. 2a. Concrete expressions of the regularizations are provided in the method section. We term the proposed method the confocal complemented signal-object collaborative regularization (CC-SOCR) due to the virtual confocal signal $\mathbf{d}$ introduced and the regularizations imposed on the signal and the target. In the following, we compare the reconstruction results of the proposed method with the Laplacian of Gaussian filtered back-projection (LOG-BP) method, LCT, D-LCT, F-K, PF and SOCR. The LCT, D-LCT and F-K methods only work directly under confocal settings. The D-LCT, SOCR and the CC-SOCR methods reconstruct both the albedo and surface normal. Performance comparisons of all these methods are shown in Table 1. To bring existing methods into comparison, we interpolate the signal with the nearest neighbor method[8,35], which generates better results than zero padding[32] in extreme cases (See Supplementary Figure 24).

**Results on synthetic data.** Instead of using an entire planar visible surface, we assume the relay to be a square box which simulates the scenario of four edges of a window. The hidden object is a regular quadrangular pyramid, whose base length and height are 1 m and 0.2 m respectively. The central axis of the pyramid is perpendicular to the plane in which the relay square box lies, and the distance of the pyramid to this plane is 0.5 m. The albedo of the pyramid is assumed to be a constant. As shown in Fig. 3a, we simulate the signal measured at 36 points with equation (1). The points are exhaustively scanned, where only one point is illuminated each time, and signals are detected at all points. The dataset contains signals measured at 36 confocal and 1260 non-confocal pairs. The time resolution is set to be 32 ps. Note that the LCT, D-LCT, F-K, PF and SOCR methods do not work directly in this scenario. We compare the reconstruction result of the proposed method with LOG-BP. The maximum intensity projections are shown in Fig. 3c and Fig. 3d. The reconstructed albedo is normalized to the range [0,1]. Albedo values that are less than 0.25 are thresholded to zero. The LOG-BP method fails to locate the target correctly, and contains misleading artifacts near the boundary of the

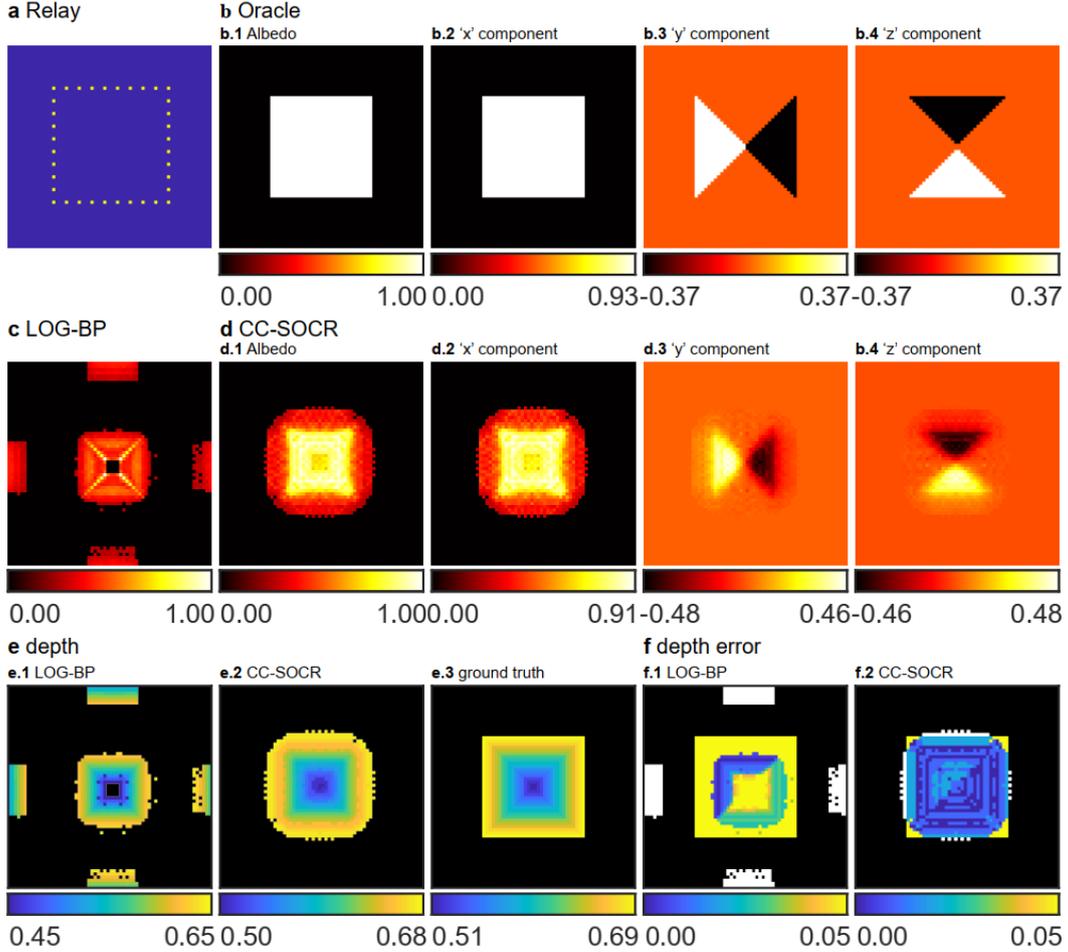

**Fig. 3 Reconstruction results of the pyramid (non-confocal, synthetic signal). a** The illumination and detection points are shown in yellow. **b** Ground truth. The 'x', 'y' and 'z' components show values of the directional albedo in the depth, horizontal and vertical directions, respectively. **c** Reconstructed albedo of the LOG-BP algorithm. **d** Reconstructed albedo and surface normal of the proposed CC-SOCR method. **e** The depth of the target of the LOG-BP, CC-SOCR reconstructions and the ground truth. **f** Absolute depth error of the LOG-BP and CC-SOCR reconstructions. The background is shown in black. Excessive voxels reconstructed are shown in white.

reconstruction domain. The proposed method locates the target correctly and does not contain noise in the background. The maximum depth error of the CC-SOCR reconstruction is 0.02 m, which is much smaller than the LOG-BP reconstruction (0.12 m). The absolute depth errors are shown in Fig. 3f. Classification error, defined as the percentage of excessive and missing voxels of the reconstruction, is used to assess the methods of locating the target. The classification error of the CC-SOCR reconstruction

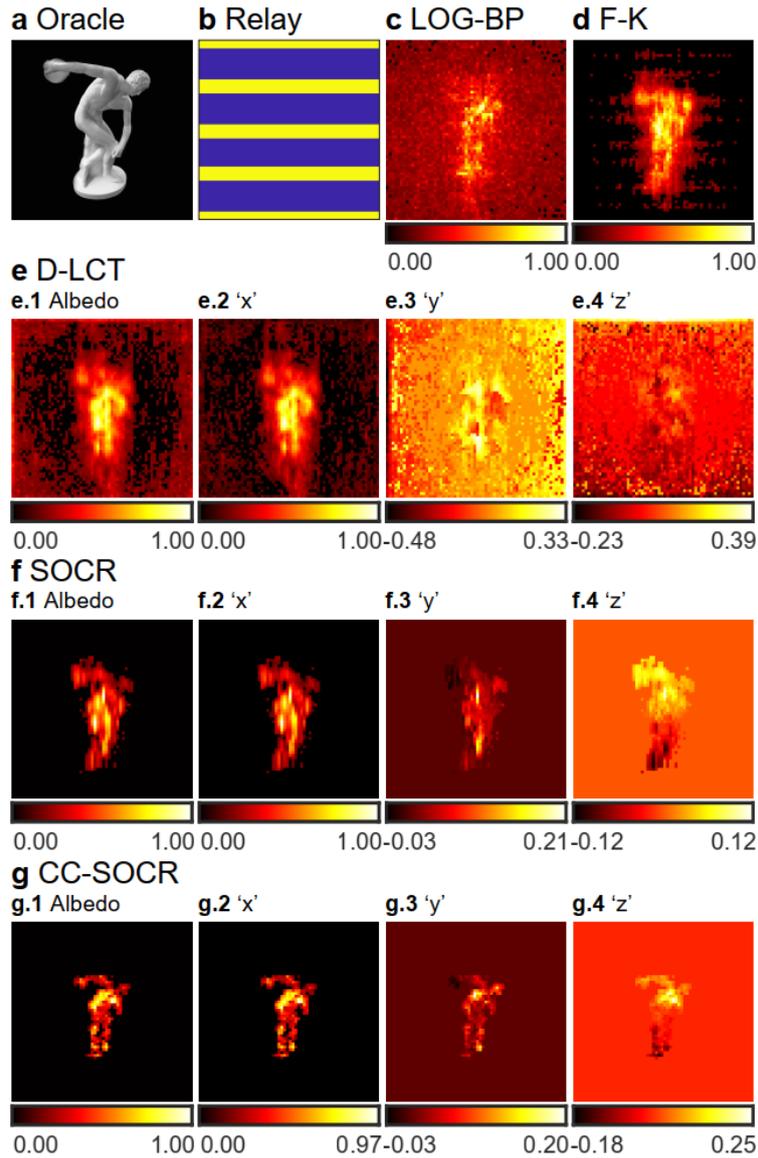

Fig. 4 Reconstructions of the statue with the relay surface in the shape of a horizontal shutter (confocal, measured signal). **a** A photo of the hidden target. **b** Confocal signals are measured in the yellow region. **c** Reconstructed albedo of the LOG-BP algorithm. **d** Reconstructed albedo of the F-K algorithm. **e** Reconstructed albedo and surface normal of the D-LCT method. **f** Reconstructed albedo and surface normal of the SOCR method. **g** Reconstructed albedo and surface normal of the CC-SOCR method.

is 2.86%, which is one order of magnitude smaller than that of the LOG-BP reconstruction (21.75%).

**Results on measured data.** For confocal experiments, we use the instance of a statue in the Stanford dataset[8] to test the performance of the proposed method. The target is 1 m away from the visible planar surface. In the original dataset, 512 × 512 focal points are raster-scanned in a square region of size 2 × 2 m$^2$. The time resolution is 32 ps and the total exposure time is 10 min. We downsample the signal to 64 × 64 in spatial dimensions. It takes 9 s to measure the downsampled signal.

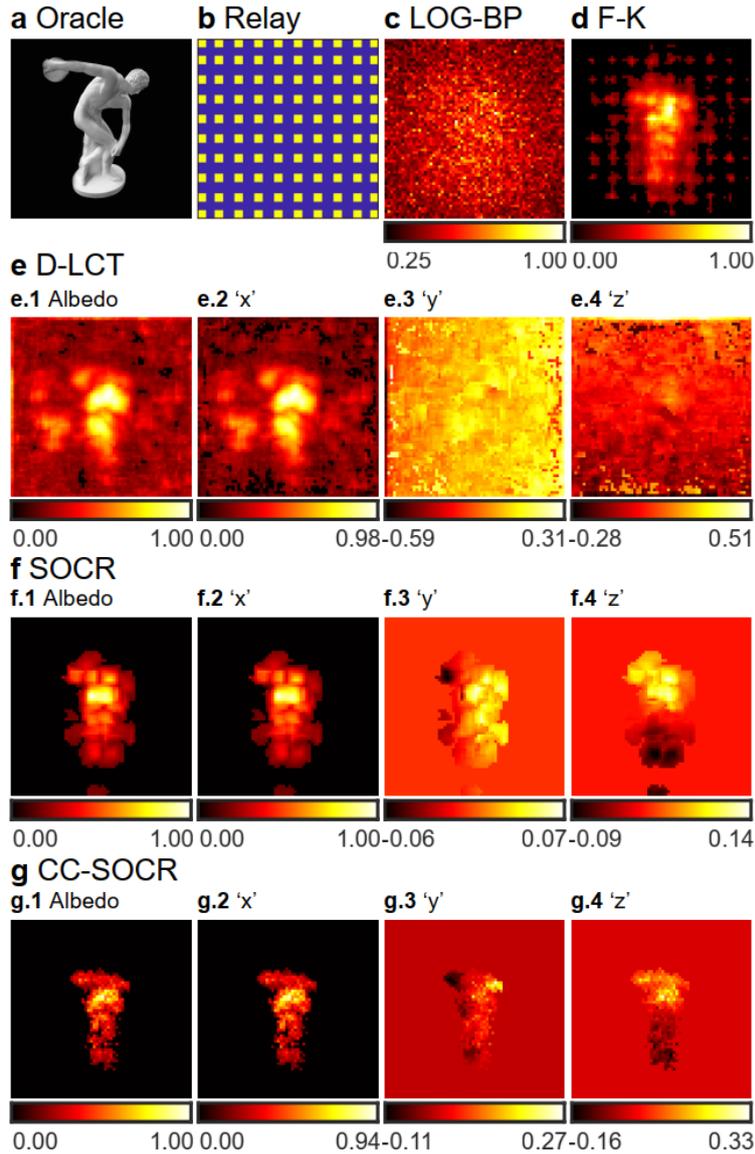

Fig. 5 Reconstructions of the statue with 10 × 10 confocal measurements (confocal, measured signal). **a** A photo of the hidden target. **b** Confocal signals are measured at the yellow points. **c** Reconstructed albedo of the LOG-BP algorithm. **d** Reconstructed albedo of the F-K algorithm. **e** Reconstructed albedo and surface normal of the D-LCT method. **f** Reconstructed albedo and surface normal of the SOCR method. **g** Reconstructed albedo and surface normal of the CC-SOCR method.

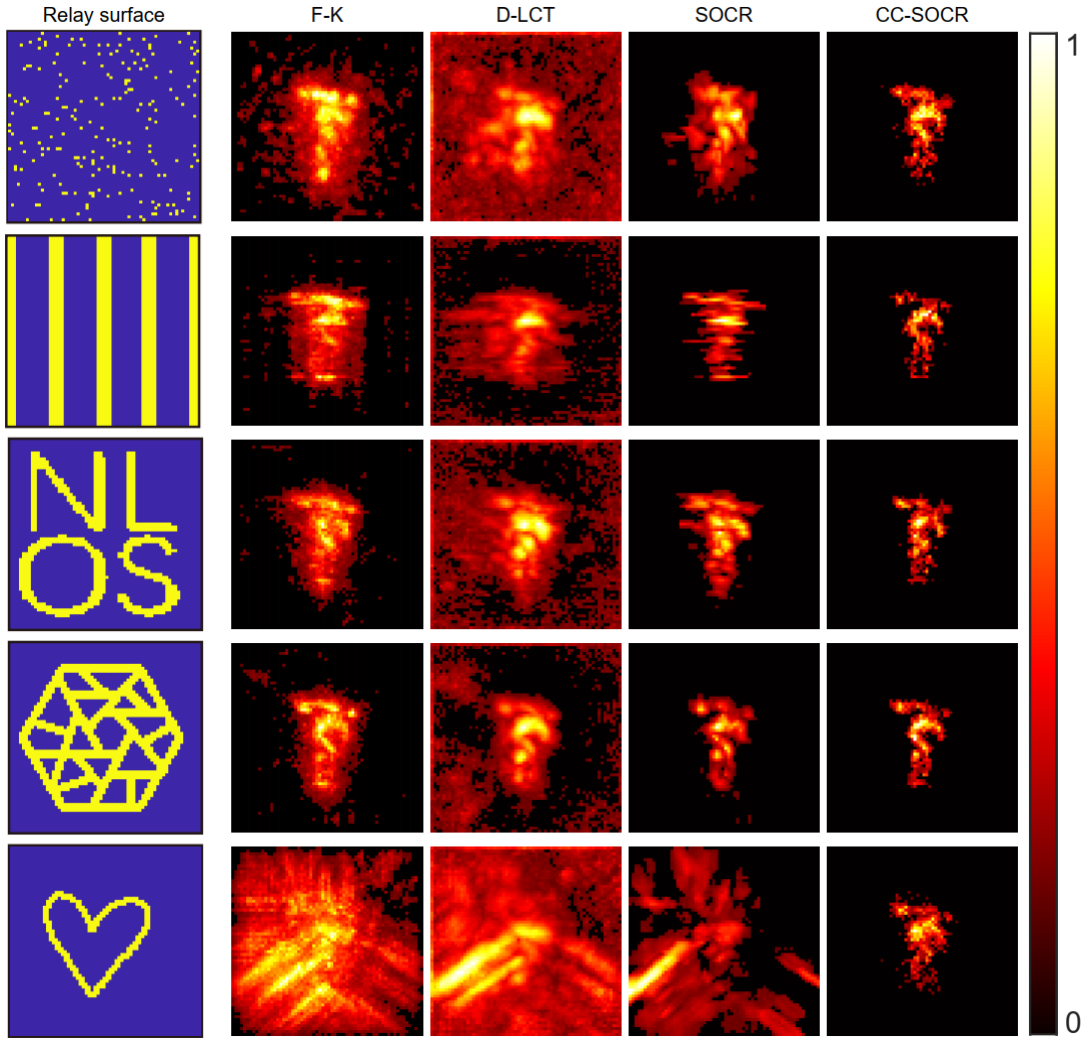

**Fig. 6 Reconstructions of the statue under representative cases with different relays (confocal, measured signal).** The illumination regions are shown in yellow in the first column. The reconstructed albedo of F-K, D-LCT, SOCR and CC-SOCR methods are compared in the second to fifth columns.

To simulate the case where the relay surface is a horizontal shutter, we only extract the signals measured at 21 rows from the downsampled data, as shown in the yellow region of Fig. 4b.

From bottom to top, the five equispaced regions contain 3, 5, 5, 5 and 3 rows of measurements, respectively. The dataset contains signals measured at 1344 focal points, which takes 3 s for data acquisition. Reconstruction results are shown in Fig. 4. The LOG-BP reconstruction is noisy. The reconstruction results of F-K, D-LCT and SOCR are blurry and contain artifacts. The proposed method reconstructs the target faithfully.

Figure 5 shows reconstruction results of the statue with signals detected at $10 \times 10$ uniformly distributed focal points in a square region of size $2 \times 2$ m$^2$, which only takes 0.23 s for the measurements. The points scanned are shown in Fig. 5b. The LOG-BP reconstruction contains heavy background noise and the target cannot be clearly identified. The F-K and D-LCT reconstructions are blurry and also contain background noise. The SOCR reconstruction contains artifacts, indicating that the error introduced in the nearest neighbor interpolation cannot be neglected. In contrast, the proposed method locates the target correctly and reconstructs more details than other methods. More reconstruction results with different numbers of uniformly distributed confocal measurements are compared in Supplementary Figures 6 – 10.

Figure 6 shows reconstruction results of the statue obtained with signals measured at different regions of the relay surface: a set of 200 randomly distributed focal points in an area of size $2 \times 2$ m$^2$; a region consisting of 5 equispaced vertical bars with 1344 focal points; a region that consists of four letters 'N', 'L', 'O' and 'S' with 825 focal points; a region made up of several sticks sparsely and randomly distributed with 1229 focal points; and a heart-shaped region with 258 focal points. These results indicate the capability of the proposed method in reconstructing the hidden target under various relay settings. For the case of the heart-shaped relay, the CC-SOCR method locates the target correctly, while all other methods fail. The measured signal, approximated signal and virtual confocal signal of the scenario with measurements at the four letters 'N', 'L', 'O' and 'S' are shown in Fig. 2b. The virtual confocal signal is smooth and plays an important role for high quality reconstruction. The three views and surface normal of the reconstructions as well as more comparisons under different relay settings are provided in Supplementary Figures 11 - 17.

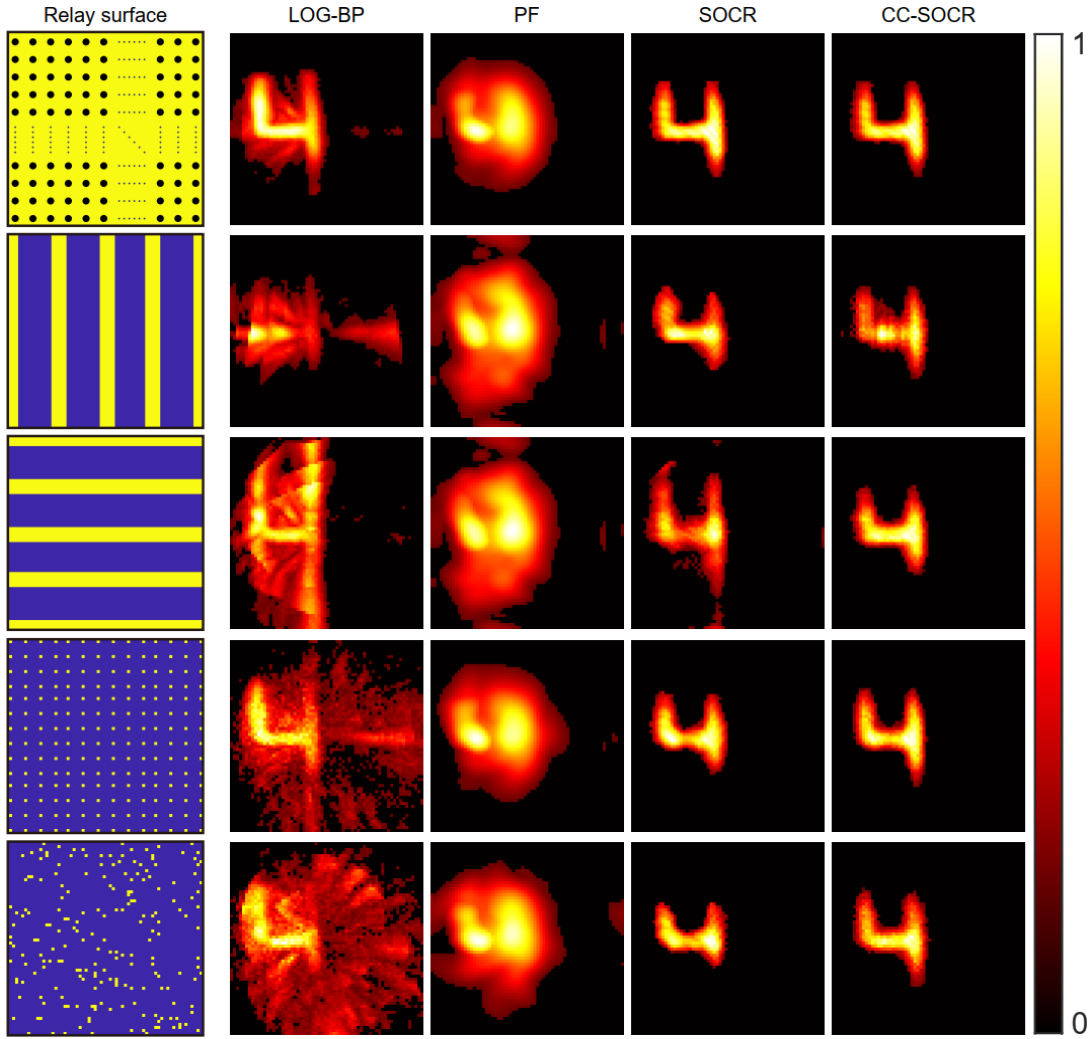

**Fig. 7 Non-confocal reconstruction results of the instance of figure '4' (non-confocal, measured signal).** The illumination regions are shown in yellow in the first column. Reconstructed albedo of the LOG-BP, PF, SOCR and CC-SOCR are compared in the second to fifth columns.

For non-confocal experiments, we use the measured data of the instance of figure '4' provided by the phasor field method[32]. The hidden object is 1 m away from the visible wall. The temporal resolution is 16 ps. We pick out the signal measured at 64 × 64 illumination points in a square region of size 1.27 × 1.27 m². The detection point is 0.64 m to the left and 0.55 m to the bottom of the illumination region. Except for the signal selected, we also use four subsets of the signal to reconstruct the target: signals measured at five equispaced vertical bars that contain 3, 5, 5, 5, and 3 columns of focal points from left to right; signals measured at five equispaced horizontal bars that contain 3, 5, 5, 5, and 3 rows of focal points from bottom to top; signals measured at $14 \times 14$ uniformly distributed focal points in an area of $1.27 \times 1.27$ m$^2$; signals measured at 200 randomly chosen focal points. To bring the PF and SOCR methods into comparison, the nearest neighbor interpolation technique is applied to extend the signal to $64 \times 64$ illuminations. As shown in Fig. 7, the LOG-BP reconstructions are noisy and contain heavy background noise. The PF reconstructions are blurry. Both SOCR and CC-SOCR methods reconstruct the target successfully. However, the SOCR reconstructions contain artifacts (the third row) or loss some details (the fourth and fifth rows). These results also indicate that the bias of the signal obtained from the nearest neighbor interpolation leads to non-negligible reconstruction error. The proposed CC-SOCR method provides faithful reconstructions in all cases. The three views and surface normal of the reconstructions are provided in Supplementary Figures 18 – 22.

## Discussion

We have proposed a novel framework towards the most general setting of NLOS imaging. In this section, we discuss its relationship with the original SOCR method, the complexity of the algorithm and possible directions for further improvements.

**The relationship between CC-SOCR and SOCR.** The SOCR method reconstructs the albedo and surface normal of the hidden targets under both confocal and non-confocal settings. However, the experimental setup is still quite limited. As demonstrated in the original paper[34], it only deals with signals measured at regular grid points. This is due to the spatial correlation of the signals in the regularization term.

The proposed CC-SOCR method generalizes the SOCR method to the most general setup, where no limitations of the measurement pairs are required. The CC-SOCR differs from SOCR in three aspects. Firstly, the introduced virtual confocal signal overcomes the rank deficiency of the measurement matrix, making it capable of reconstructing the targets under more general settings. Secondly, CC-SCOR does not include spatial correlations of the measured signal in the regularization term. As discussed in the method section, in CC-SOCR, the Wiener filter is applied only to the temporal dimension of the measured signal. Thirdly, the priors imposed on the target are related not only to the measured data but also to the introduced virtual confocal signal. Concrete expressions of these regularization terms are provided in the method section.

**Convergence of the CC-SOCR algorithm.** The proposed optimization problem can be solved efficiently using the alternative iteration method. In Supplementary Note 2, we decompose the problem into several sub-problems and discuss in detail the solutions to each sub-problem. We also provide a guide for choosing parameters in Supplementary Note 3. Convergences of all sub-problems are guaranteed, as discussed in the work of the SOCR method[34] and Supplementary Note 2. However, global convergence is not guaranteed because the sub-problem of updating the reconstructed target is solved approximately. Nonetheless, extensive results in Supplementary Note 1 have demonstrated the capability of the proposed method in providing high-quality

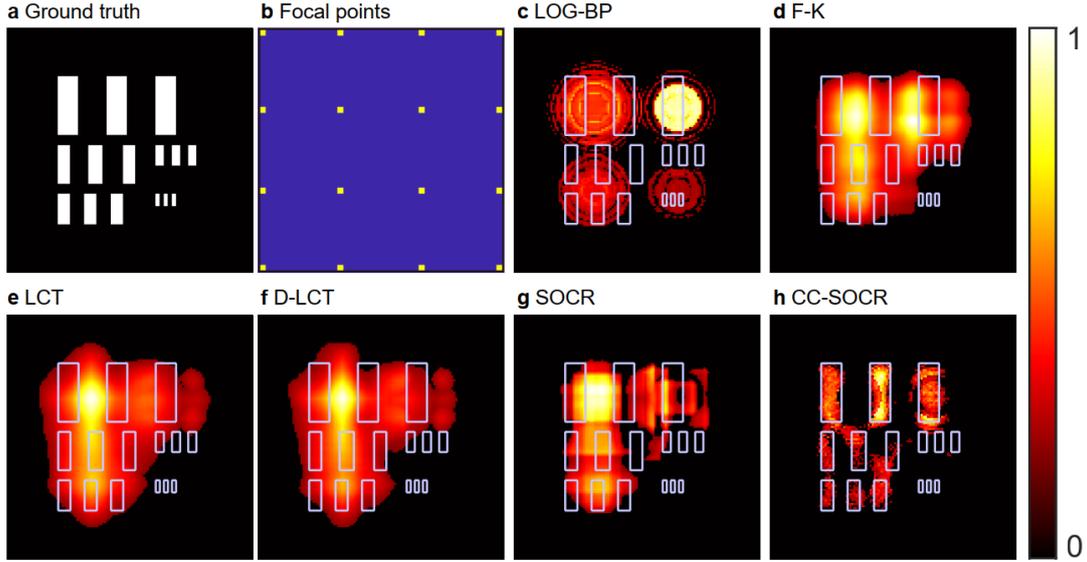

**Fig. 8 Reconstructions of the USAF resolution chart with 4 × 4 focal points (confocal, synthetic signal). a** Ground truth. **b** The confocal signal is measured at the yellow points. **c** Reconstructed albedo of the LOG-BP algorithm. **d** Reconstructed albedo of the F-K algorithm. **e** Reconstructed albedo of the LCT algorithm. **f** Reconstructed albedo of the D-LCT algorithm. **g** Reconstructed albedo of the SOCR algorithm. **h** Reconstructed albedo of the CC-SOCR algorithm.

reconstructions in various scenarios.

**Time and memory complexity.** When the reconstruction domain is discretized with $N \times N \times N$ voxels and the signal is detected at $M$ measurement pairs, the time complexity and memory complexity of the CC-SOCR method are $\mathcal{O}(\max\{N^5, MN^3\})$ and $\mathcal{O}(\max\{N^3, MN\})$ respectively. In Supplementary Note 4, we provide a detailed discussion of the complexity and report the running time for the instance of the statue with 200 randomly distributed confocal measurements. For the spatial case of $N \times N$ confocal measurements, the time complexity and memory complexity are $\mathcal{O}(N^5)$ and $\mathcal{O}(N^3)$, which is the same as the SOCR algorithm. If $N \times N$ points are exhaustively scanned, the corresponding time complexity and memory complexity are $\mathcal{O}(N^7)$ and $\mathcal{O}(N^5)$ respectively.

**Other types of virtual signals.** In CC-SOCR, virtual confocal signals observed at planar rectangular grid points are used to complement the reconstruction process in the case of incomplete measurements. It is also possible to consider virtual non-confocal signals for stronger regularizations. Besides, virtual confocal signals at several planes may be introduced to use the spatial correlation better. However, the time and memory complexities will also increase.

**Virtual confocal signals at coarse grids.** In the CC-SOCR method, the time complexity is still $\mathcal{O}(N^5)$ due to the virtual signal introduced, even when the number of measurement pairs $M=\mathcal{O}(1)$. To accelerate the reconstruction process, the virtual signal at coarser grids may be used. If the virtual confocal signal is considered at $\sqrt{N}\times\sqrt{N}$ points, the time complexity reduces to $\mathcal{O}(N^4)$. In Supplementary Figure 23, we compare the reconstruction results of the statue with virtual signals of different sizes. The excution time is provided in Supplementary Tables 4 - 7.

**The necessity of the virtual confocal signal.** With sufficient measurements, both the SOCR method and the CC-SOCR method provide high-quality reconstructions (See Supplementary Figures 1, 11, 18). However, when the number of measurement pairs is small, the reconstruction problem is ill-posed. Although the complete signal can be obtained with zero padding or nearest neighbor interpolation techniques, existing methods still fail due to the bias introduced in the signal (Supplementary Figure 24). The introduced virtual confocal signal benefits from the regularization guided by the simulated signal of the target and provides faithfull reconstructions. In the absence of the virtual confocal signal, the reconstructions may be blurry (Supplementary Figure 25) or contain artifacts in the background (Supplementary Figure 26).

Table 2: Reconstruction errors of the USAF resolution chart with 4 × 4 confocal measurements

| Methods | Missing voxels | Excessive voxels | Classification error | albedo PSNR | albedo SSIM |
|---------|----------------|------------------|----------------------|-------------|-------------|
| LOG-BP  | 496            | 2204             | 18.44%               | 10.6        | 0.59        |
| FK      | 152            | 2916             | 20.95%               | 10.05       | 0.61        |
| LCT     | **130**        | 3359             | 23.83%               | 10.41       | 0.57        |
| D-LCT   | 221            | 3112             | 22.76%               | 10.4        | 0.58        |
| SOCR    | 646            | 2027             | 18.26%               | 9.89        | 0.63        |
| CC-SOCR | 518            | **482**          | **6.83%**            | **11.72**   | **0.71**    |

**Limitations on the number of measurement pairs.** Although the proposed method does not impose limitations on the spatial pattern or the number of measurement pairs, the target cannot be clearly reconstructed with very few measurements. Figure 8 shows reconstruction results of the USAF resolution chart from the Zaragoza dataset[38] with only 4×4 focal points. All methods fail to reconstruct the whole target, while the proposed method reconstructs more details than other algorithms. The number of missing and excessive voxels in the reconstructed target, classification error, albedo PSNR and SSIM are compared in Table 2. The CC-SOCR reconstruction has the highest classification accuracy, as well as PSNR and SSIM values. In this extreme case, stronger regularizations of the target and the signal may be used for better reconstructions.

## Materials and methods

**The joint regularizations.** In equation (7), we formulate the CC-SOCR framework as an optimization problem. Here we show how the regularization terms $\Gamma(\mathbf{u},\mathbf{b})$, $\Xi(\mathbf{u},\mathbf{d})$ and $\Upsilon(\mathbf{u},\mathbf{b},\tilde{\mathbf{b}})$ are designed.

$\Gamma(\mathbf{u},\mathbf{b})$ describes the prior distribution of the reconstructed target and the approximated signal of the measurement pairs. For the reconstructed target, we consider the sparsity and non-local self-similarity priors and directly follow the SOCR method[34,39,40]. We also use the zero norm to impose sparseness on the approximated signal $\mathbf{b}$. We set

$$\Gamma(\mathbf{u},\mathbf{b}) = s_u \|\mathbf{L}\|_1 + \lambda_u \sum_i \left[ \|B_i(\mathbf{L}) - D_s C_i D_n^T\|^2 + \lambda_{pu} |C_i|_0 \right] + s_b |\mathbf{b}|_0 \qquad (8)$$

in which $s_u$, $\lambda_u$, $\lambda_{pu}$ and $s_b$ are fixed parameters. $\mathbf{L}$ is the albedo of $\mathbf{u}$, $B_i$ is the block matching operator, with $i$ the index of a reference block. The summation is made over all possible blocks. $D_s$ and $D_n$ are two orthogonal matrices that capture the local structure and non-local correlations of the 3D albedo block. $C_i$ is the matrix consisting of transform coefficients of the $i^{th}$ block. $|\cdot|_0$ denotes the zero norm, which represents the number of nonzero values of a tensor.

For the term $\Upsilon(\mathbf{u},\mathbf{b},\tilde{\mathbf{b}})$, we also follow the original SOCR method and set

$$\Upsilon(\mathbf{u},\mathbf{b},\tilde{\mathbf{b}}) = \sum_i \|P_i(\tilde{\mathbf{b}}) - DS_i\|^2 + \sum_{i,j} \left( \frac{\sigma_\mathbf{b}}{d_j^T P_i(A_\mathbf{b}\mathbf{u})} S_i(j) \right)^2 + \lambda_{sb} \sum_i \|P_i(\mathbf{b}) - DS_i\|^2 \qquad (9)$$

in which $\lambda_{sb}$ is a fixed parameter, $P_i$ is the patch extracting operator, with $i$ the index of a local patch. Noting that the signals may not be measured at regular grid points, the patch extracting operator $P_i$ only applies to the temporal direction of the signals. $\tilde{\mathbf{b}}$ is the measured signal. $D$ is the matrix of discrete cosine transform. The $j^{th}$ filter of $D$ is denoted by $d_j$. $A_\mathbf{b}$ is the measurement matrix. $S_i$ is the vector that consists of Wiener coefficients of the $i^{th}$ patch, with its $j^{th}$ element denoted by $S_i(j)$. $\sigma_\mathbf{b}$ is the noise level. The summations are made over all possible patches and filters of the discrete cosine matrix.

For the regularization term $\Xi(\mathbf{u},\mathbf{d})$, the prior of the virtual confocal signal $\mathbf{d}$ is constructed under the guidance of the target $\mathbf{u}$ and the physical model $A_\mathbf{d}$. Noting that the confocal signal $\mathbf{d}$ is considered at rectangular grid points, both the spatial and temporal correlations can be used. Let $P_i$ be the 3D patch extracting operator (2D in space and 1D in time), we seek a data-driven orthogonal dictionary $\Psi$ that sparsely represents the local patches of both the approximated signal $\mathbf{d}$ and the simulated signal $A_\mathbf{d}\mathbf{u}$. For simplicity, we abuse the notation $P_i$ to represent either a 1D patch of the measured signal $\mathbf{b}$ in the temporal direction or a 3D patch of the virtual confocal signal. The meaning can be made clear from the variable to which it applies. Let $Q_i$ be the matrix of transform coefficients of the $i^{th}$ patch, the regularization term is given by

$$\Xi(\mathbf{u},\mathbf{d}) = s_d |\mathbf{d}|_0 + \sum_i \left[ \|Q_i - \Psi^T P_i(\mathbf{d})\|^2 + \lambda_{sd} \|Q_i - \Psi^T P_i(A_\mathbf{d}\mathbf{u})\|^2 + \lambda_{fd} |Q_i|_0 \right] \quad (10)$$

in which $\lambda_{sd}$ and $\lambda_{fd}$ are two fixed parameters that control the weight of the simulated signal and the sparsity of the representation, respectively.

**The CC-SOCR optimization problem.** By substituting equations (8), (9), (10) into equation (7) and introducing weights, we obtain the concrete optimization problem of the proposed CC-SOCR framework as follows.

$$\min_{\mathbf{u},\mathbf{b},\mathbf{d},D_s,D_n,\mathbf{C},\mathbf{S},\Psi,\mathbf{Q}} \|A_\mathbf{b}\mathbf{u}-\mathbf{b}\|^2 + s_u\|\mathbf{L}\|_1 + s_b|\mathbf{b}|_0$$

$$+ \lambda_u \sum_i \left[\|B_i(\mathbf{L}) - D_s C_i D_n^T\|^2 + \lambda_{pu}|C_i|_0\right]$$

$$+ \lambda_b \|\mathbf{b}-\tilde{\mathbf{b}}\|^2 + \lambda_b\lambda_{pb} \sum_i \|P_i(\tilde{\mathbf{b}}) - DS_i\|^2$$

$$+ \lambda_b\lambda_{pb} \sum_{i,j} \left[\frac{\sigma_\mathbf{b}}{d_j^T P_i(A_\mathbf{b}\mathbf{u})} S_i(j)\right]^2$$

$$+ \lambda_b\lambda_{pb}\lambda_{sb} \sum_i \|P_i(\mathbf{b}) - DS_i\|^2$$

$$+ \lambda_d \|A_\mathbf{d}\mathbf{u}-\mathbf{d}\|^2 + s_d|\mathbf{d}|_0$$

$$+ \lambda_d\lambda_{pd} \sum_i \|Q_i - \Psi^T P_i(\mathbf{d})\|^2$$

$$+ \lambda_d\lambda_{pd}\lambda_{sd} \sum_i \|Q_i - \Psi^T P_i(A_\mathbf{d}\mathbf{u})\|^2$$

$$+ \lambda_d\lambda_{pd}\lambda_{fd} \sum_i |Q_i|_0$$

$$+ \lambda_{bd} \|R_\mathbf{b}(\mathbf{b},\mathbf{d}) - R_\mathbf{d}(\mathbf{b},\mathbf{d})\|^2$$

$$\text{s.t.} \quad \mathbf{L} = \text{albedo}(\mathbf{u}),$$
$$D_s^T D_s = I[p_x p_y p_z],\ D_n^T D_n = I[r], \quad (11)$$
$$\Psi^T \Psi = I[q_y q_z q_t]$$

in which $\mathbf{C}$, $\mathbf{S}$ and $\mathbf{Q}$ represent the collections of the transform-domain coefficients $\{C_i\}$, $\{S_i\}$ and $\{Q_i\}$ respectively. $I[n]$ represents the identity matrix of order $n$. $p_x$, $p_y$ and $p_z$ are the patch sizes of the albedo in the depth, horizontal and vertical directions. $r$ is the number of neighboring blocks of each reference albedo block. $q_y$, $q_z$ and $q_t$ are the patch sizes of the virtual confocal signal $\mathbf{d}$ in the horizontal, vertical and temporal directions. $\sigma_\mathbf{b}$ is a parameter related to the noise level of the measured signal. The fixed parameters $s_u$, $s_b$, $s_d$, $\lambda_u$, $\lambda_b$, $\lambda_d$, $\lambda_{pu}$, $\lambda_{pb}$, $\lambda_{pd}$, $\lambda_{sb}$, $\lambda_{sd}$, $\lambda_{fd}$, $\lambda_{bd}$ balance the data-fitting terms and the regularization terms. Detailed explainations of the variables and parameters are provided in Supplementary Tables 1 - 3. The data fitting and regularization terms of equation (11) are illustrated in Fig. 2a. The solution to the optimization problem is provided in Supplementary Note 2.

**Data availability**

The Zaragoza dataset is available in *Zaragoza NLOS synthetic dataset*

[http://graphics.unizar.es/nlos_dataset.html].

The Stanford dataset can be downloaded at the project page

[http://www.computationalimaging.org/publications/nlos-fk/].

The dataset provided by the phasor field method is available at the project page

[https://biostat.wisc.edu/~compoptics/phasornlos20/fastnlos.html].

Synthetic data of the instance of the pyramid are attached to the code.

**Code availability**

The code will be made freely available in the future.

**Acknowledgements**

This work was supported by the National Natural Science Foundation of China (61975087, 12071244, 11971258).

**Author Contributions**

X.L. conceived the idea of the CC-SOCR method and implemented the code. X.L. and J.W. ran the experiments. All authors discussed the results and contributed to the writing of the manuscript.

**Conflict of Interest**

The authors declare that they have no conflict of interest.

**Supplementary information** accompanies the manuscript.

# References

1. Velten, A. *et al.* Recovering three-dimensional shape around a corner using ultrafast time-of-flight imaging. *Nat. Commun.* **3**, 745 (2012).

2. Arellano, V., Gutierrez, D. & Jarabo, A. Fast Back-Projection for Non-Line of Sight Reconstruction. in *ACM SIGGRAPH 2017 Posters* (Association for Computing Machinery, 2017). doi:10.1145/3102163.3102241.

3. Jarabo, A., Masia, B., Marco, J. & Gutierrez, D. Recent advances in transient imaging: A computer graphics and vision perspective. *Vis. Inform.* **1**, 65–79 (2017).

4. Heide, F. *et al.* Non-Line-of-Sight Imaging with Partial Occluders and Surface Normals. *ACM Trans Graph* **38**, (2019).

5. Thrampoulidis, C. *et al.* Exploiting Occlusion in Non-Line-of-Sight Active Imaging. *IEEE Trans. Comput. Imaging* **4**, 419–431 (2018).

6. Ahn, B., Dave, A., Veeraraghavan, A., Gkioulekas, I. & Sankaranarayanan, A. Convolutional Approximations to the General Non-Line-of-Sight Imaging Operator. in *2019 IEEE/CVF International Conference on Computer Vision (ICCV)* 7888–7898 (IEEE, 2019). doi:10.1109/ICCV.2019.00798.

7. Chen, W., Daneau, S., Brosseau, C. & Heide, F. Steady-State Non-Line-Of-Sight Imaging. in *2019 IEEE/CVF Conference on Computer Vision and Pattern Recognition (CVPR)* 6783–6792 (IEEE, 2019). doi:10.1109/CVPR.2019.00695.

8. Lindell, D. B., Wetzstein, G. & O'Toole, M. Wave-based non-line-of-sight imaging using fast *f-k* migration. *ACM Trans. Graph.* **38**, 1–13 (2019).

9. Liu, X. *et al.* Non-line-of-sight imaging using phasor-field virtual wave optics. *Nature* **572**, 620–623 (2019).

10. Pediredla, A., Dave, A. & Veeraraghavan, A. SNLOS: Non-line-of-sight Scanning through Temporal Focusing. in *2019 IEEE International Conference on*

*Computational Photography (ICCP)* 1–13 (IEEE, 2019). doi:10.1109/ICCPHOT.2019.8747336.

11. Chen, W., Wei, F., Kutulakos, K. N., Rusinkiewicz, S. & Heide, F. Learned feature embeddings for non-line-of-sight imaging and recognition. *ACM Trans. Graph.* **39**, 1–18 (2020).

12. Chopite, J. G., Hullin, M. B., Wand, M. & Iseringhausen, J. Deep Non-Line-of-Sight Reconstruction. in *Proceedings of the IEEE/CVF Conference on Computer Vision and Pattern Recognition (CVPR)* (2020).

13. Isogawa, M., Yuan, Y., O'Toole, M. & Kitani, K. Optical Non-Line-of-Sight Physics-Based 3D Human Pose Estimation. in *2020 IEEE/CVF Conference on Computer Vision and Pattern Recognition (CVPR)* 7011–7020 (IEEE, 2020). doi:10.1109/CVPR42600.2020.00704.

14. La Manna, M., Nam, J.-H., Azer Reza, S. & Velten, A. Non-line-of-sight-imaging using dynamic relay surfaces. *Opt. Express* **28**, 5331 (2020).

15. Liu, X. & Velten, A. The role of Wigner Distribution Function in Non-Line-of-Sight Imaging. in *2020 IEEE International Conference on Computational Photography (ICCP)* 1–12 (IEEE, 2020). doi:10.1109/ICCP48838.2020.9105266.

16. Liao, Z. *et al.* FPGA Accelerator for Real-Time Non-Line-of-Sight Imaging. *IEEE Trans. Circuits Syst. Regul. Pap.* 1–14 (2021) doi:10.1109/TCSI.2021.3122309.

17. Metzler, C. A., Lindell, D. B. & Wetzstein, G. Keyhole Imaging:Non-Line-of-Sight Imaging and Tracking of Moving Objects Along a Single Optical Path. *IEEE Trans. Comput. IMAGING* **7**, 12 (2021).

18. Pei, C. *et al.* Dynamic non-line-of-sight imaging system based on the optimization of point spread functions. *Opt Express* **29**, 32349–32364 (2021).

19. Wu, C. *et al.* Non–line-of-sight imaging over 1.43 km. *Proc. Natl. Acad. Sci.* **118**, e2024468118 (2021).

20. Yang, W., Zhang, C., Jiang, W., Zhang, Z. & Sun, B. None-line-of-sight imaging enhanced with spatial multiplexing. *Opt. Express* **30**, 5855 (2022).

21. Liu, X., Bauer, S. & Velten, A. Analysis of Feature Visibility in Non-Line-Of-Sight Measurements. in *2019 IEEE/CVF Conference on Computer Vision and Pattern Recognition (CVPR)* 10132–10140 (2019). doi:10.1109/CVPR.2019.01038.

22. Feng, X. & Gao, L. Ultrafast light field tomography for snapshot transient and non-line-of-sight imaging. *Nat. Commun.* **12**, 2179 (2021).

23. Geng, R. *et al.* Passive Non-Line-of-Sight Imaging Using Optimal Transport. *IEEE Trans. Image Process.* **31**, 110–124 (2022).

24. Sasaki, T., Hashemi, C. & Leger, J. R. Passive 3D location estimation of non-line-of-sight objects from a scattered thermal infrared light field. *Opt. Express* **29**, 43642 (2021).

25. La Manna, M. *et al.* Error Backprojection Algorithms for Non-Line-of-Sight Imaging. *IEEE Trans. Pattern Anal. Mach. Intell.* **41**, 1615–1626 (2019).

26. Li, Z. *et al.* Fast non-line-of-sight imaging based on first photon event stamping. *Opt. Lett.* **47**, (2022).

27. Buttafava, M., Zeman, J., Tosi, A., Eliceiri, K. & Velten, A. Non-line-of-sight imaging using a time-gated single photon avalanche diode. *Opt. Express* **23**, 20997

(2015).

28. Xin, S., Nousias, S., Kutulakos, K. N., Sankaranarayanan, A. C. & Gkioulekas, I. A Theory of Fermat Paths for Non-Line-Of-Sight Shape Reconstruction. in *2019 IEEE/CVF Conference on Computer Vision and Pattern Recognition (CVPR)* (2019).

29. Tsai, C.-Y., Sankaranarayanan, A. C. & Gkioulekas, I. Beyond Volumetric Albedo — A Surface Optimization Framework for Non-Line-Of-Sight Imaging. in *2019 IEEE/CVF Conference on Computer Vision and Pattern Recognition (CVPR)* 1545–1555 (IEEE, 2019). doi:10.1109/CVPR.2019.00164.

30. Matthew O'Toole, Lindell, D. B. & Wetzstein, G. Confocal non-line-of-sight imaging based on the light-cone transform. *Nature* **555**, 338–341 (2018).

31. Young, S. I., Lindell, D. B., Girod, B., Taubman, D. & Wetzstein, G. Non-Line-of-Sight Surface Reconstruction Using the Directional Light-Cone Transform. in *2020 IEEE/CVF Conference on Computer Vision and Pattern Recognition (CVPR)* 1404–1413 (IEEE, 2020). doi:10.1109/CVPR42600.2020.00148.

32. Liu, X., Bauer, S. & Velten, A. Phasor field diffraction based reconstruction for fast non-line-of-sight imaging systems. *Nat. Commun.* **11**, 1645 (2020).

33. Laurenzis, M. & Velten, A. Feature selection and back-projection algorithms for nonline-of-sight laser–gated viewing. *J. Electron. Imaging* **23**, 1–6 (2014).

34. Liu, X. *et al.* Non-line-of-sight reconstruction with signal–object collaborative regularization. *Light Sci. Appl.* **10**, 198 (2021).

35. Xu, F., Ye, J., Huang, X. & Li, Z.-P. Compressed sensing for non-line-of-sight imaging. *Opt. Express* **29**, (2020).

36. Feng, X. & Gao, L. Improving non-line-of-sight image reconstruction with weighting factors. *Opt. Lett.* **45**, 3921 (2020).

37. Nam, J. H. *et al.* Low-latency time-of-flight non-line-of-sight imaging at 5 frames per second. *Nat. Commun.* **12**, 6526 (2021).

38. Galindo, M., Marco, J., O'Toole, M., Wetzstein, G. & Jarabo, A. A dataset for benchmarking time-resolved non-line-of-sight imaging. in *ACM SIGGRAPH 2019 Posters* (2019).

39. Goldstein, T. & Osher, S. The Split Bregman Method for L1-Regularized Problems. *SIAM J. Imaging Sci.* **2**, 323–343 (2009).

40. D Abov, K., Foi, A., Katkovnik, V. & Egiazarian, K. Image denoising with block-matching and 3D filtering. in *Proceedings of SPIE - The International Society for Optical Engineering* 354–365 (2006).

# Supplementary Information

# Non-line-of-sight imaging with arbitrary illumination and detection pattern

**Authors**

Xintong Liu[1], Jianyu Wang[1], Leping Xiao[2,3], Zuoqiang Shi[1,4], Xing Fu[2,3,*], Lingyun Qiu[1,4,*]

**Affiliations**

[1] Yau Mathematical Sciences Center, Tsinghua University, Beijing, China, 100084

[2] State Key Laboratory of Precision Measurement Technology and Instruments, Department of Precision Instrument, Tsinghua University, Beijing, China, 100084

[3] Key Laboratory of Photonic Control Technology (Tsinghua University), Ministry of Education, Beijing, China, 100084

[4] Yanqi Lake Beijing Institute of Mathematical Sciences and Applications, Beijing, China, 101408

[*] Correspondence and requests for materials should be addressed to fuxing@tsinghua.edu.cn (Xing Fu) and lyqiu@tsinghua.edu.cn (Lingyun Qiu).

# Contents

**Supplementary Note 1: Additional experimental results**
Reconstruction results of different experimental setups are shown.

**Supplementary Note 2: The CC-SOCR algorithm**
An iterative algorithm is designed to solve the proposed CC-SOCR optimization problem.

**Supplementary Note 3: The choice of parameters**
A self-adaptive scheme is used to determine the parameters in the CC-SOCR algorithm.

**Supplementary Note 4: Time and memory complexity**
A theoretical analysis of time and memory complexity is provided.

# Supplementary Note 1   Additional experimental results

For all experiments, we interpolate the signals with the nearest neighbor method where necessary to bring F-K[1], LCT[2], D-LCT[3], PF[4] and SOCR[5] methods into comparison. The coordinates of the focal points of all experiments are provided in the code.

Supplementary Figures 1 - 5 compare the reconstruction results of the bunny under different relay settings with the synthetic confocal signal provided in the Zaragoza dataset[6]. These results indicate the capability of the proposed CC-SOCR method in providing clear reconstructions of the hidden targets, even in cases with highly irregular relay settings (See Supplementary Figures 4 and 5).

Supplementary Figures 6 - 10 show the reconstruction results of the statues with the confocal measured signal provided in the Stanford dataset[1]. The signals are measured at $10 \times 10$, $14 \times 14$, $16 \times 16$, $20 \times 20$, and $24 \times 24$ focal points in an area of $2 \times 2$ m2. The reconstruction quality of all methods increases with the number of measurements. The proposed CC-SCOR reconstructions are of high quality and do not contain artifacts or background noise.

Supplementary Figures 11 - 17 compare the complete reconstruction results of several methods for the instance of statue with different settings of the confocal measurements: $64 \times 64$ uniformly distributed focal points in an area of $2 \times 2$ m2; a cross-shaped region that contains 1392 focal points; a set consisting of 200 randomly chosen focal points; a region consisting of 5 vertical bars with 1344 focal points; a region that consists of four letters 'N', 'L', 'O' and 'S' with 825 focal points; a region made up of several sticks sparsely and randomly distributed with 1229 focal points; and a heart-shaped region consisting of 258 focal points. As shown in Supplementary Figures 11 b.5 and b.6, the CC-SOCR reconstruction is less noisy than the SOCR reconstruction compared with the SOCR reconstruction with complete measurements. For the case of a heart-shaped relay (Supplementary Figure 17), the CC-SOCR method locates the target correctly, while all existing methods.

Supplementary Figures 18 - 22 show reconstruction results of the figure '4' with the measured non-confocal signal provided by the phasor field method[4]. The detection point is fixed, which is 0.64 m to the left and 0.55 m to the bottom of the illumination region. We show reconstruction results under five different settings of the illumination regions: $64 \times 64$ uniformly distributed points in a square of size $1.27 \times 1.27$ m2; five equispaced vertical bars that contain 3, 5, 5, 5, and 3 columns of illumination points from left to right; five equispaced horizontal bars that contain 3, 5, 5, 5, and 3 rows of illumination points from bottom to top; $14 \times 14$ uniformly distributed points; and 200 randomly distributed points. Both the SOCR and CC-SOCR methods provide noiseless reconstructions of the target under all scenarios, while the results of the CC-SOCR method contain fewer artifacts (Supplementary Figure 20) and are more accurate than the SOCR reconstructions (Supplementary Figures 21 - 22).

Supplementary Figure 23 shows the reconstruction results of the statues with different sizes of the virtual confocal signals introduced. The confocal signal is measured at 200 randomly distributed focal points in a square region of $2 \times 2$ m2. The reconstruction quality decreases with the size of the virtual confocal signal, which indicates the necessity of the dense virtual signal introduced. However, sparser virtual signals result in shorter execution time (See Supplementary Tables 4 - 7), which shows the trade off between the reconstruction quality and computation runtime.

Supplementary Figure 24 compares the F-K, LCT, D-LCT and SOCR reconstructions of the statue with confocal signal measured at a heart-shaped region consisting of 258 focal points. The signal is preprocessed with zero padding and nearest neighbor interpolation techniques. It is shown that existing methods fail in this extreme case.

Supplementary Figures 25 and 26 show reconstruction results of the statue with confocal signals measured at the letters 'N', 'L', 'O' and 'S' and a heart-shaped region. It is shown that the least squares reconstruction without regularizations is of poor quality. When the sparsity and non-local self-similarity priors of the target are introduced, the quality of the reconstruction enhances, but is still blurry or contains artifacts. The CC-SOCR method reconstructs the target faithfully.

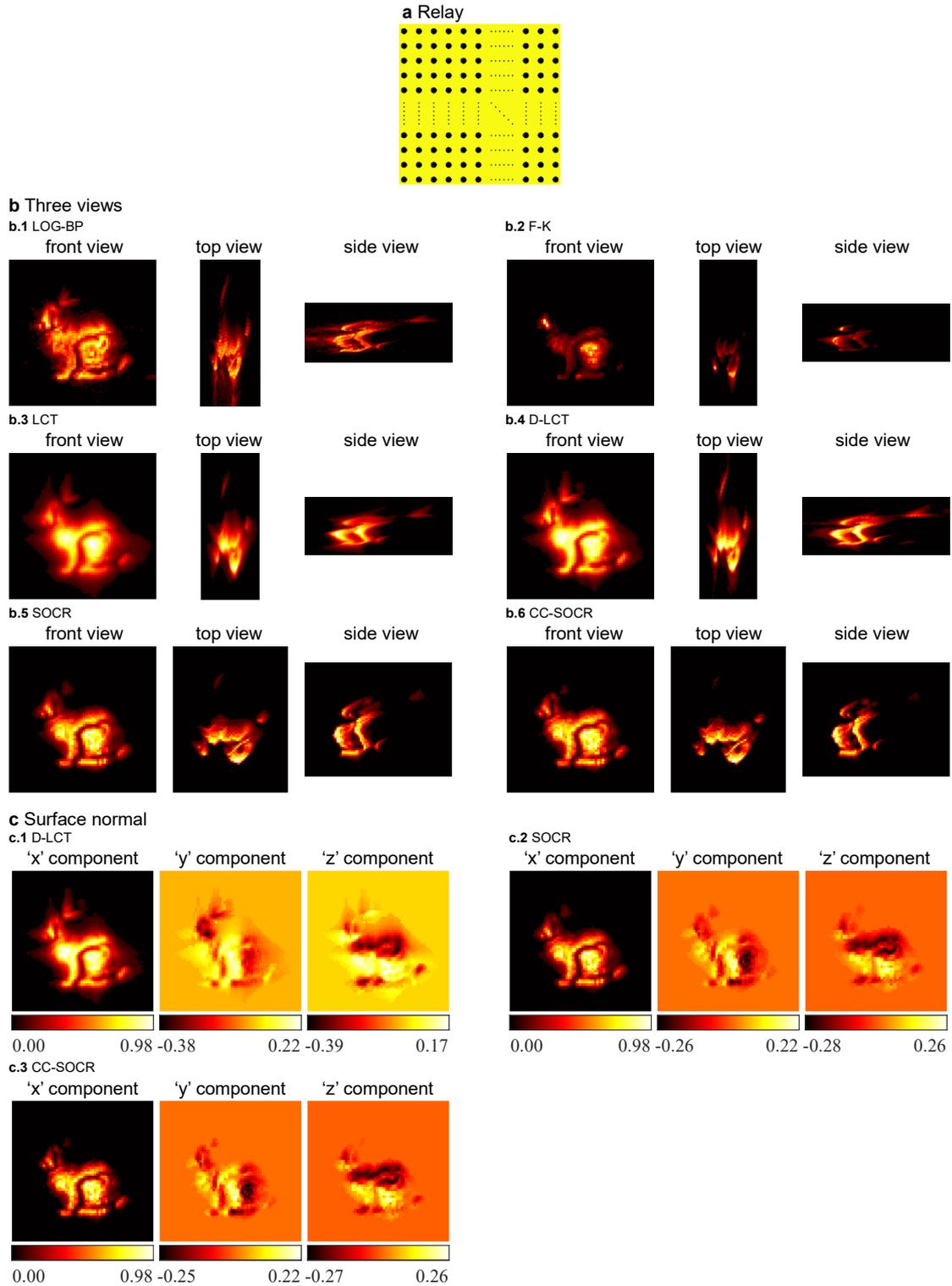

**Supplementary Figure 1 Reconstructions of the bunny with 64 × 64 measurements (confocal, synthetic signal). a** The relay is a square region. **b** Three views of the reconstructions. For all methods, the same reconstruction domain is shown. For the LOG-BP, F-K, LCT, and D-LCT methods, the length of the voxels in the depth direction is 0.125 cm. For the SOCR and CC-SOCR methods, the length of voxels in the depth direction is 0.25 cm. **c** The reconstructed surface normal of the D-LCT, SOCR and CC-SOCR methods is shown in the form of three components. The 'x', 'y' and 'z' components show values of the directional albedo in the depth, horizontal and vertical directions, respectively.

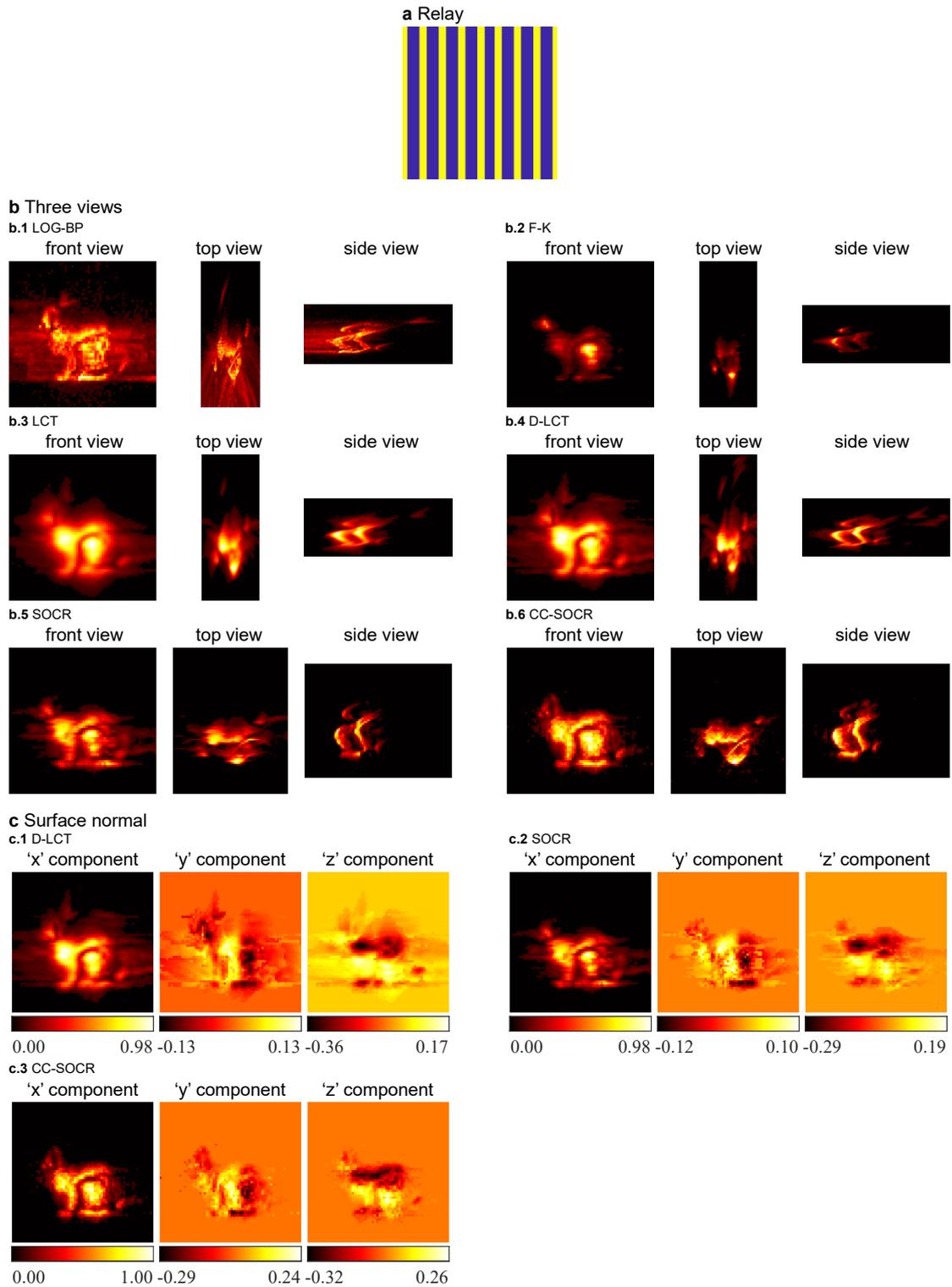

**Supplementary Figure 2 Reconstructions of the bunny with vertical bars as the relay (confocal, synthetic signal).** **a** The relay is made up of 9 equispaced vertical bars. **b** Three views of the reconstructions. For all methods, the same reconstruction domain is shown. For the LOG-BP, F-K, LCT, and D-LCT methods, the length of the voxels in the depth direction is 0.125 cm. For the SOCR and CC-SOCR methods, the length of voxels in the depth direction is 0.25 cm. **c** The reconstructed surface normal of the D-LCT, SOCR and CC-SOCR methods is shown in the form of three components. The 'x', 'y' and 'z' components show values of the directional albedo in the depth, horizontal and vertical directions, respectively.

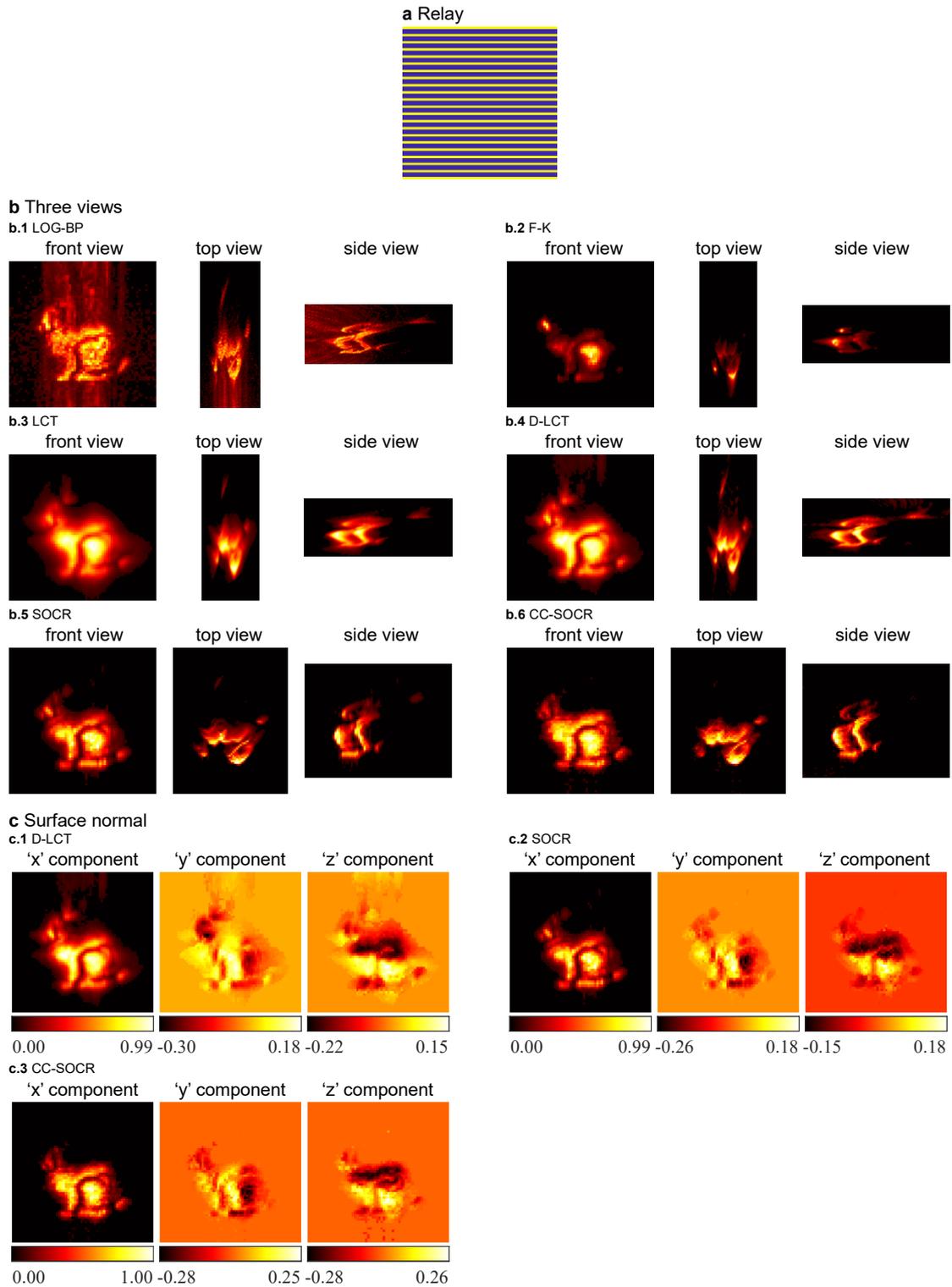

**Supplementary Figure 3 Reconstructions of the bunny with horizontal bars as the relay (confocal, synthetic signal). a** The relay is made up of 22 equispaced horizontal bars. **b** Three views of the reconstructions. For all methods, the same reconstruction domain is shown. For the LOG-BP, F-K, LCT, and D-LCT methods, the length of the voxels in the depth direction is 0.125 cm. For the SOCR and CC-SOCR methods, the length of voxels in the depth direction is 0.25 cm. **c** The reconstructed surface normal of the D-LCT, SOCR and CC-SOCR methods is shown in the form of three components. The 'x', 'y' and 'z' components show values of the directional albedo in the depth, horizontal and vertical directions, respectively.

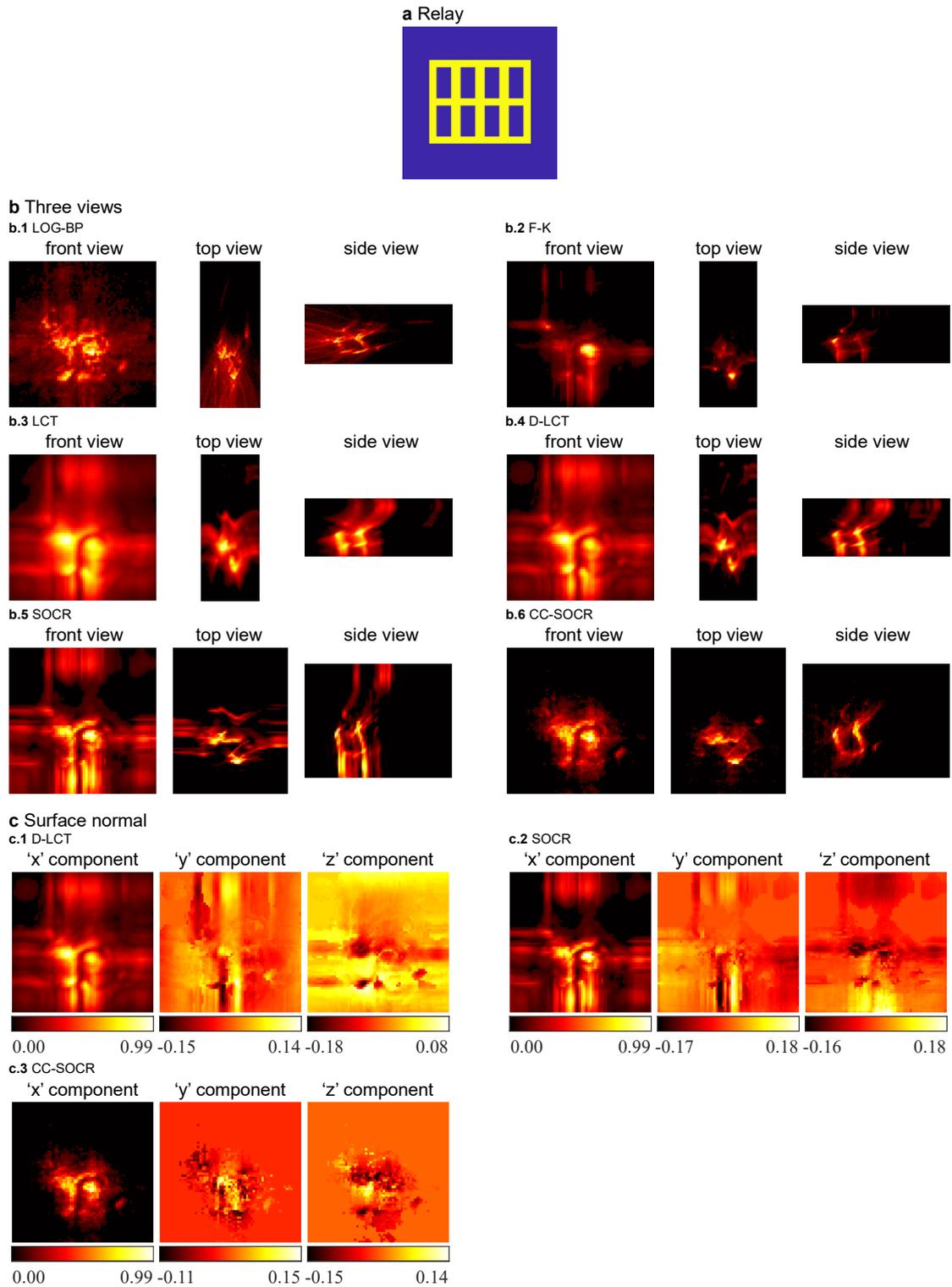

**Supplementary Figure 4 Reconstructions of the bunny with an array of window edges as the relay (confocal, synthetic signal). a** The relay is an array of window edges. **b** Three views of the reconstructions. For all methods, the same reconstruction domain is shown. For the LOG-BP, F-K, LCT, and D-LCT methods, the length of the voxels in the depth direction is 0.125 cm. For the SOCR and CC-SOCR methods, the length of voxels in the depth direction is 0.25 cm. **c** The reconstructed surface normal of the D-LCT, SOCR and CC-SOCR methods is shown in the form of three components. The 'x', 'y' and 'z' components show values of the directional albedo in the depth, horizontal and vertical directions, respectively.

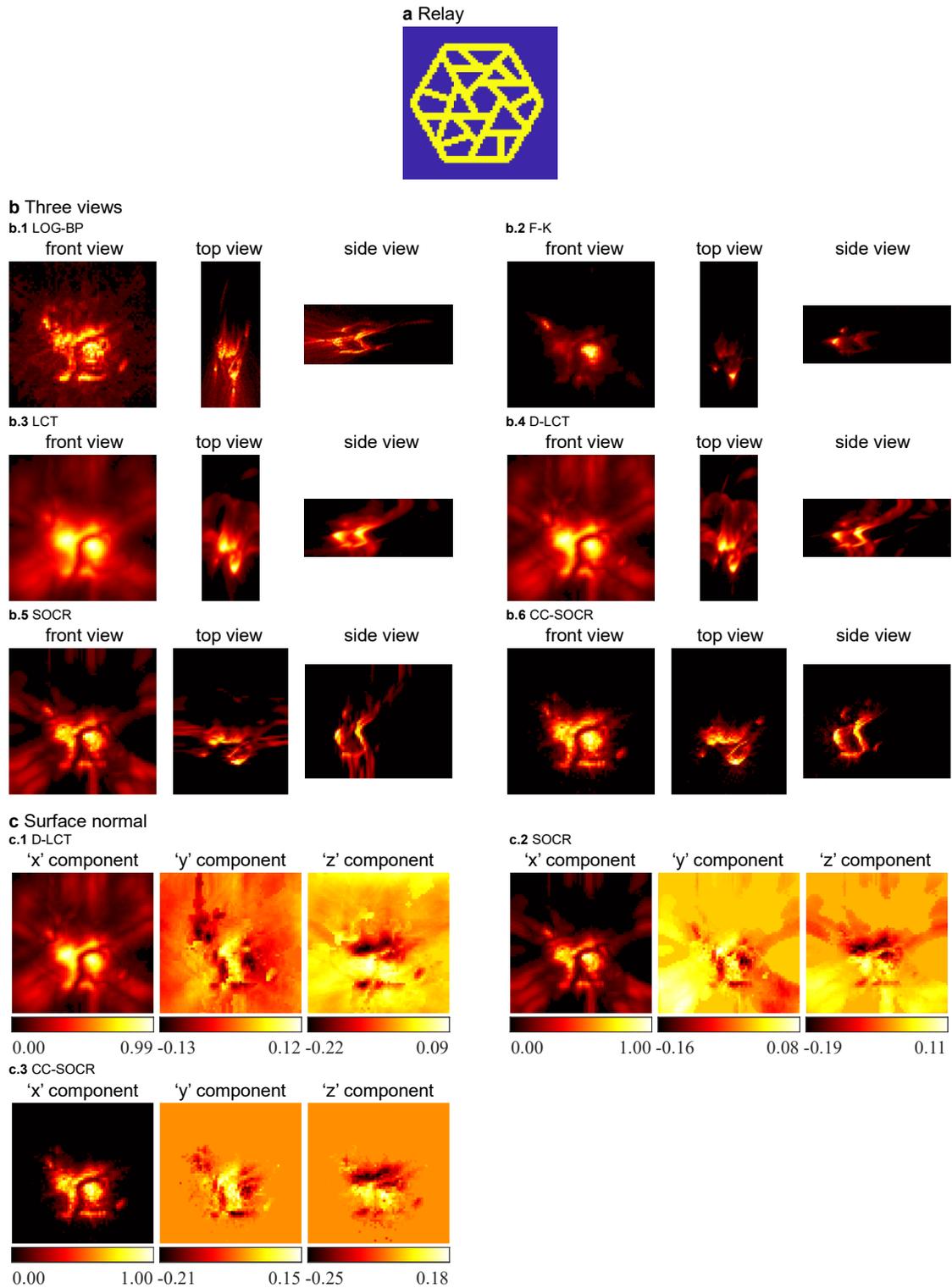

**Supplementary Figure 5 Reconstructions of the bunny with an irregular relay surface (confocal, synthetic signal). a** The relay is a set of several sticks sparsely and randomly distributed. **b** Three views of the reconstructions. For all methods, the same reconstruction domain is shown. For the LOG-BP, F-K, LCT, and D-LCT methods, the length of the voxels in the depth direction is 0.125 cm. For the SOCR and CC-SOCR methods, the length of voxels in the depth direction is 0.25 cm. **c** The reconstructed surface normal of the D-LCT, SOCR and CC-SOCR methods is shown in the form of three components. The 'x', 'y' and 'z' components show values of the directional albedo in the depth, horizontal and vertical directions, respectively.

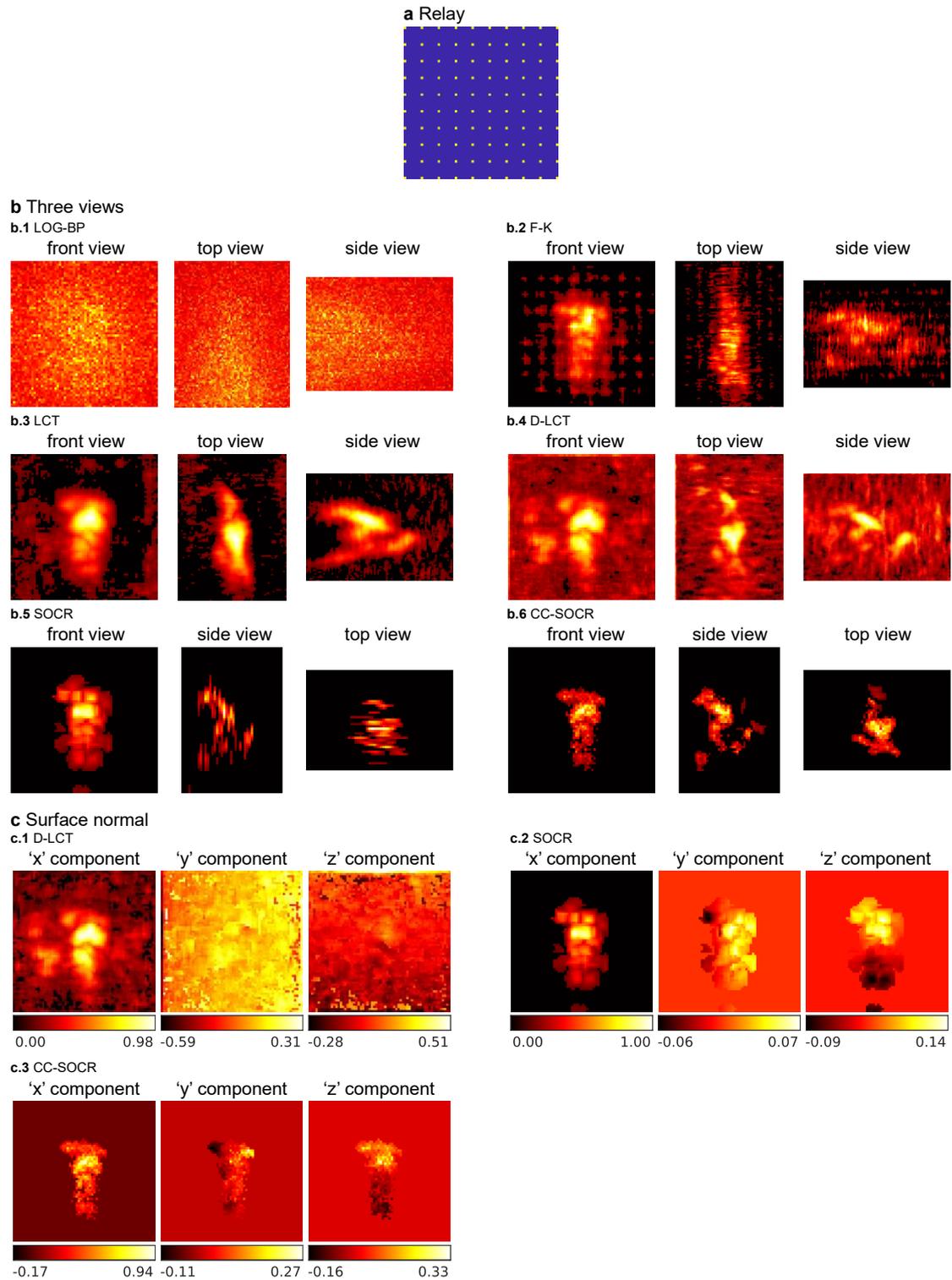

**Supplementary Figure 6 Reconstructions of the statue with 10 × 10 measurements (confocal, measured signal).** **a** Confocal signals are measured at 10 × 10 focal points. **b** Three views of the reconstructions. For all methods, the same reconstruction domain is shown. For the LOG-BP, F-K, LCT, and D-LCT methods, the length of the voxels in the depth direction is 0.48 cm. For the SOCR and CC-SOCR methods, the length of voxels in the depth direction is 0.96 cm. **c** The reconstructed surface normal of the D-LCT, SOCR and CC-SOCR methods is shown in the form of three components. The 'x', 'y' and 'z' components show values of the directional albedo in the depth, horizontal and vertical directions, respectively.

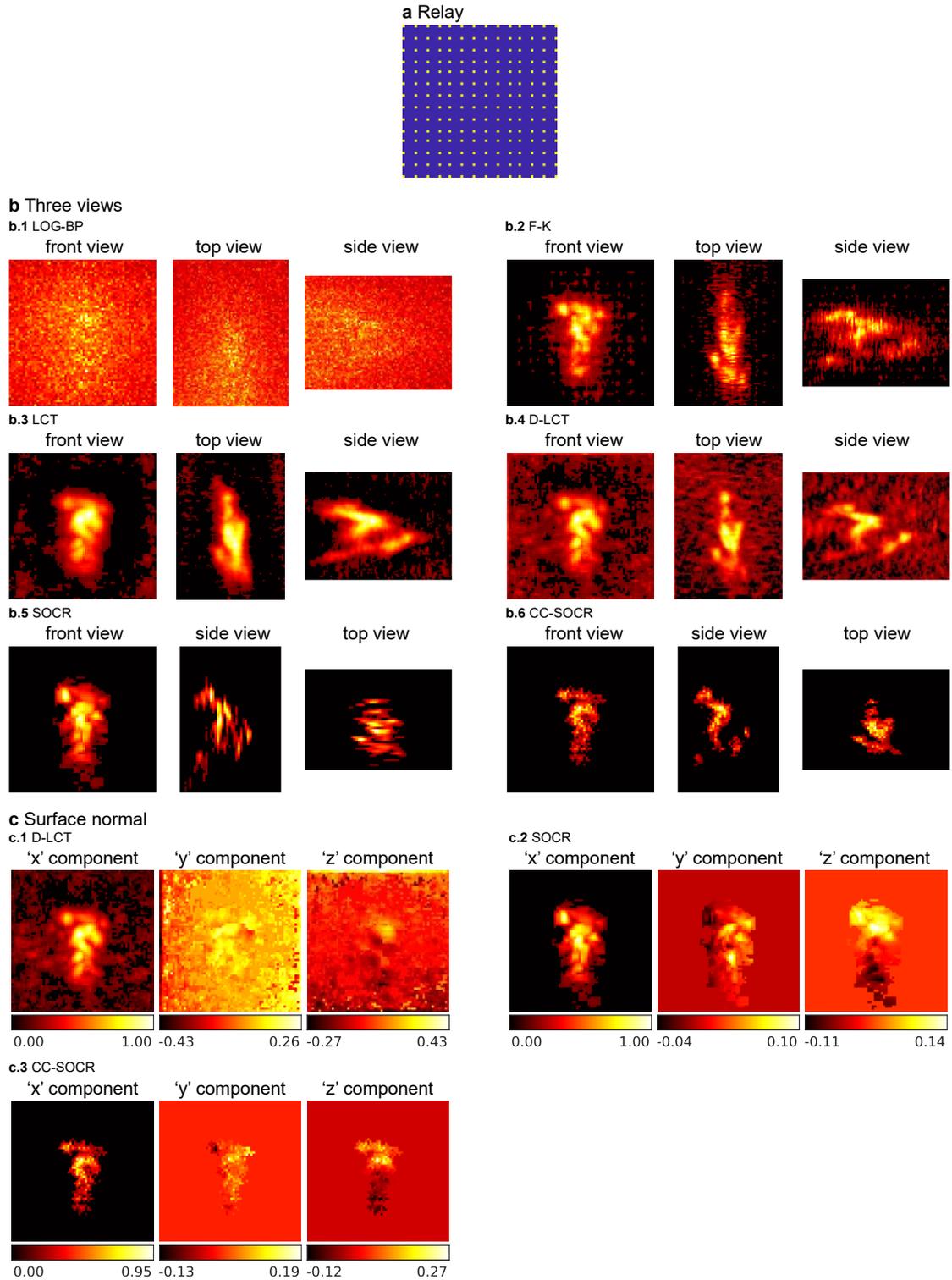

**Supplementary Figure 7 Reconstructions of the statue with 14 × 14 measurements (confocal, measured signal).** **a** Confocal signals are measured at 14 × 14 focal points. **b** Three views of the reconstructions. For all methods, the same reconstruction domain is shown. For the LOG-BP, F-K, LCT, and D-LCT methods, the length of the voxels in the depth direction is 0.48 cm. For the SOCR and CC-SOCR methods, the length of voxels in the depth direction is 0.96 cm. **c** The reconstructed surface normal of the D-LCT, SOCR and CC-SOCR methods is shown in the form of three components. The 'x', 'y' and 'z' components show values of the directional albedo in the depth, horizontal and vertical directions, respectively.

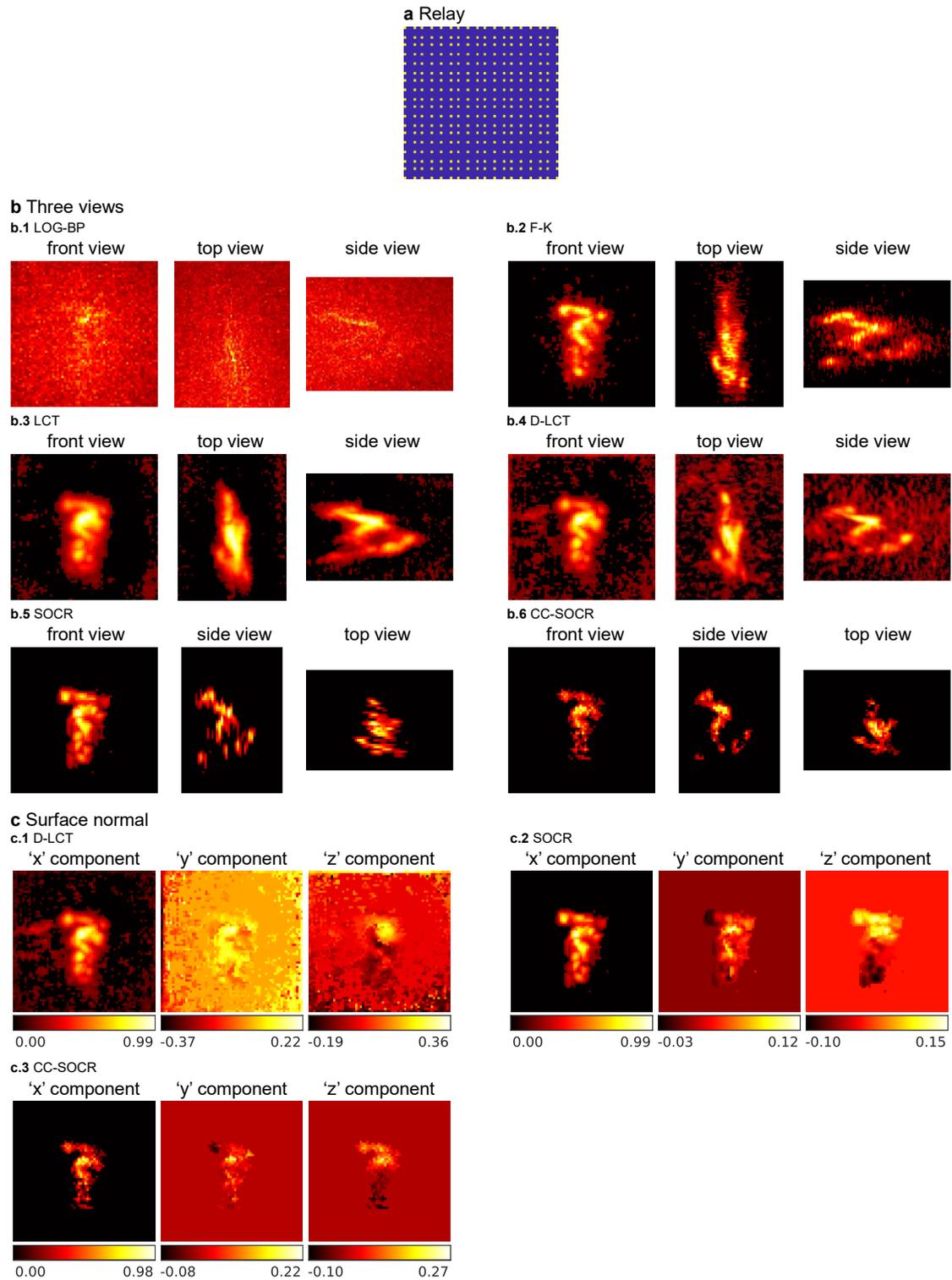

**Supplementary Figure 8 Reconstructions of the statue with 18 × 18 measurements (confocal, measured signal).** **a** Confocal signals are measured at 18 × 18 focal points. **b** Three views of the reconstructions. For all methods, the same reconstruction domain is shown. For the LOG-BP, F-K, LCT, and D-LCT methods, the length of the voxels in the depth direction is 0.48 cm. For the SOCR and CC-SOCR methods, the length of voxels in the depth direction is 0.96 cm. **c** The reconstructed surface normal of the D-LCT, SOCR and CC-SOCR methods is shown in the form of three components. The 'x', 'y' and 'z' components show values of the directional albedo in the depth, horizontal and vertical directions, respectively.

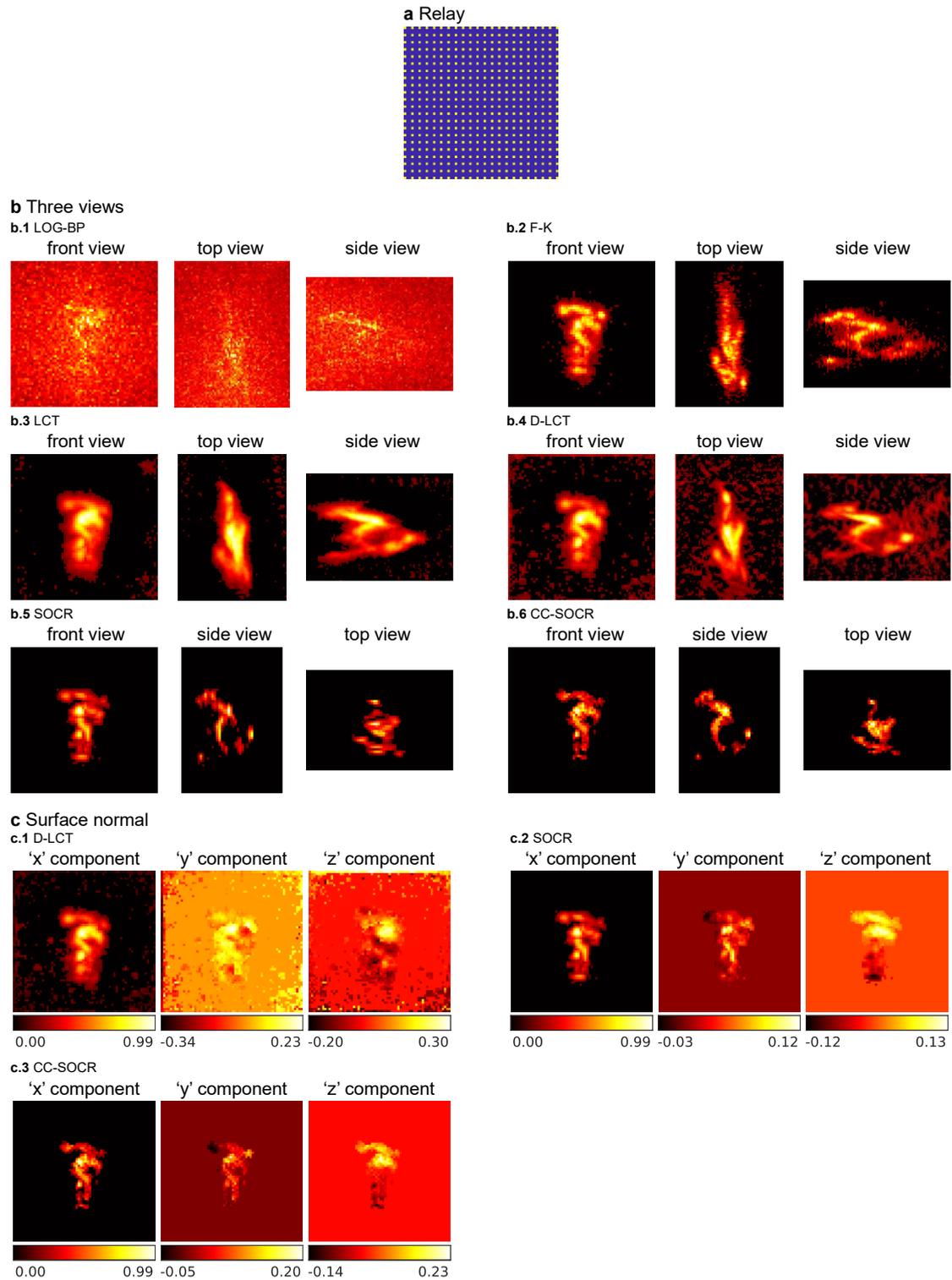

**Supplementary Figure 9 Reconstructions of the statue with 22 × 22 measurements (confocal, measured signal).** **a** Confocal signals are measured at 22 × 22 focal points. **b** Three views of the reconstructions. For all methods, the same reconstruction domain is shown. For the LOG-BP, F-K, LCT, and D-LCT methods, the length of the voxels in the depth direction is 0.48 cm. For the SOCR and CC-SOCR methods, the length of voxels in the depth direction is 0.96 cm. **c** The reconstructed surface normal of the D-LCT, SOCR and CC-SOCR methods is shown in the form of three components. The 'x', 'y' and 'z' components show values of the directional albedo in the depth, horizontal and vertical directions, respectively.

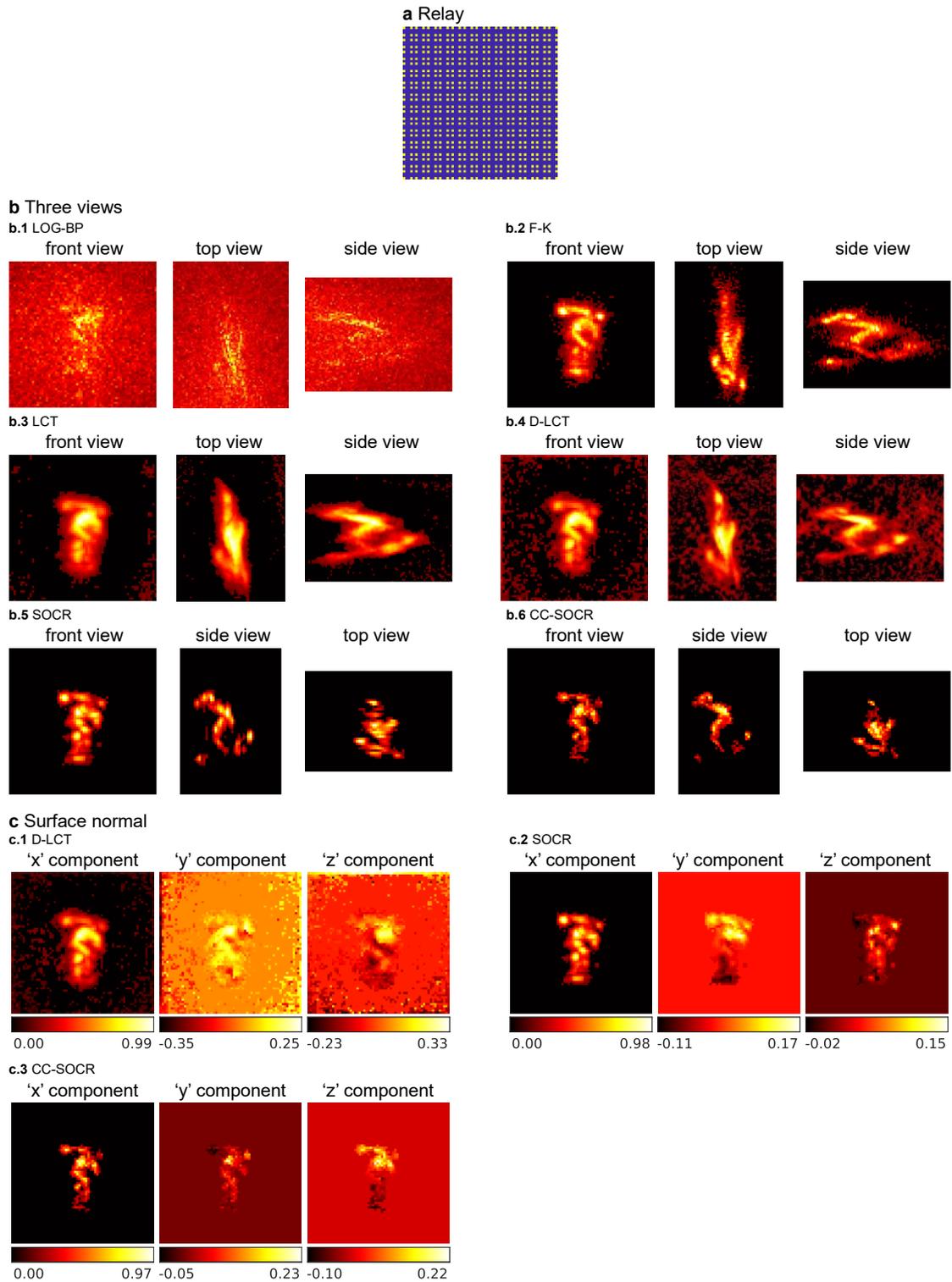

**Supplementary Figure 10 Reconstructions of the statue with 26 × 26 measurements (confocal, measured signal).** **a** Confocal signals are measured at 26 × 26 focal points. **b** Three views of the reconstructions. For all methods, the same reconstruction domain is shown. For the LOG-BP, F-K, LCT, and D-LCT methods, the length of the voxels in the depth direction is 0.48 cm. For the SOCR and CC-SOCR methods, the length of voxels in the depth direction is 0.96 cm. **c** The reconstructed surface normal of the D-LCT, SOCR and CC-SOCR methods is shown in the form of three components. The 'x', 'y' and 'z' components show values of the directional albedo in the depth, horizontal and vertical directions, respectively.

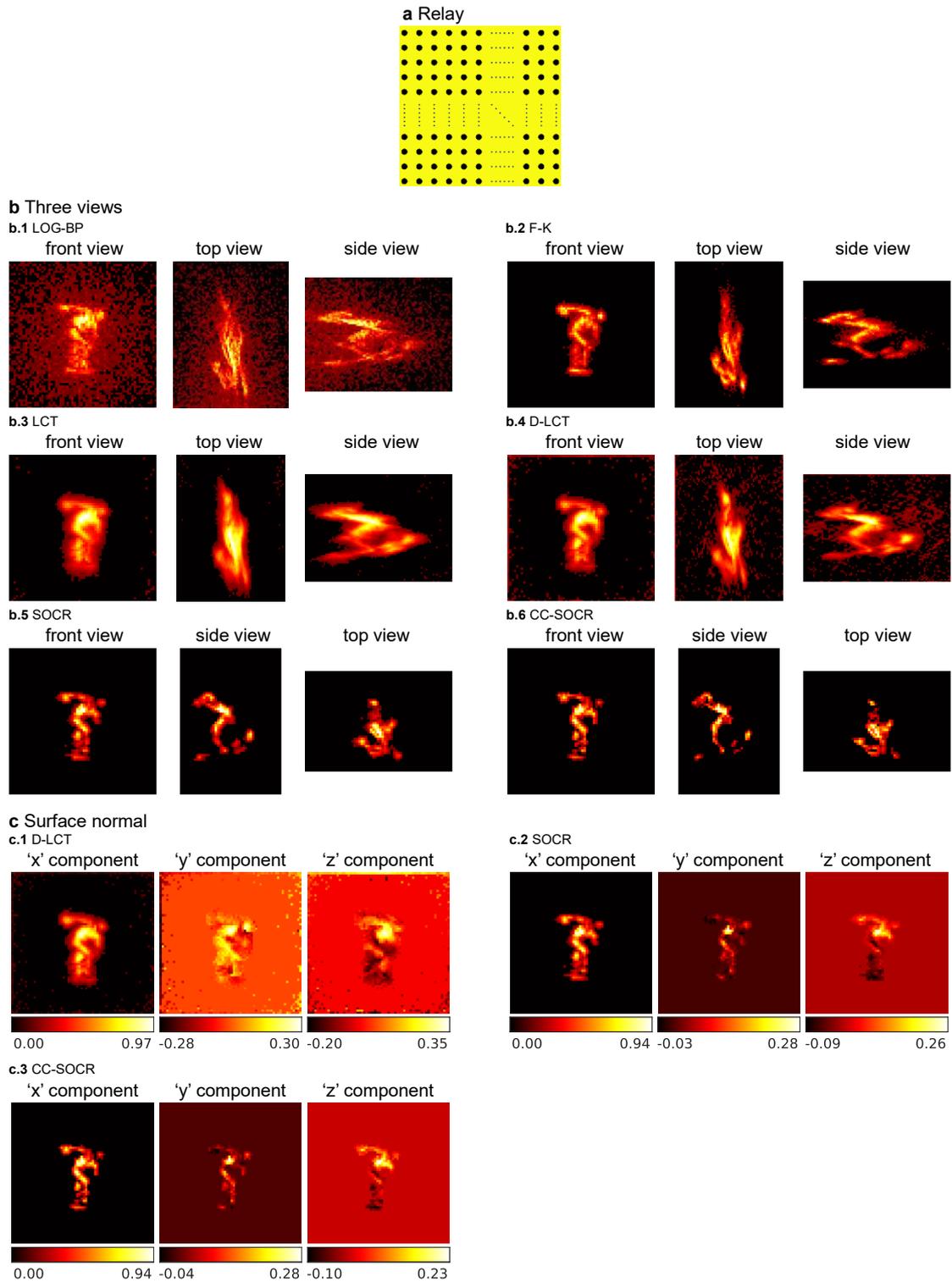

**Supplementary Figure 11 Reconstructions of the statue with 64 × 64 measurements (confocal, measured signal). a** Confocal signals are measured at 64 × 64 focal points. **b** Three views of the reconstructions. For all methods, the same reconstruction domain is shown. For the LOG-BP, F-K, LCT, and D-LCT methods, the length of the voxels in the depth direction is 0.48 cm. For the SOCR and CC-SOCR methods, the length of voxels in the depth direction is 0.96 cm. **c** The reconstructed surface normal of the D-LCT, SOCR and CC-SOCR methods is shown in the form of three components. The 'x', 'y' and 'z' components show values of the directional albedo in the depth, horizontal and vertical directions, respectively.

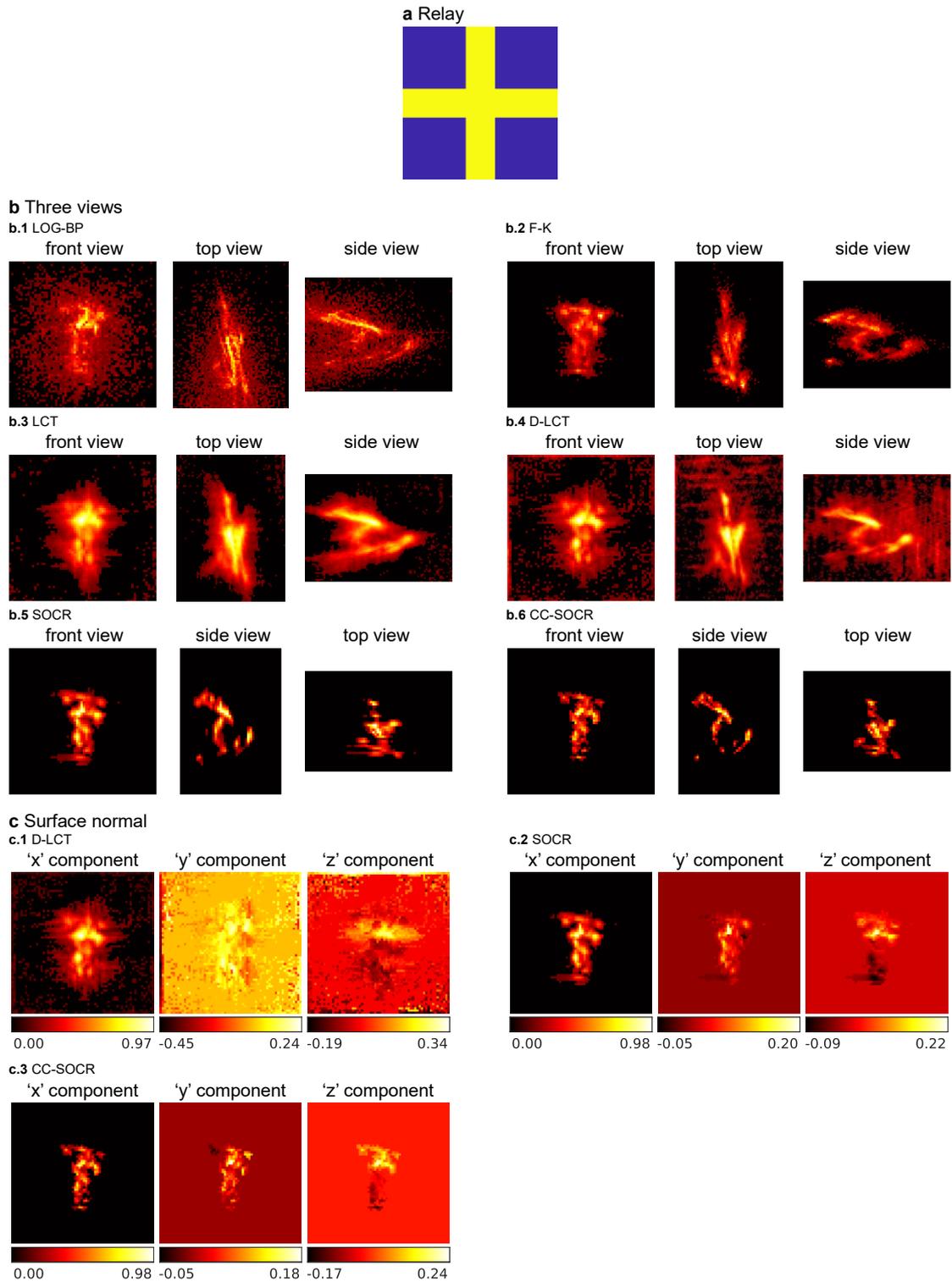

**Supplementary Figure 12 Reconstructions of the statue with a cross-shaped relay surface (confocal, measured signal). a** Confocal signals are measured in the yellow region, which consists of 1392 focal points. **b** Three views of the reconstructions. For all methods, the same reconstruction domain is shown. For the LOG-BP, F-K, LCT, and D-LCT methods, the length of the voxels in the depth direction is 0.48 cm. For the SOCR and CC-SOCR methods, the length of voxels in the depth direction is 0.96 cm. **c** The reconstructed surface normal of the D-LCT, SOCR and CC-SOCR methods is shown in the form of three components. The 'x', 'y' and 'z' components show values of the directional albedo in the depth, horizontal and vertical directions, respectively.

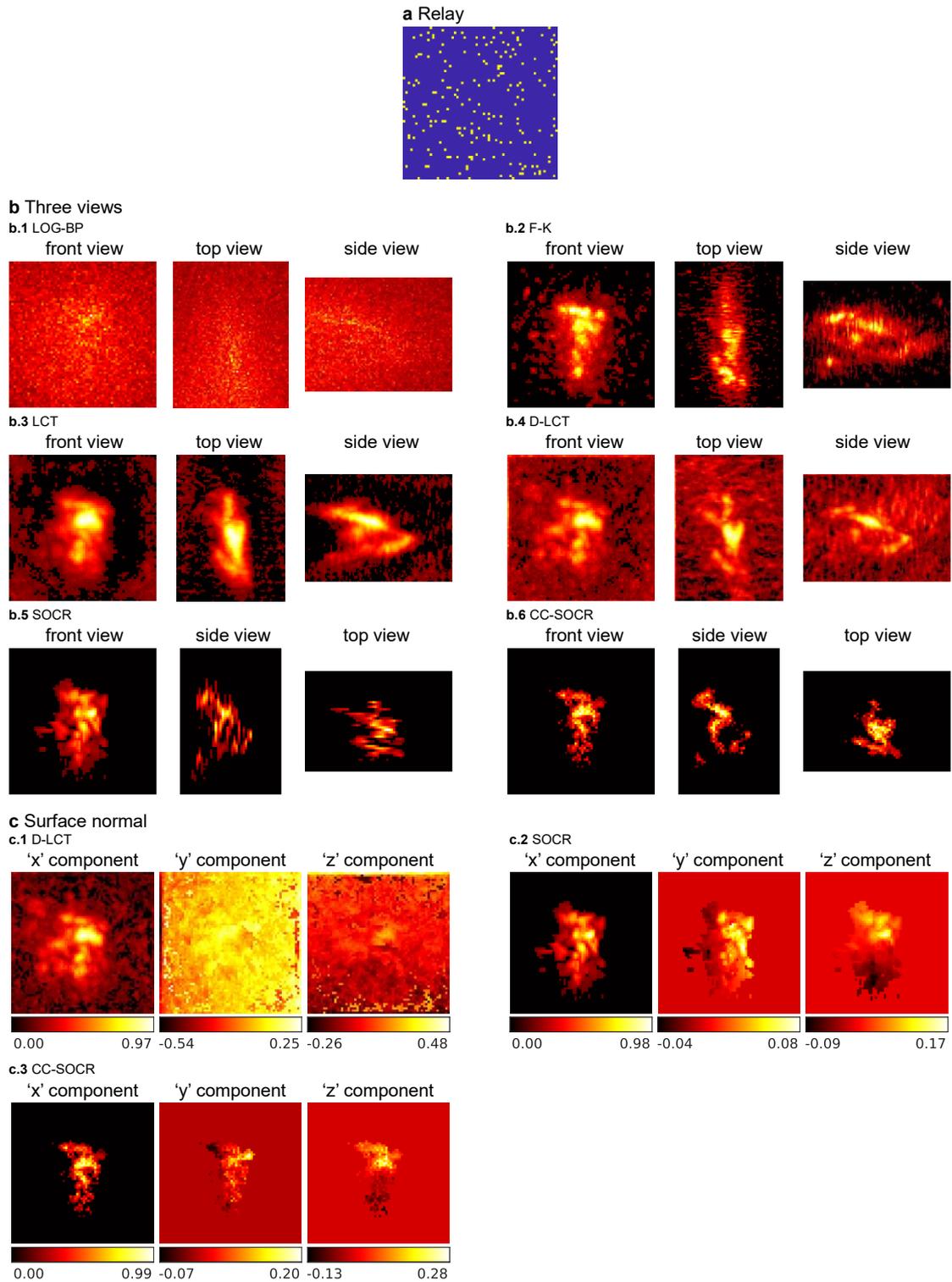

**Supplementary Figure 13 Reconstructions of the statue with signals measured at 200 randomly chosen focal points (confocal, measured signal).** **A** The confocal signal is measured at 200 randomly chosen focal points. **b** Three views of the reconstructions. For all methods, the same reconstruction domain is shown. For the LOG-BP, F-K, LCT, and D-LCT methods, the length of the voxels in the depth direction is 0.48 cm. For the SOCR and CC-SOCR methods, the length of voxels in the depth direction is 0.96 cm. **c** The reconstructed surface normal of the D-LCT, SOCR and CC-SOCR methods is shown in the form of three components. The 'x', 'y' and 'z' components show values of the directional albedo in the depth, horizontal and vertical directions, respectively.

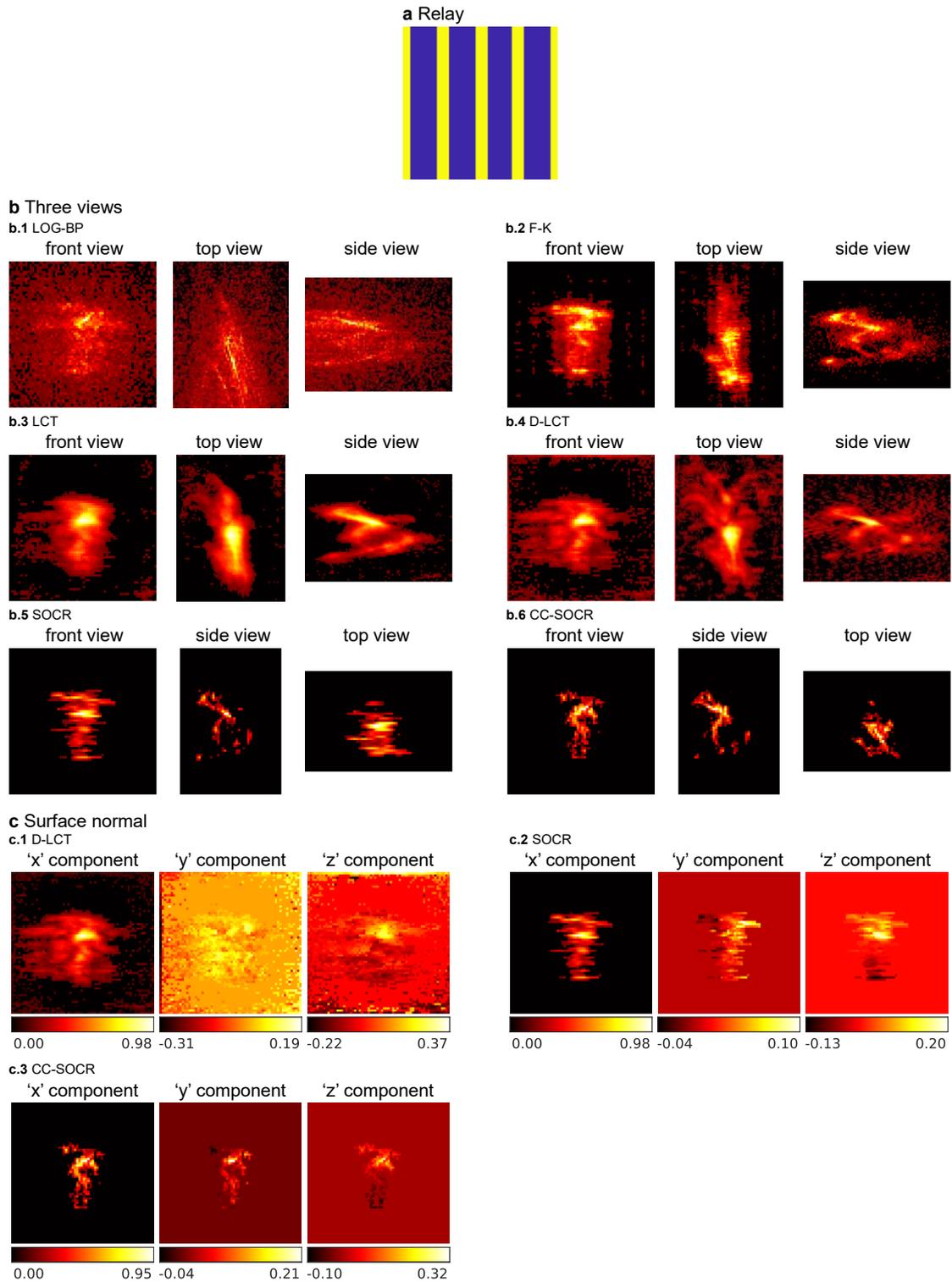

**Supplementary Figure 14 Reconstructions of the statue with signals measured at vertical bars (confocal, measured signal). a** The confocal signal is measured at 5 equispaced vertical bars, which contains 1344 focal points. **b** Three views of the reconstructions. For all methods, the same reconstruction domain is shown. For the LOG-BP, F-K, LCT, and D-LCT methods, the length of the voxels in the depth direction is 0.48 cm. For the SOCR and CC-SOCR methods, the length of voxels in the depth direction is 0.96 cm. **c** The reconstructed surface normal of the D-LCT, SOCR and CC-SOCR methods is shown in the form of three components. The 'x', 'y' and 'z' components show values of the directional albedo in the depth, horizontal and vertical directions, respectively.

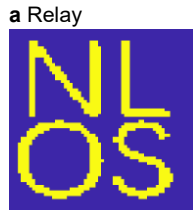

**a** Relay

**b** Three views

**b.1** LOG-BP

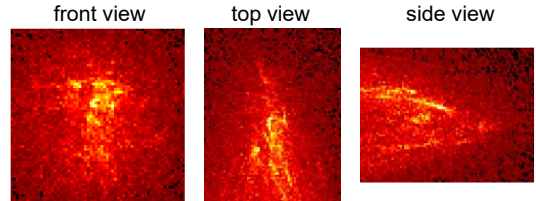

front view　top view　side view

**b.2** F-K

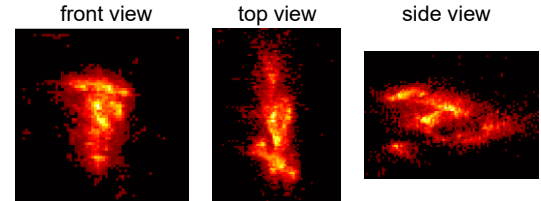

front view　top view　side view

**b.3** LCT

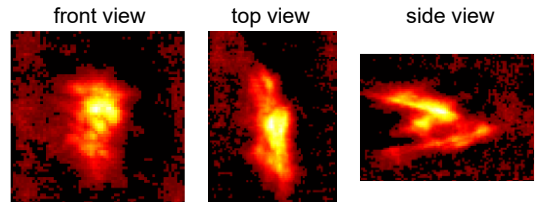

front view　top view　side view

**b.4** D-LCT

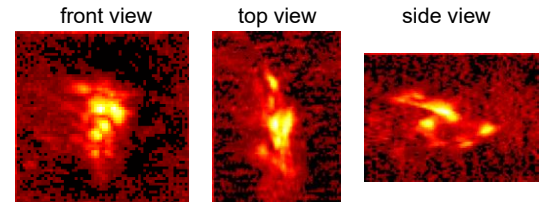

front view　top view　side view

**b.5** SOCR

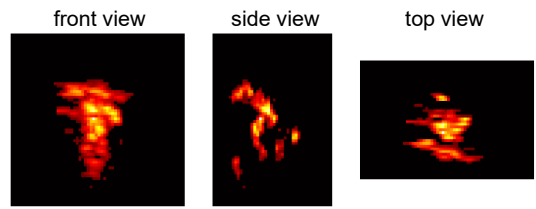

front view　side view　top view

**b.6** CC-SOCR

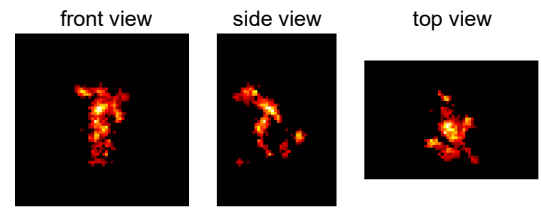

front view　side view　top view

**c** Surface normal

**c.1** D-LCT

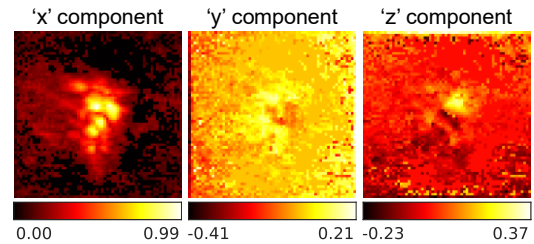

'x' component　'y' component　'z' component

**c.2** SOCR

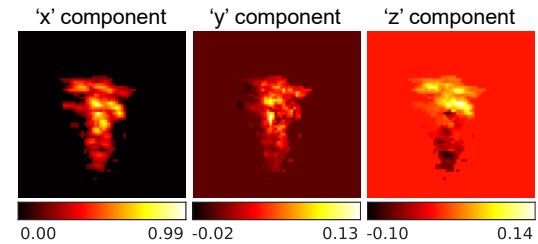

'x' component　'y' component　'z' component

**c.3** CC-SOCR

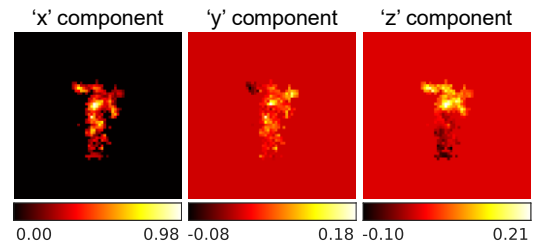

'x' component　'y' component　'z' component

**Supplementary Figure 15 Reconstructions of the statue with confocal signals measured at the letters 'N', 'L', 'O' and 'S'. a** The relay is the region consisting of four letters 'N', 'L', 'O' and 'S', which includes 825 focal points. **b** Three views of the reconstructions. For all methods, the same reconstruction domain is shown. For the LOG-BP, F-K, LCT, and D-LCT methods, the length of the voxels in the depth direction is 0.48 cm. For the SOCR and CC-SOCR methods, the length of voxels in the depth direction is 0.96 cm. **c** The reconstructed surface normal of the D-LCT, SOCR and CC-SOCR methods is shown in the form of three components. The 'x', 'y' and 'z' components show values of the directional albedo in the depth, horizontal and vertical directions, respectively.

**a** Relay

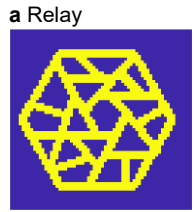

**b** Three views

**b.1** LOG-BP

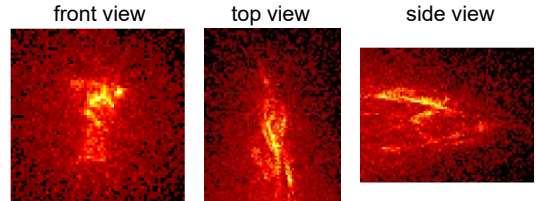

**b.2** F-K

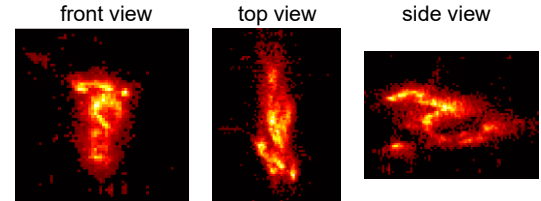

**b.3** LCT

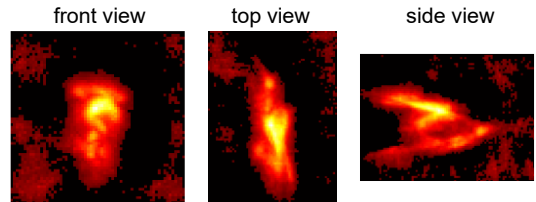

**b.4** D-LCT

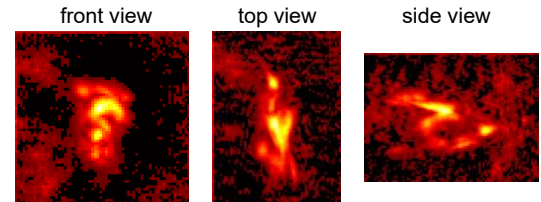

**b.5** SOCR

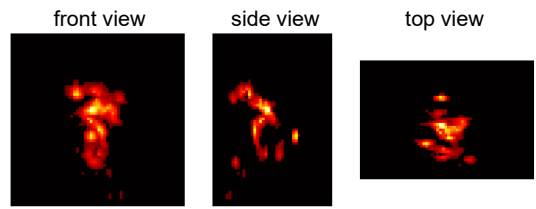

**b.6** CC-SOCR

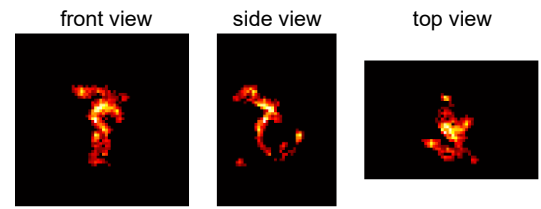

**c** Surface normal

**c.1** D-LCT

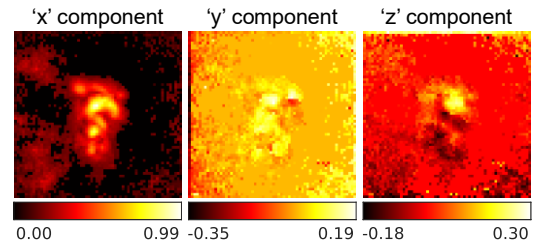

**c.2** SOCR

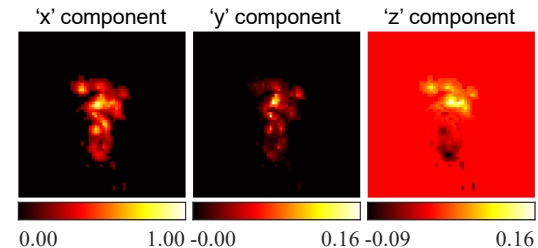

**c.3** CC-SOCR

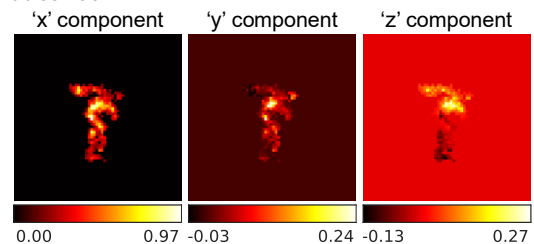

**Supplementary Figure 16 Reconstructions of the statue with an irregular relay. a** The relay is a set of several sticks sparsely and randomly distributed, which contains 1229 focal points. **b** Three views of the reconstructions. For all methods, the same reconstruction domain is shown. For the LOG-BP, F-K, LCT, and D-LCT methods, the length of the voxels in the depth direction is 0.48 cm. For the SOCR and CC-SOCR methods, the length of voxels in the depth direction is 0.96 cm. **c** The reconstructed surface normal of the D-LCT, SOCR and CC-SOCR methods is shown in the form of three components. The 'x', 'y' and 'z' components show values of the directional albedo in the depth, horizontal and vertical directions, respectively.

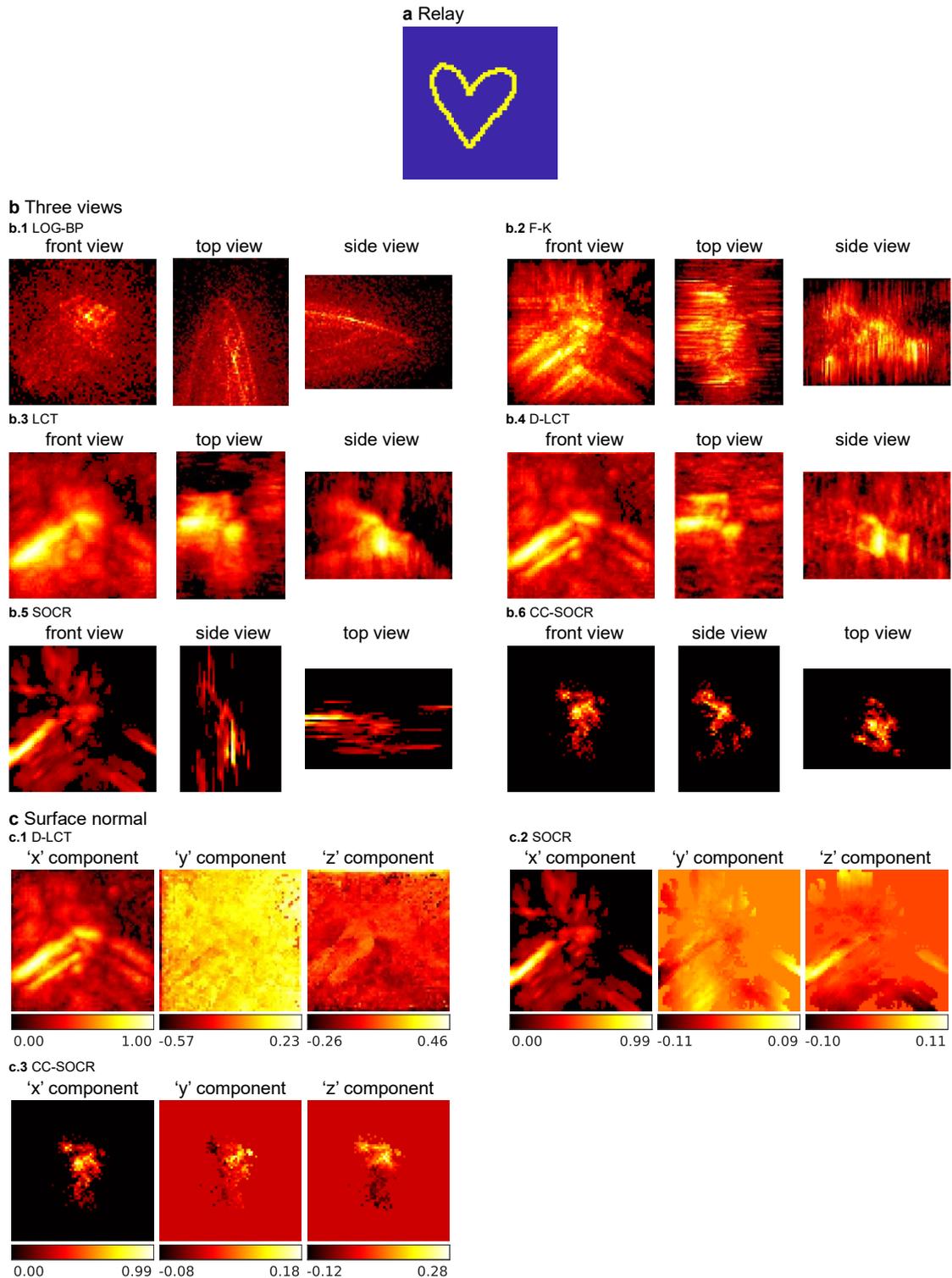

**Supplementary Figure 17 Reconstructions of the statue with a heart-shaped relay surface. a** The relay is a heart-shaped region, which contains 258 focal points. **b** Three views of the reconstructions. For all methods, the same reconstruction domain is shown. For the LOG-BP, F-K, LCT, and D-LCT methods, the length of the voxels in the depth direction is 0.48 cm. For the SOCR and CC-SOCR methods, the length of voxels in the depth direction is 0.96 cm. **c** The reconstructed surface normal of the D-LCT, SOCR and CC-SOCR methods is shown in the form of three components. The 'x', 'y' and 'z' components show values of the directional albedo in the depth, horizontal and vertical directions, respectively.

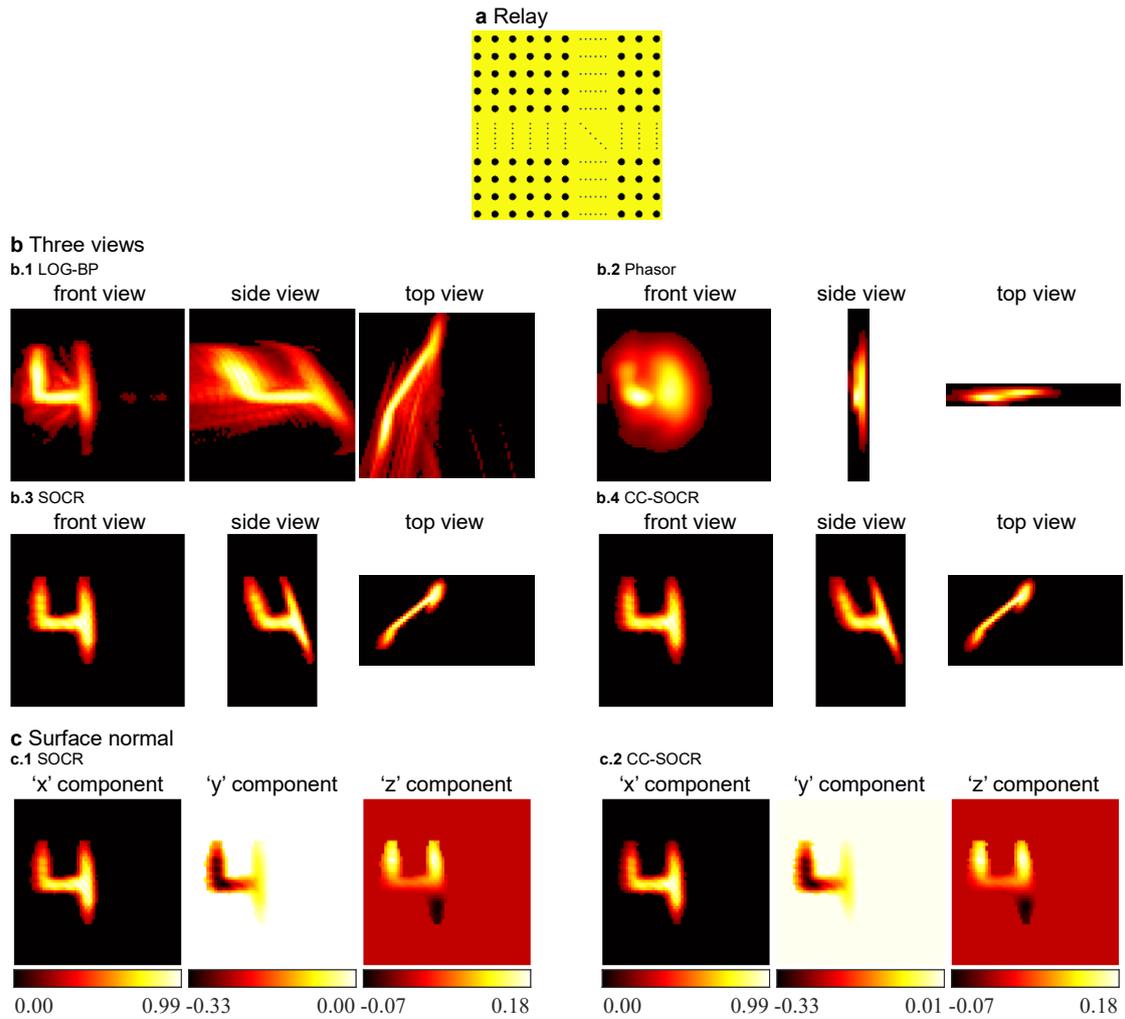

**Supplementary Figure 18 Reconstructions of the figure '4' with 64 × 64 measurements (non-confocal, measured signal). a** 64 × 64 focal points are raster scanned. **b** Three views of the reconstructions. For all methods, the same reconstruction domain is shown. For the LOG-BP, PF, SOCR and CC-SOCR methods, the lengths of the voxels in the depth direction are 0.24 cm, 1.87 cm, 0.48 cm and 0.48 cm. **c** The reconstructed surface normal of the SOCR and CC-SOCR methods are shown in the form of three components. The 'x', 'y' and 'z' components show values of the directional albedo in the depth, horizontal and vertical directions, respectively.

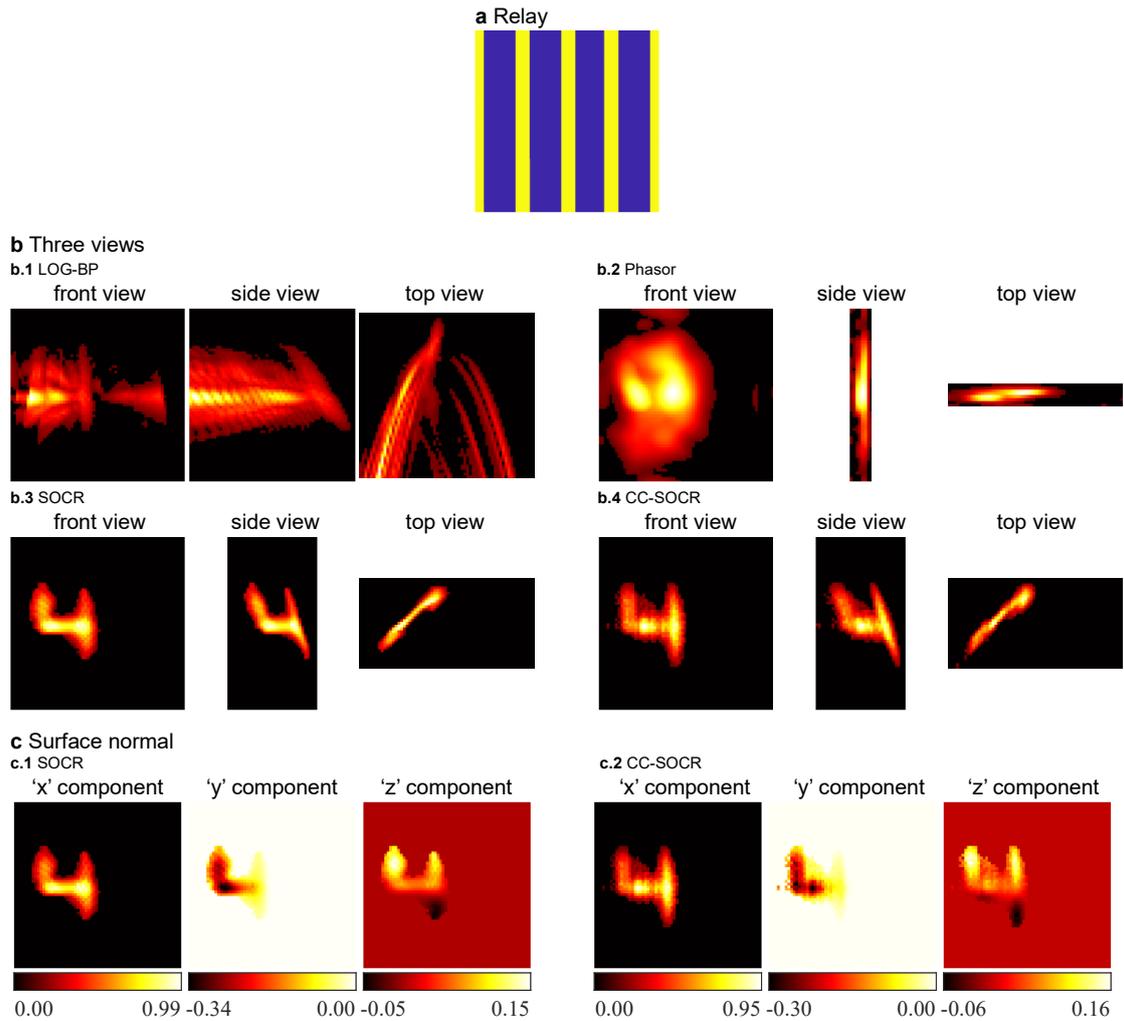

**Supplementary Figure 19 Reconstructions of the figure '4' with a vertical illumination pattern (non-confocal, measured signal).** **a** The illumination points are in the yellow region. **b** Three views of the reconstructions. For all methods, the same reconstruction domain is shown. For the LOG-BP, PF, SOCR and CC-SOCR methods, the lengths of the voxels in the depth direction are 0.24 cm, 1.87 cm, 0.48 cm and 0.48 cm. **c** The reconstructed surface normal of the SOCR and CC-SOCR methods are shown in the form of three components. The 'x', 'y' and 'z' components show values of the directional albedo in the depth, horizontal and vertical directions, respectively.

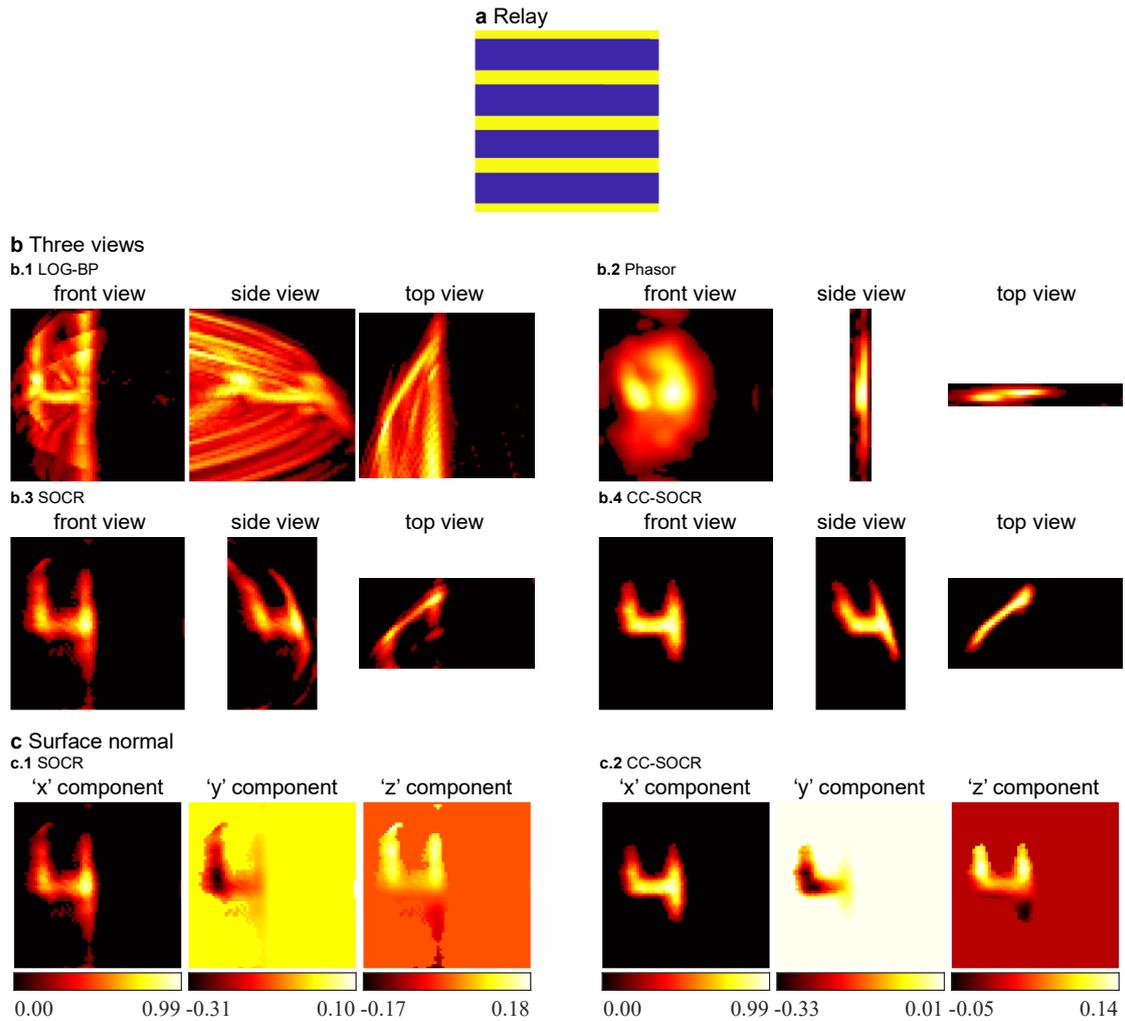

**Supplementary Figure 20 Reconstructions of the figure '4' with a horizontal illumination pattern (non-confocal, measured signal). a** The illumination points are in the yellow region. **b** Three views of the reconstructions. For all methods, the same reconstruction domain is shown. For the LOG-BP, PF, SOCR and CC-SOCR methods, the lengths of the voxels in the depth direction are 0.24 cm, 1.87 cm, 0.48 cm and 0.48 cm. **c** The reconstructed surface normal of the SOCR and CC-SOCR methods are shown in the form of three components. The 'x', 'y' and 'z' components show values of the directional albedo in the depth, horizontal and vertical directions, respectively.

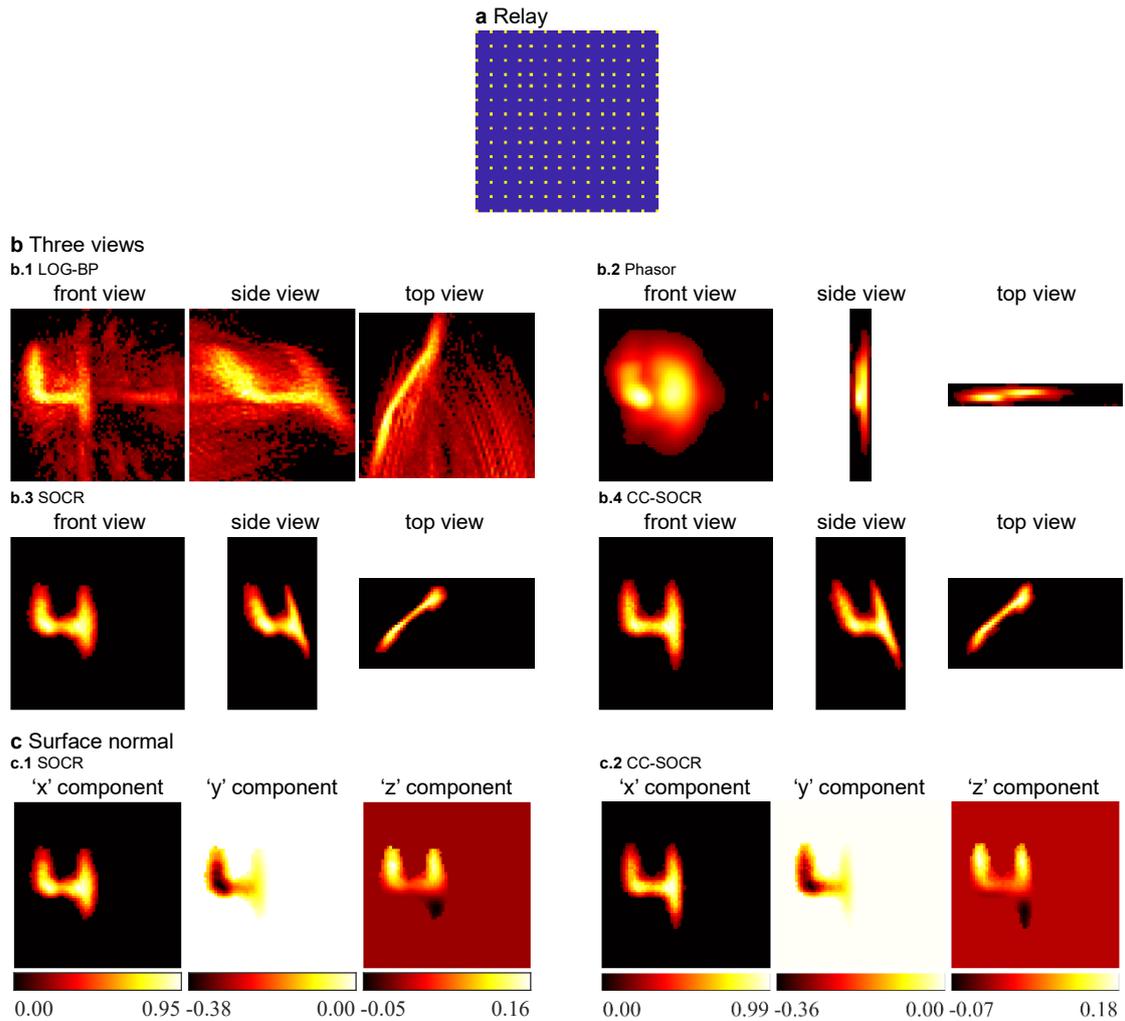

**Supplementary Figure 21 Reconstructions of the figure '4' with 14 × 14 illumination points (non-confocal, measured signal). a** The illumination points are shown in yellow. **b** Three views of the reconstructions. For all methods, the same reconstruction domain is shown. For the LOG-BP, PF, SOCR and CC-SOCR methods, the lengths of the voxels in the depth direction are 0.24 cm, 1.87 cm, 0.48 cm and 0.48 cm. **c** The reconstructed surface normal of the SOCR and CC-SOCR methods are shown in the form of three components. The 'x', 'y' and 'z' components show values of the directional albedo in the depth, horizontal and vertical directions, respectively.

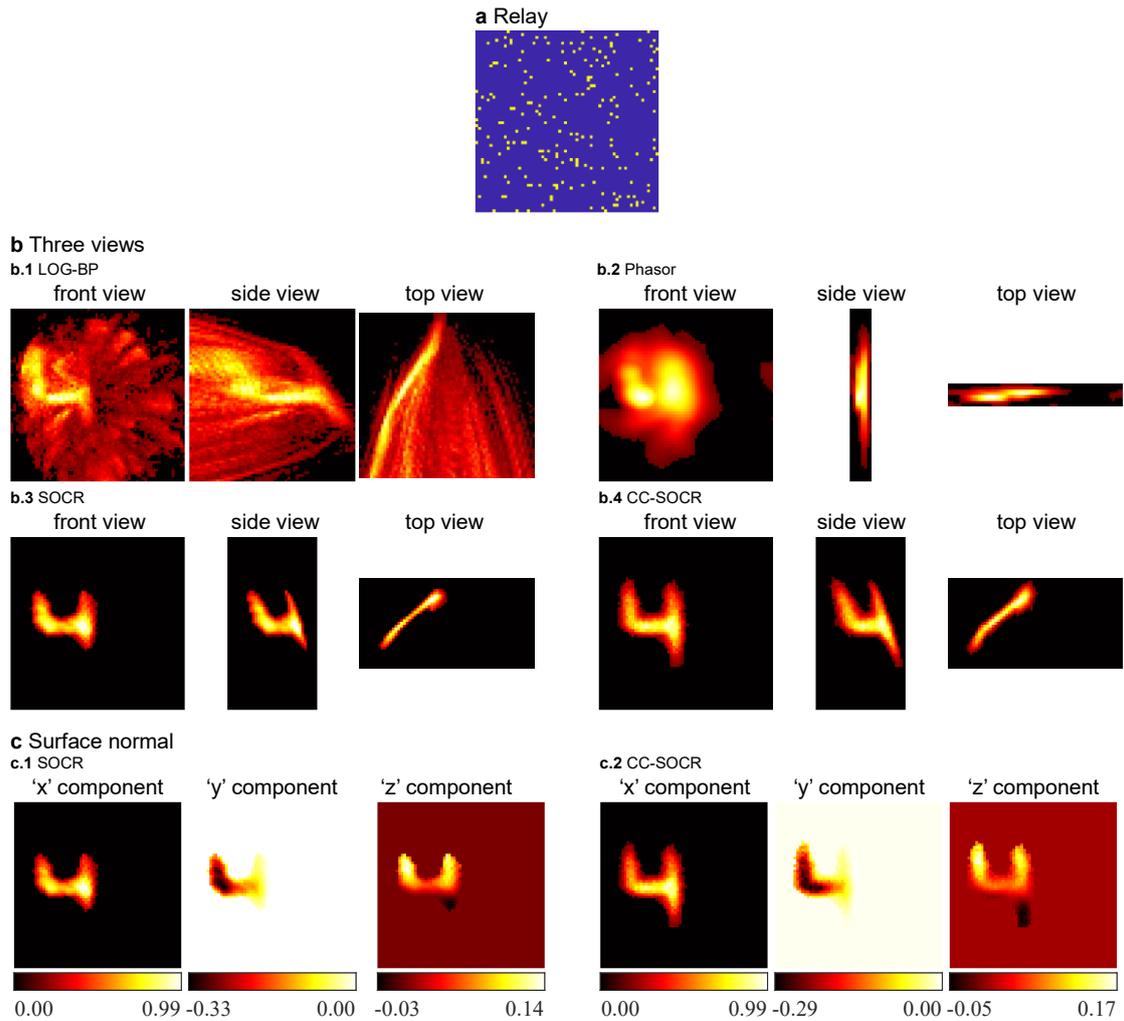

**Supplementary Figure 22 Reconstructions of the figure '4' with 200 randomly distributed illumination points (non-confocal, measured signal). a** The illumination points are shown in yellow. **b** Three views of the reconstructions. For all methods, the same reconstruction domain is shown. For the LOG-BP, PF, SOCR and CC-SOCR methods, the lengths of the voxels in the depth direction are 0.24 cm, 1.87 cm, 0.48 cm and 0.48 cm. **c** The reconstructed surface normal of the SOCR and CC-SOCR methods are shown in the form of three components. The 'x', 'y' and 'z' components show values of the directional albedo in the depth, horizontal and vertical directions, respectively.

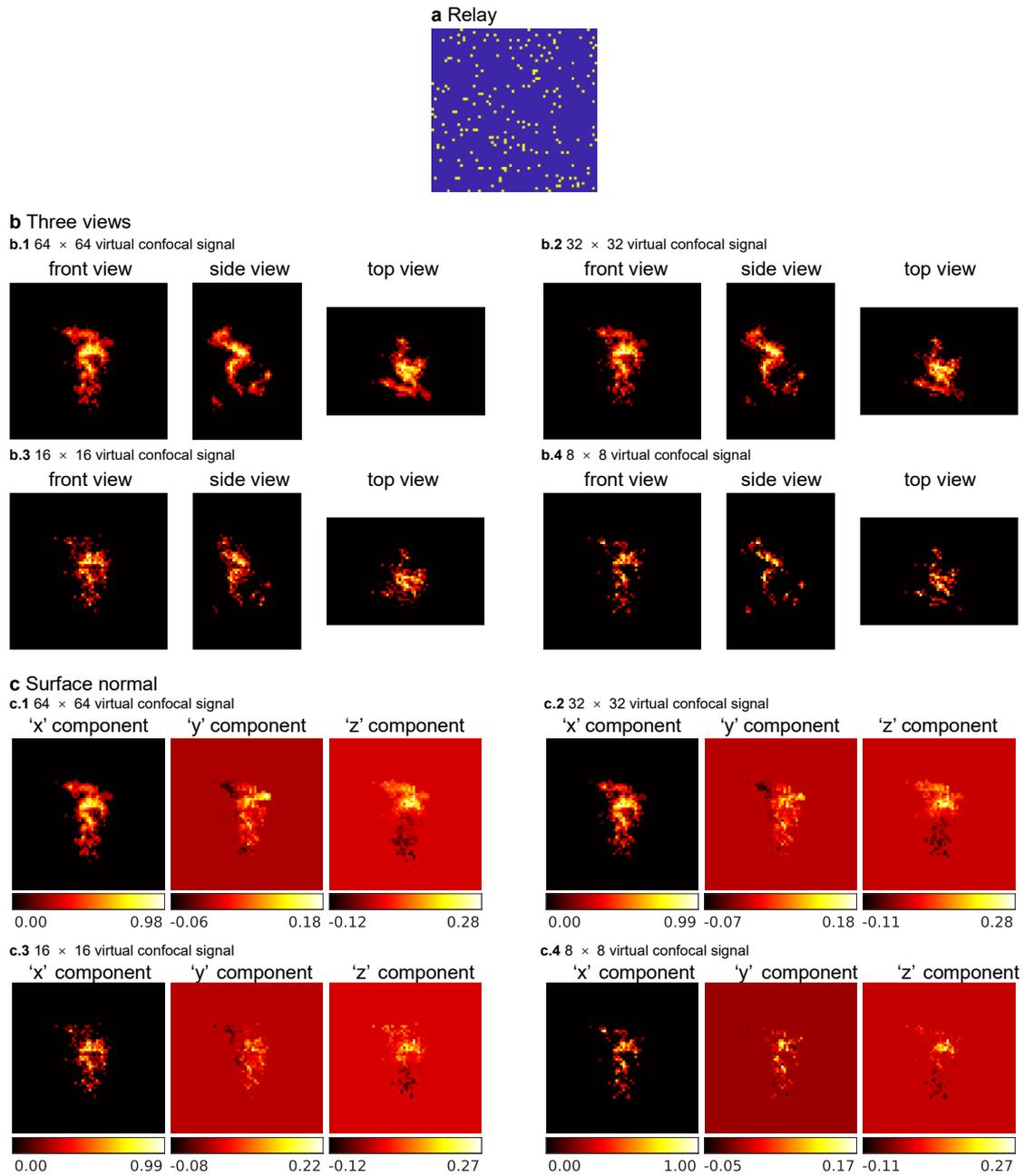

**Supplementary Figure 23 Reconstructions of the statue with different sizes of the virtual confocal signal. (confocal, measured signal). a** Confocal signals are measured at 200 randomly distributed focal points. **b** Three views of the reconstructions. The virtual confocal signal of different sizes are introduced. The reconstruction quality decreases with the size of the virtual confocal signal. **c** The reconstructed surface normal is shown in the form of three components. The 'x', 'y' and 'z' components show values of the directional albedo in the depth, horizontal and vertical directions, respectively.

**a** Relay

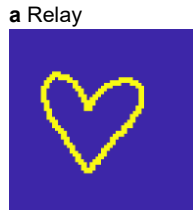

**b** Three views

**b.1** F-K (zero padding)

front view    top view    side view

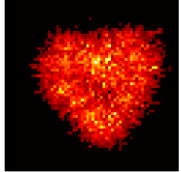 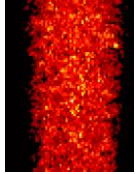 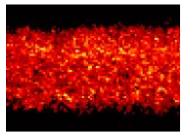

**b.2** F-K (nearest neighbor)

front view    top view    side view

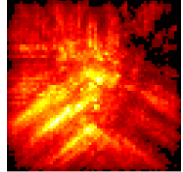 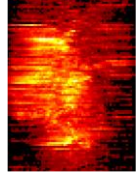 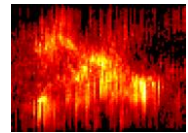

**b.3** LCT (zero padding)

front view    top view    side view

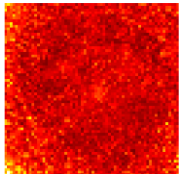 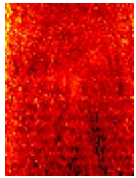 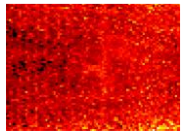

**b.4** LCT (nearest neighbor)

front view    top view    side view

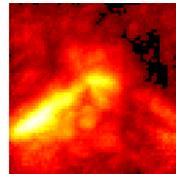 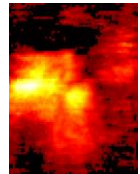 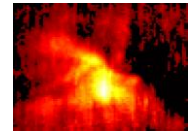

**b.5** D-LCT (zero padding)

front view    top view    side view

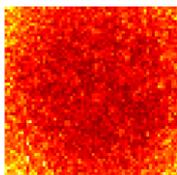 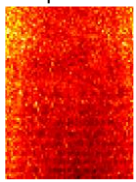 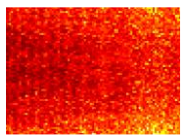

**b.6** D-LCT (nearest neighbor)

front view    top view    side view

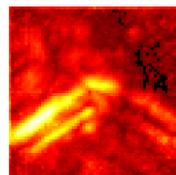 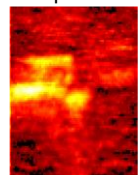 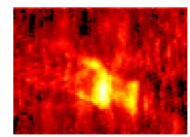

**b.7** SOCR (zero padding)

front view    side view    top view

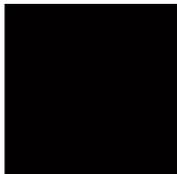 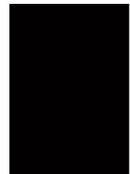 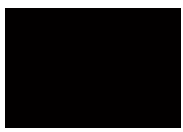

**b.8** SOCR (nearest neighbor)

front view    side view    top view

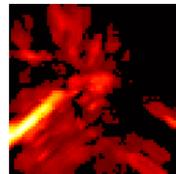 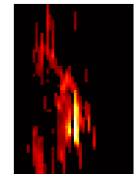 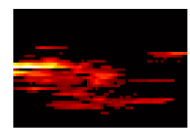

**Supplementary Figure 24 Reconstructions of the statue with a heart-shaped relay surface.** The signal is preprocessed with two techniques: zero padding and the nearest neighbor interpolation. **a** The relay is a heart-shaped region, which contains 258 focal points. **b** Three views of the F-K, LCT, D-LCT and SOCR reconstructions. For the F-K, LCT, and D-LCT methods, the length of the voxels in the depth direction is 0.48 cm. For the SOCR method, the length of voxels in the depth direction is 0.96 cm. These methods fail to reconstruct the target (See also Supplementary Figure 17).

**a** Relay

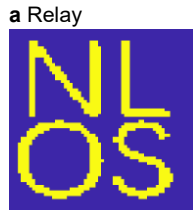

**b** Three views

**b.1** Least squares without regularization
front view  side view  top view

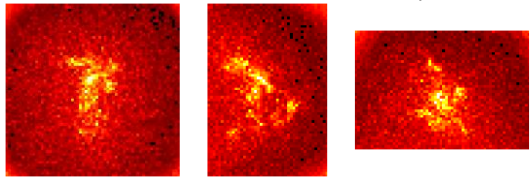

**b.2** Sparsity regularization
front view  side view  top view

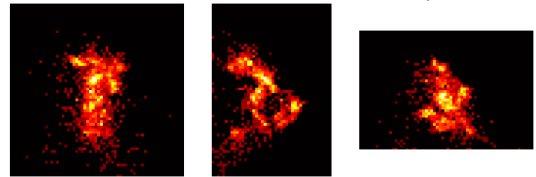

**b.3** Sparsity and non-local self-similarity regularizations
front view  side view  top view

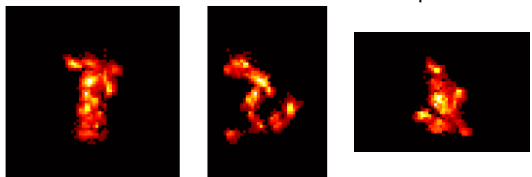

**b.4** CC-SOCR (1 iteration)
front view  side view  top view

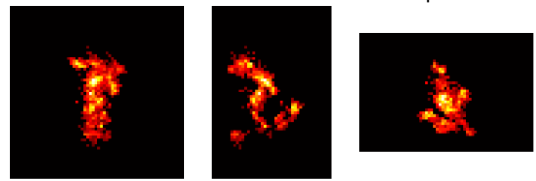

**b.5** CC-SOCR (2 iterations)
front view  side view  top view

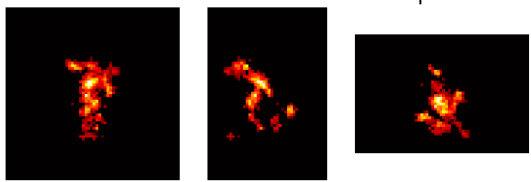

**Supplementary Figure 25 Reconstructions of the statue with confocal signals measured at the letters 'N', 'L', 'O' and 'S' under different regularization settings. a** The relay is the region consisting of four letters 'N', 'L', 'O' and 'S', which includes 825 focal points. **b** Three views of the reconstructions. The reconstruction result without regularizations is of poor quality. When the sparsity and non-local self-similarity priors of the target are introduced, the quality of the reconstruction enhances, but is still blurry. The CC-SOCR method reconstructs the target faithfully.

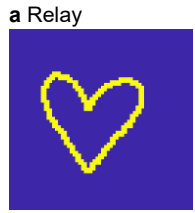

**a** Relay

**b** Three views

**b.1** Least squares without regularization
front view    side view    top view
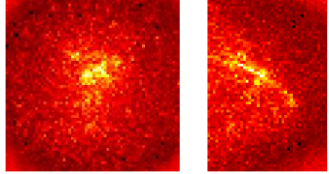 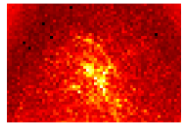

**b.2** Sparsity regularization
front view    side view    top view
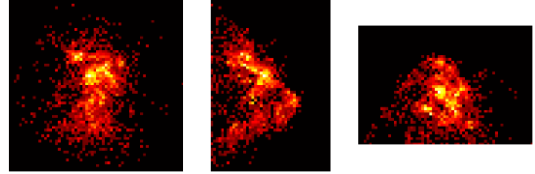

**b.3** Sparsity and non-local self-similarity regularizations
front view    side view    top view
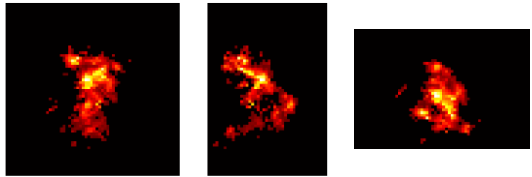

**b.4** CC-SOCR (1 iteration)
front view    side view    top view
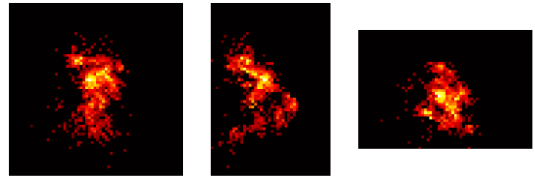

**b.5** CC-SOCR (2 iterations)
front view    side view    top view
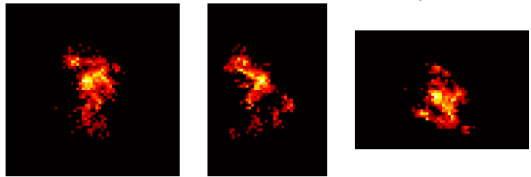

**Supplementary Figure 26 Reconstructions of the statue with a heart-shaped relay surface under different regularization settings. a** The relay is a heart-shaped region, which contains 258 focal points. **b** Three views of the reconstructions. The reconstruction result without regularizations is of poor quality. When the sparsity and non-local self-similarity priors of the target are introduced, the quality of the reconstruction enhances, but still contains artifacts in the background. The CC-SOCR method reconstructs the target faithfully.

# Supplementary Note 2  The CC-SOCR algorithm

## 2.1 Notation

Scalars are represented with lowercase letters. Vectors or matrices are represented with uppercase letters. Uppercase letters in bold are used to represent higher-order tensors. Let $a_i$ be a positive integer. We use $\mathbb{R}[a_i]$ to denote the Euclidean space of dimension $a_i$. The identity matrix of order $a_i$ is denoted by $I[a_i]$. For a tensor $\mathbf{A}$, we denote by $|\mathbf{A}|_0$ the number of non-zero elements of $\mathbf{A}$. We also denote by $|\mathbf{A}|_1$ the summation of the absolute value of elements of $\mathbf{A}$. The reconstruction domain is discretized with $n_x \times n_y \times n_z$ voxels. Let $\mathbf{L} \in \mathbb{R}[n_x \times n_y \times n_z]$ and $\mathbf{n} \in \mathbb{R}[n_x \times n_y \times n_z \times 3]$ be the albedo and unit surface normal of the target. The directional albedo $\mathbf{u} \in \mathbb{R}[n_x \times n_y \times n_z \times 3]$ is defined by $\mathbf{u}(i_1, i_2, i_3, j) = \mathbf{L}(i_1, i_2, i_3) \mathbf{n}(i_1, i_2, i_3, j)$ for all $i_1 = 1, ..., n_x$, $i_2 = 1, ..., n_y$, $i_3 = 1, ..., n_z$ and $j = 1, 2, 3$. Noting that the vector field $\mathbf{n}$ has unit length at each voxel, the albedo is given by

$$\mathbf{L}(i_1, i_2, i_3) = \sqrt{\sum_{j=1}^{3} \mathbf{u}(i_1, i_2, i_3, j)^2} \tag{S.1}$$

In the following, we use $\mathbf{L} = \mathrm{albedo}(\mathbf{u})$ as a shorthand of equation (S.1). The summation of elements of $\mathbf{L}$ is denoted by $\|\mathbf{u}\|_{2,1}$. Let $\tilde{\mathbf{b}} \in \mathbb{R}[M \times n_t]$ be the signal measured at $M$ points on the relay surface, in which $n_t$ is the maximum number of time bins used. The approximated signal and virtual confocal signal are denoted by $\mathbf{b} \in \mathbb{R}[M \times n_t]$ and $\mathbf{d} \in \mathbb{R}[n_y \times n_z \times n_t]$.

## 2.2 The CC-SOCR algorithm

The proposed CC-SOCR optimization problem for NLOS reconstruction writes

$$\min_{\mathbf{u},\mathbf{b},\mathbf{d},D_s,D_n,\mathbf{C},\mathbf{S},\Psi,\mathbf{Q}} \|A_\mathbf{b}\mathbf{u}-\mathbf{b}\|^2 + s_u\|\mathbf{L}\|_1 + s_b|\mathbf{b}|_0$$

$$+ \lambda_u \sum_i \left[\|B_i(\mathbf{L}) - D_s C_i D_n^T\|^2 + \lambda_{pu}|C_i|_0\right]$$

$$+ \lambda_b \|\mathbf{b} - \tilde{\mathbf{b}}\|^2 + \lambda_b \lambda_{pb} \sum_i \|P_i(\tilde{\mathbf{b}}) - DS_i\|^2$$

$$+ \lambda_b \lambda_{pb} \sum_{i,j} \left[\frac{\sigma_\mathbf{b}}{d_j^T P_i(A_\mathbf{b}\mathbf{u})} S_i(j)\right]^2$$

$$+ \lambda_b \lambda_{pb} \lambda_{sb} \sum_i \|P_i(\mathbf{b}) - DS_i\|^2 \qquad (S.2)$$

$$+ \lambda_d \|A_\mathbf{d}\mathbf{u} - \mathbf{d}\|^2 + s_d|\mathbf{d}|_0$$

$$+ \lambda_d \lambda_{pd} \sum_i \|Q_i - \Psi^T P_i(\mathbf{d})\|^2$$

$$+ \lambda_d \lambda_{pd} \lambda_{sd} \sum_i \|Q_i - \Psi^T P_i(A_\mathbf{d}\mathbf{u})\|^2$$

$$+ \lambda_d \lambda_{pd} \lambda_{fd} \sum_i |Q_i|_0$$

$$+ \lambda_{bd} \|R_\mathbf{b}(\mathbf{b},\mathbf{d}) - R_\mathbf{d}(\mathbf{b},\mathbf{d})\|^2$$

$$\text{s.t.} \quad \mathbf{L} = \text{albedo}(\mathbf{u}),$$
$$D_s^T D_s = I[p_x p_y p_z],\ D_n^T D_n = I[r],$$
$$\Psi^T \Psi = I[q_y q_z q_t]$$

in which $i$ is the index of a local patch, $P_i$ is the patch generating operator, $B_i$ is the block matching operator. $D$ represents the matrix of the discrete cosine transform, with its $j^{th}$ filter denoted by $d_j$. $S_i(j)$ represents the $j^{th}$ element of the vector $S_i$. $\mathbf{C}$, $\mathbf{S}$ and $\mathbf{Q}$ represent the collections of the transform-domain coefficients $\{C_i\}$, $\{S_i\}$ and $\{Q_i\}$ respectively. $R_\mathbf{b}(\mathbf{b},\mathbf{d})$ and $R_\mathbf{d}(\mathbf{b},\mathbf{d})$ are the subsets of the approximated signals $\mathbf{b}$ and $\mathbf{d}$ that share the same measurement pairs. $p_x$, $p_y$ and $p_z$ are sizes of the local patches of the albedo. $r$ is the maximum number of neighbors kept in the block matching process. $q_y$, $q_z$ and $q_t$ are the patch sizes of the virtual confocal signal in the horizontal, vertical and temporal directions. $s_u$, $s_b$, $s_d$, $\sigma_d$, $\lambda_u$, $\lambda_b$, $\lambda_d$, $\lambda_{pu}$, $\lambda_{pb}$, $\lambda_{pd}$, $\lambda_{sb}$, $\lambda_{sd}$, $\lambda_{fd}$ and $\lambda_{bd}$ are fixed parameters. See Supplementary Tables 1 - 3 for detailed explanations of the parameters and variables.

In Supplementary Algorithm 1 we present an iterative scheme to solve the NLOS reconstruction problem (S.2). The only input is the measured signal $\tilde{\mathbf{b}}$. The algorithm contains three stages. In the first stage, all variables are initialized by solving the sub-problems. In the second stage, the approximated signal $\mathbf{b}$, the reconstructed target $\mathbf{u}$ and the virtual confocal signal $\mathbf{d}$ are updated sequentially. In the third stage, the albedo and surface normal are computed directly from $\mathbf{u}$. Solutions to the sub-problems (1.2), (1.3), (1.4), (2.2), (2.4) and (2.6) have been studied in the supplementary information of the work of SOCR method[5]. In the following, we provide a detailed discussion of the remaining sub-problems. We begin with the definition of the hard-thresholding operator. For a real number $a$ and a positive real number $y$, the hard-thresholding operator $\mathcal{T}$ is given by

$$\mathcal{T}(a,y) = \begin{cases} 0, & |a| < y \\ a, & |a| \geq y \end{cases} \quad (S.3)$$

For a tensor $\mathbf{A}$, this operator is applied elementwise. With this definition, the solution to the sub-problem (1.1) is expressed as

$$\mathbf{b}^0 = \mathcal{T}\left(\tilde{\mathbf{b}}, \sqrt{\frac{s_b}{\lambda_b}}\right) \quad (S.4)$$

For the sub-problem (1.5), if the $j^{th}$ measurement pair of the virtual confocal signal does not appear in the measured signal, the solution is given by

$$\mathbf{d}^0[j] = \mathcal{T}\left(A_\mathbf{d}\mathbf{u}^0[j], \sqrt{\frac{s_d}{\lambda_d}}\right) \quad (S.5)$$

in which $A_\mathbf{d}\mathbf{u}^0[j]$ is the simulated signal of the $j^{th}$ measurement pair. Otherwise, the solution writes

$$\mathbf{d}^0[j] = \mathcal{T}\left(\frac{\lambda_d A_\mathbf{d}\mathbf{u}^0[j] + \lambda_{bd}\mathbf{b}_j^0}{\lambda_d + \lambda_{bd}}, \sqrt{\frac{s_d}{\lambda_d + \lambda_{bd}}}\right) \quad (S.6)$$

in which $A_\mathbf{d}\mathbf{u}^0[j]$ and $\mathbf{b}_j^0$ are the simulated signal of the $j^{th}$ measurement pair and the corresponding signal of $\mathbf{b}^0$, respectively.

The sub-problem (1.6) is related to frequency domain sparse approximation with orthonormal dictionary atoms[7]. Note that the objective function can be written equivalently as

$$J_1(\Psi, \mathbf{Q}) = (1 + \lambda_{sd}) \sum_i \left[ \left\| Q_i - \frac{\Psi^T P_i (\mathbf{d}^0 + \lambda_{sd} A_\mathbf{d}\mathbf{u}^0)}{1 + \lambda_{sd}} \right\|^2 + \frac{\lambda_{fd}}{1 + \lambda_{sd}} |Q_i|_0 \right] \quad (S.7)$$

in which $\Psi$ is an orthogonal matrix. In order to solve this problem with convergence guarantee, it suffices to generalize the data-driven tight frame image denoising algorithm[7] to three dimensions and apply it to $(\mathbf{d}^0 + \lambda_{sd} A_\mathbf{d}\mathbf{u}^0)/(1 + \lambda_{sd})$ with the regularization parameter $\lambda_{fd}/(1 + \lambda_{sd})$.

For the sub-problem (2.1), if the $j^{th}$ measurement pair of the measured signal does not appear in the virtual confocal signal, the solution is given by

$$\mathbf{b}^{k+1}[j] = \mathcal{T}\left(\frac{A_\mathbf{b}\mathbf{u}^k[j] + \lambda_b \tilde{\mathbf{b}}[j] + \lambda_b \lambda_{pb} \lambda_{sb} P^*(D\mathbf{S}^k)[j]}{\lambda_{b,pb,sb}}, \sqrt{\frac{s_b}{\lambda_{b,pb,sb}}}\right) \quad (S.8)$$

in which $\lambda_{b,pb,sb} = 1 + \lambda_b + \lambda_b \lambda_{pb} \lambda_{sb}$. Otherwise,

$$\mathbf{b}^{k+1}[j] = \mathcal{T}\left(\frac{A_\mathbf{b}\mathbf{u}^k[j] + \lambda_b \tilde{\mathbf{b}}[j] + \lambda_b \lambda_{pb} \lambda_{sb} P^*(D\mathbf{S}^k)[j] + \lambda_{bd}\mathbf{d}_j^k}{\lambda_{b,pb,sb,bd}}, \sqrt{\frac{s_b}{\lambda_{b,pb,sb,bd}}}\right) \quad (S.9)$$

in which $\lambda_{b,pb,sb,bd} = 1 + \lambda_b + \lambda_b \lambda_{pb} \lambda_{sb} + \lambda_{bd}$. $A_\mathbf{b} \mathbf{u}^k$ and $\mathbf{d}_j^k$ are the simulated signal of the $j^{th}$ measurement pair and the corresponding signal of $\mathbf{d}^k$, respectively. $P^*$ represents the operator that aggregates the patches back to the signal. $\mathbf{S}$ is the collection of the Wiener coefficients in the frequency domain. $D\mathbf{S}^k$ is an abbreviation of $\{DS_i^k\}_{i \in I}$, where $I$ is the set of indices of the spatial patches.

The reconstructed target is updated by solving the sub-problem (2.3). In this sub-problem, the term $\phi(\mathbf{u}) = \sum_{i,j} \left( \dfrac{\sigma_d}{d_j^T P_i(A_\mathbf{b}\mathbf{u})} S_i^{k+1}(j) \right)^2$ is omitted. Otherwise, the problem will be non-linear and difficult to solve. This problem contains a $L_1$-regularization term, which can be solved efficiently with the split Bregman method[8]. Convergence is guaranteed if the sub-problems are solved accurately. The scheme is shown in Supplementary Algorithm 2, in which the sub-problem of updating $\mathbf{v}_{j+1}^{k+1}$ has a closed-form solution as follows

$$\mathbf{v}_{j+1}^{k+1}(i_1, i_2, i_3, h) = \max\left(0, 1 - \dfrac{s_u}{2\mu \|\mathbf{u}_j^{k+1}(i_1,i_2,i_3,h) - \mathbf{q}_j^{k+1}(i_1,i_2,i_3,h)\|}\right) \quad \text{(S.10)}$$
$$\cdot (\mathbf{u}_j^{k+1}(i_1,i_2,i_3,h) - \mathbf{q}_j^{k+1}(i_1,i_2,i_3,h))$$

in which $i_1$, $i_2$ and $i_3$ are indices of the voxel in three directions and $h = 1, 2, 3$. To update $\mathbf{u}_{j+1}^{k+1}$, we minimize the following objective function

$$\begin{aligned}
J_2(\mathbf{u}) = &\|A_\mathbf{b}\mathbf{u} - \mathbf{b}^{k+1}\|^2 \\
&+ (\lambda_d + \lambda_d \lambda_{pd} \lambda_{sd}) \left\| A_\mathbf{d}\mathbf{u} - \dfrac{\lambda_d \mathbf{d}^k + \lambda_d \lambda_{pd} \lambda_{sd} P^*(\Psi^k \mathbf{Q}^k)}{\lambda_d + \lambda_d \lambda_{pd} \lambda_{sd}} \right\|^2 \\
&+ (\lambda_u + \mu) \left\| \mathbf{u} - \dfrac{\lambda_u B^*(D_s^k \mathbf{C}^k (D_n^k)^T) + \mu(\mathbf{v}_{j+1}^{k+1} + \mathbf{q}_j^{k+1})}{\lambda_u + \mu} \right\|^2
\end{aligned} \quad \text{(S.11)}$$

in which $\mathbf{Q}^k$ and $\mathbf{C}^k$ are the collections of transform-domain coefficients, $P^*$ and $B^*$ are the operators that aggregate the patch dataset or block dataset back to the signal and albedo. $\Psi^k \mathbf{Q}^k$ and $D_s^k \mathbf{C}^k (D_n^k)^T$ are understood as $\{\Psi^k Q_i^k | i \in I\}$ and $\{D_s^k C_i^k (D_n^k)^T | i \in I\}$. Noting that $B^*(D_s^k \mathbf{C}^k (D_n^k)^T)$ is a three-dimensional volume that does not contain information of the surface normal, we use the technique introduced in SOCR[5] to construct a directional albedo with the surface normal provided by $\mathbf{u}^k$. See the supplement of SOCR[5] for more detail. Here, we abuse the notation and also use $B^*(D_s^k \mathbf{C}^k (D_n^k)^T)$ to represent this directional albedo. Minimizing the objective function (S.10) yields a least-squares problem without constraint, which can be solved with the conjugate gradient method.

We remark that this sub-problem is solved approximately due to the omitted term $\phi(\mathbf{u})$ and the treatment of $B^*(D_s^k \mathbf{C}^k (D_n^k)^T)$. Nonetheless, extensive experimental results in Supplementary Note 1 indicate that high-quality reconstructions are obtained with these tricks.

For the sub-problem (2.5), if the $j^{th}$ measurement pair of the virtual confocal signal does not appear in the measured signal, we have

$$\mathbf{d}^{k+1}[j] = \mathcal{T}\left(\frac{A_{\mathbf{d}}\mathbf{u}^{k+1}[j] + \lambda_{pd}P^*(\Psi^k\mathbf{Q}^k)[j]}{1+\lambda_{pd}}, \sqrt{\frac{s_d}{\lambda_d + \lambda_d\lambda_{pd}}}\right) \quad \text{(S.12)}$$

Otherwise, the solution writes

$$\mathbf{d}^{k+1}[j] = \mathcal{T}\left(\frac{\lambda_d A_{\mathbf{d}}\mathbf{u}^{k+1}[j] + \lambda_d\lambda_{pd}P^*(\Psi^k\mathbf{Q}^k)[j] + \lambda_{bd}\mathbf{b}_j^{k+1}}{\lambda_d + \lambda_d\lambda_{pd} + \lambda_{bd}}, \sqrt{\frac{s_d}{\lambda_d + \lambda_d\lambda_{pd} + \lambda_{bd}}}\right) \quad \text{(S.13)}$$

in which $A_{\mathbf{d}}\mathbf{u}^{k+1}[j]$ and $\mathbf{b}_j^{k+1}$ are the simulated signal of the $j^{th}$ measurement pair and the corresponding signal of $\mathbf{b}^{k+1}$, respectively.

The sub-problem (2.6) is of the same type with (1.6) and can be solved using the same method discussed above.

# Supplementary Note 3    The choice of parameters

There are several parameters in the constrained optimization problem (S.2), namely $s_u$, $s_b$, $s_d$, $\sigma_{\mathbf{b}}$, $\lambda_u$, $\lambda_b$, $\lambda_d$, $\lambda_{pu}$, $\lambda_{pb}$, $\lambda_{pd}$, $\lambda_{sb}$, $\lambda_{sd}$, $\lambda_{fd}$ and $\lambda_{bd}$. In this section, we show how these parameters are chosen adaptively. Supplementary Table 3 summarizes the sub-problems that involves these parameters and in which steps these parameters are fixed.

## 3.1 Implicit and explicit parameters in the initialization stage

The input signal $\tilde{\mathbf{b}}$ is normalized into the range $[0, 255]$ in advance. In step (1.1), the approximated signal $\mathbf{b}$ is initialized by applying the hard-thresholding operator to the input signal. The threshold value $\sqrt{s_b/\lambda_b}$ depends on two parameters, and we set $s_b^{imp} = \sqrt{s_b/\lambda_b} = 0.01 \times 255$. In step (1.2), a sparse reconstructed target is initialized. The parameter $s_u$ and an additional parameter $\mu$ used for split Bregman iteration are initialized as $s_u^{init}$ and $\mu^{init}$ by directly following the work of SOCR[5]. Noting that the sub-problem (1.2) only contains two terms, we will adjust these two parameters in step (2.3). In step (1.3), the parameter $\lambda_{pu}$ can also be chosen implicitly following the method presented in the work of SOCR[5]. In step (1.4), we choose $\lambda_{sb} = 0.25$ and $\sigma_{\mathbf{b}} = 40$ to avoid large bias of the signal. The sub-problem (1.5) involves three parameters $\lambda_d$, $\lambda_{bd}$ and $s_d$, which play an important role in updating the virtual confocal signal. The parameter $\lambda_{bd}$ controls the weight of the difference between the portion of measured signal and the virtual confocal signal that share the same measurement pairs. In this sub-problem, we don't consider the relationship between these signals and simply set $\sqrt{s_d/\lambda_d} = \sqrt{s_d/(\lambda_d + \lambda_{bd})} = 0.01 \times 255$. The parameters $\lambda_{bd}$, $\lambda_d$ and $s_d$ will be fixed in steps (2.1), (2.3) and (2.5), respectively. In step (1.6), we choose $\lambda_{sd} = 1$ to assign equal weight to $\mathbf{d}^0$ and $A_{\mathbf{d}}\mathbf{u}^0$ in the process of dictionary learning. The parameter $\lambda_{fd}$ is implicitly chosen such that the virtual noise level equals 40.

## 3.2 Parameters determined in the first iteration

In the first iteration, all parameters can be fixed. In step (2.1), noting that the hard-thresholding operator applies to a convex combination of $A_{\mathbf{b}}\mathbf{u}^0$, $\tilde{\mathbf{b}}$, $P^*(D\mathbf{S}^0)$ and $\mathbf{d}^0$ with weights $1:\lambda_b:\lambda_b\lambda_{pb}\lambda_{sb}:\lambda_{bd}$, we set $\lambda_b = 1$, $\lambda_{pb} = 16$ and $\lambda_{bd} = 4$. In step (2.3), the reconstructed target is updated. By setting the terms

$$\|A_{\mathbf{b}}\mathbf{u}^0 - \mathbf{b}^1\|^2 : \|\mathbf{u}^0 - B^*(D_s^0 \mathbf{C}^0 (D_n^0)^T)\|^2 : \|A_{\mathbf{d}}\mathbf{u}^0 - \mathbf{d}^0\|^2 = 1:\lambda_u^{imp}:\lambda_b^{imp} \quad (S.14)$$

The parameters $\lambda_u$ and $\lambda_d$ are adaptively chosen as

$$\lambda_u = \lambda_u^{imp} \frac{\|A_{\mathbf{b}}\mathbf{u}^0 - \mathbf{b}^1\|^2}{\|\mathbf{u}^0 - B^*(D_s^0 \mathbf{C}^0 (D_n^0)^T)\|^2} \quad (S.15)$$

and

$$\lambda_d = \lambda_d^{imp} \frac{\|A_{\mathbf{b}}\mathbf{u}^0 - \mathbf{b}^1\|^2}{\|A_{\mathbf{d}}\mathbf{u}^0 - \mathbf{d}^0\|^2} \tag{S.16}$$

We set $\lambda_u^{imp} = 5$ and $\lambda_d^{imp} = 2$ to benefit from the contributions of the sparse representation of the target and the virtual confocal signal. We also set $\lambda_{pd} = 4$ to assign a large weight to the contribution of $P^*(\Psi^0 \mathbf{Q}^0)$. Besides, comparing sub-problems (1.2) and (2.3), we fix $s_u$ and $\mu$ as

$$s_u = s_u^{init} \left(1 + \lambda_u^{imp} + \lambda_d^{imp}(1 + \lambda_{pd} + \lambda_{sd})\right) \tag{S.17}$$

$$\mu = \mu^{init} \left(1 + \lambda_u^{imp} + \lambda_d^{imp}(1 + \lambda_{pd} + \lambda_{sd})\right) \tag{S.18}$$

The virtual confocal signal is updated in step (2.5). We set the truncation value in equation (S.12) to be $s_d^{imp} = \sqrt{s_d/(\lambda_d + \lambda_d \lambda_{pd})} = 0.01 \times 255$ and fix $s_d$ as

$$s_d = (s_d^{imp})^2 \lambda_d (1 + \lambda_{pd}) \tag{S.19}$$

# Supplementary Note 4  Time and memory complexity

Consider a typical setting where the reconstruction domain is discretized with $N \times N \times N$ voxels and the signal is detected at $M$ measurement pairs. When $p_x$, $p_y$, $p_z$, $q_y$, $q_z$, $q_t$, $s$, $r \leq N^{1/4}$ and $w \leq N^{1/3}$ (See Supplementary Table 1 for the meaning of these parameters), the time complexity and memory complexity of the CC-SOCR method are $\mathcal{O}(\max\{N^5, MN^3\})$ and $\mathcal{O}(\max\{N^3, MN\})$, respectively.

## 4.1 Time complexity

The overall time complexity of sub-problems (1.2), (1.3), (1.4), (2.2) and (2.4) is $\mathcal{O}(\max\{N^5, MN^3\})$, as discussed in section 4 of the supplement of the work of SOCR[5]. For the rest of the sub-problems, it takes $\mathcal{O}(MN)$ to apply the elementwise hard-thresholding in step (1.1). In step (1.5), it takes $\mathcal{O}(N^5)$ to generate the simulated signal $A_d \mathbf{u}^0$ and $\mathcal{O}(N^3)$ to initialize $\mathbf{d}$ according to equations (S.5) and (S.6). By writing the objective function of sub-problem (1.6) equivalently as equation (S.7), we deduce that the time complexity of this step is no more than that of step (1.3), because it does not contain the block matching process. The sub-problem (2.1) can be solved in $\mathcal{O}(MN)$ with equations (S.8) and (S.9). The sub-problem (2.3) is solved with Supplementary Algorithm 2, which takes $\mathcal{O}(\max\{N^5, MN^3\})$ to solve the least-squares problem and $\mathcal{O}(N^3)$ to update $\mathbf{v}_{k+1}^{j+1}$ and $\mathbf{q}_{k+1}^{j+1}$. The time complexity of step (2.5) is the same as that of step (1.5), which requires $\mathcal{O}(N^5)$. The sub-problem (2.6) is exactly of the same type with the sub-problem (1.6). To sum up, the overall time complexity of the CC-SOCR algorithm is $\mathcal{O}(\max\{N^5, MN^3\})$.

## 4.2 Memory complexity

To store the input signal $\tilde{\mathbf{b}}$, $\mathcal{O}(MN)$ memory is needed. It takes $\mathcal{O}(N^3)$ to store the physical linear operator due to the repetition of elements in the measurement matrices $A_\mathbf{b}$ and $A_\mathbf{d}$. Generating the simulated signal with the target takes $\mathcal{O}(\max\{MN, N^3\})$ memory with linear operator-based implementation of the physical model. The steps (1.1), (1.5), (2.1) and (2.5) take $\mathcal{O}(\max\{MN, N^3\})$ memory because the solutions are obtained with pointwise truncation. The sub-problems (1.2) and (2.3) also take $\mathcal{O}(\max\{MN, N^3\})$ storage, where solutions of the linear systems are obtained with the conjugate gradient method. For the sub-problems (1.3) and (2.4), it suffices to compute $B^*(D_s \mathbf{C} D_n^T)$, which can be implemented block by block with $\mathcal{O}(N^3)$ memory[9]. In steps (1.4) and (2.2), $\mathcal{O}(N^3)$ is not enough to store the whole dataset of the Wiener coefficients. However, storing this dataset explicitly is not necessary for the final reconstruction. It suffies to compute $P^*(D\mathbf{S})$, which can be realized using the sliding window Wiener filtering technique with $\mathcal{O}(N^3)$ storage[9]. For the sub-problems (1.6) and (2.6), it suffies to compute $P^*(\Psi \mathbf{Q})$, which takes $\mathcal{O}(N^3)$ memory with a patch by patch implementation of the data-driven tight frame denoising algorithm[7]. In all, the memory complexity of the proposed method is

$\mathcal{O}(\max\{MN, N^3\})$.

**4.3 Execution time**

Execution time of the CC-SOCR algorithm for the instance of statue with 200 randomly distributed confocal measurements and virtual confocal signals of different sizes are shown in Supplementary Tables 4 - 7. The code was run on an AMD EPYC 7452 server with 64 cores. It is shown that sparser virtual confocal signal result in shorter excution time. However, the reconstruction quality decreases with the size of the virtual confocal signal (See Supplementary Figure 23).

# Reference

1. Lindell, D. B., Wetzstein, G. & O'Toole, M. Wave-based non-line-of-sight imaging using fast *f-k* migration. *ACM Trans. Graph.* **38**, 1–13 (2019).
2. Matthew O'Toole, Lindell, D. B. & Wetzstein, G. Confocal non-line-of-sight imaging based on the light-cone transform. *Nature* **555**, 338–341 (2018).
3. Young, S. I., Lindell, D. B., Girod, B., Taubman, D. & Wetzstein, G. Non-Line-of-Sight Surface Reconstruction Using the Directional Light-Cone Transform. in *2020 IEEE/CVF Conference on Computer Vision and Pattern Recognition (CVPR)* 1404–1413 (IEEE, 2020). doi:10.1109/CVPR42600.2020.00148.
4. Liu, X., Bauer, S. & Velten, A. Phasor field diffraction based reconstruction for fast non-line-of-sight imaging systems. *Nat. Commun.* **11**, 1645 (2020).
5. Liu, X. *et al.* Non-line-of-sight reconstruction with signal–object collaborative regularization. *Light Sci. Appl.* **10**, 198 (2021).
6. Galindo, M., Marco, J., O'Toole, M., Wetzstein, G. & Jarabo, A. A dataset for benchmarking time-resolved non-line-of-sight imaging. in *ACM SIGGRAPH 2019 Posters* (2019).
7. Cai, J.-F., Ji, H., Shen, Z. & Ye, G.-B. Data-driven tight frame construction and image denoising. *Appl. Comput. Harmon. Anal.* **37**, 89–105 (2014).
8. Goldstein, T. & Osher, S. The Split Bregman Method for L1-Regularized Problems. *SIAM J. Imaging Sci.* **2**, 323–343 (2009).
9. D Abov, K., Foi, A., Katkovnik, V. & Egiazarian, K. Image denoising with block-matching and 3D filtering. in *Proceedings of SPIE - The International Society for Optical Engineering* 354–365 (2006).

## Supplementary Algorithm 1 Solving the CC-SOCR optimization problem

**Stage 1: Initialization**

(1.1) Initialize the approximated signal.
$$\mathbf{b}^0 = \underset{\mathbf{b}}{\operatorname{argmin}} \ \lambda_b \|\mathbf{b} - \tilde{\mathbf{b}}\|^2 + s_b |\mathbf{b}|_0$$

(1.2) Initialize the reconstructed target.
$$\mathbf{u}^0 = \underset{\mathbf{u}}{\operatorname{argmin}} \ \|A_{\mathbf{b}} \mathbf{u} - \mathbf{b}^0\|^2 + s_u \|\mathbf{u}\|_{2,1}$$

(1.3) Initialize the dictionaries of the albedo.
$$(D_s^0, D_n^0, \mathbf{C}^0) = \underset{D_s, D_n, \mathbf{C}}{\operatorname{argmin}} \ \sum_i \left( \|B_i(\mathbf{L}^0) - D_s C_i D_n^T\|^2 + \lambda_{pu} |C_i|_0 \right)$$
$$\text{s.t.} \quad \mathbf{L}^0 = \operatorname{albedo}(\mathbf{u}^0), \quad D_s^T D_s = I[p_x p_y p_z], \quad D_n^T D_n = I[r]$$

(1.4) Initialize the Wiener coefficients.
$$\mathbf{S}^0 = \underset{\mathbf{S}}{\operatorname{argmin}} \ \sum_i \left[ \|P_i(\tilde{\mathbf{b}}) - DS_i\|^2 + \lambda_{sb} \|P_i(\mathbf{b}^0) - DS_i\|^2 + \sum_j \left( \frac{\sigma_{\mathbf{b}}}{d_j^T P_i(A_{\mathbf{b}} \mathbf{u}^0)} S_i(j) \right)^2 \right]$$

(1.5) Initialize the virtual confocal signal.
$$\mathbf{d}^0 = \underset{\mathbf{d}}{\operatorname{argmin}} \ \lambda_d \|A_{\mathbf{d}} \mathbf{u}^0 - \mathbf{d}\|^2 + \lambda_{bd} \|R_{\mathbf{b}^0}(\mathbf{b}^0, \mathbf{d}) - R_{\mathbf{d}}(\mathbf{b}^0, \mathbf{d})\|^2 + s_d |\mathbf{d}|_0$$

(1.6) Initialize the dictionary of the virtual confocal signal.
$$(\Psi^0, \mathbf{Q}^0) = \underset{\Psi, \mathbf{Q}}{\operatorname{argmin}} \ \sum_i \left( \|Q_i - \Psi^T P_i(\mathbf{d}^0)\|^2 + \lambda_{sd} \|Q_i - \Psi^T P_i(A_{\mathbf{d}} \mathbf{u}^0)\|^2 + \lambda_{fd} |Q_i|_0 \right)$$
$$\text{s.t.} \quad \Psi^T \Psi = I[q_y q_z q_t]$$

**Stage 2: Iteration**

**For** $k = 0, 1, \ldots, K$

(2.1) Update the approximated signal.
$$\mathbf{b}^{k+1} = \underset{\mathbf{b}}{\operatorname{argmin}} \ \|A_{\mathbf{b}} \mathbf{u}^k - \mathbf{b}\|^2 + \lambda_b \|\mathbf{b} - \tilde{\mathbf{b}}\|^2 + s_b |\mathbf{b}|_0$$
$$+ \lambda_b \lambda_{pb} \lambda_{sb} \|P_i(\mathbf{b}) - DS_i^k\|^2 + \lambda_{bd} \|R_{\mathbf{b}}(\mathbf{b}, \mathbf{d}^k) - R_{\mathbf{d}^k}(\mathbf{b}, \mathbf{d}^k)\|^2$$

(2.2) Update the Wiener coefficients.
$$\mathbf{S}^{k+1} = \underset{\mathbf{S}}{\operatorname{argmin}} \ \sum_i \left[ \|P_i(\tilde{\mathbf{b}}) - DS_i\|^2 + \lambda_{sb} \|P_i(\mathbf{b}^{k+1}) - DS_i\|^2 + \sum_j \left( \frac{\sigma_{\mathbf{b}}}{d_j^T P_i(A_{\mathbf{b}} \mathbf{u}^k)} S_i(j) \right)^2 \right]$$

(2.3) Update the reconstructed target.
$$\mathbf{u}^{k+1} = \underset{\mathbf{u}}{\operatorname{argmin}} \ \|A_{\mathbf{b}} \mathbf{u} - \mathbf{b}^{k+1}\|^2 + \lambda_u \sum_i \|B_i(\mathbf{L}) - D_s^k C_i^k (D_n^k)^T\|^2 + s_u \|\mathbf{u}\|_{2,1}$$
$$+ \lambda_d \|A_{\mathbf{d}} \mathbf{u} - \mathbf{d}^k\|^2 + \lambda_d \lambda_{pd} \lambda_{sd} \sum_i \|Q_i^k - (\Psi^k)^T P_i(A_{\mathbf{d}} \mathbf{u})\|^2$$
$$\text{s.t.} \quad \mathbf{L} = \operatorname{albedo}(\mathbf{u})$$

(2.4) Update the dictionaries of the albedo.

$$(D_s^{k+1}, D_n^{k+1}, \mathbf{C}^{k+1}) = \underset{D_s, D_n, \mathbf{C}}{\operatorname{argmin}} \sum_i \left( \|B_i(\mathbf{L}^{k+1}) - D_s C_i D_n^T\|^2 + \lambda_{pu} |C_i|_0 \right)$$

$$\text{s.t.} \quad \mathbf{L}^{k+1} = \operatorname{albedo}(\mathbf{u}^{k+1}), \quad D_s^T D_s = I[p_x p_y p_z], \quad D_n^T D_n = I[r]$$

(2.5) Update the virtual confocal signal.

$$\mathbf{d}^{k+1} = \underset{\mathbf{d}}{\operatorname{argmin}} \ \lambda_d \|A_\mathbf{d} \mathbf{u}^{k+1} - \mathbf{d}\|^2 + \lambda_d \lambda_{pd} \sum_i \|Q_i^k - (\Psi^k)^T P_i(\mathbf{d})\|^2$$
$$+ \lambda_{bd} \|R_{\mathbf{b}^{k+1}}(\mathbf{b}^{k+1}, \mathbf{d}) - R_\mathbf{d}(\mathbf{b}^{k+1}, \mathbf{d})\|^2 + s_d |\mathbf{d}|_0$$

(2.6) Update the dictionary of the virtual confocal signal.

$$(\Psi^{k+1}, \mathbf{Q}^{k+1}) = \underset{\Psi, \mathbf{Q}}{\operatorname{argmin}} \sum_i \left( \|Q_i - \Psi^T P_i(\mathbf{d}^{k+1})\|^2 + \lambda_{sd} \|Q_i - \Psi^T P_i(A_\mathbf{d} \mathbf{u}^{k+1})\|^2 + \lambda_{fd} |Q_i|_0 \right)$$

$$\text{s.t.} \quad \Psi^T \Psi = I[q_y q_z q_t]$$

**End**
**Stage 3: Output results**
(3.1) The reconstructed albedo is given by $\mathbf{L}^* = \operatorname{albedo}(\mathbf{u}^K)$.

(3.2) The reconstructed surface normal is given by $\mathbf{n}^* = \mathbf{u}^K / \mathbf{L}^*$.

**Supplementary Algorithm 2** Updating the reconstructed target

$\mathbf{q}_0^{k+1} = \mathbf{0}$

$\mathbf{u}_0^{k+1} = \mathbf{u}^k$

**For** $j = 0, 1, ..., J-1$

$\mathbf{v}_{j+1}^{k+1} = \underset{\mathbf{v}}{\operatorname{argmin}} \; s_u \|\mathbf{v}\|_{2,1} + \mu \|\mathbf{v} - \mathbf{u}_j^{k+1} + \mathbf{q}_j^{k+1}\|^2$

$\mathbf{u}_{j+1}^{k+1} = \underset{\mathbf{u}}{\operatorname{argmin}} \|A_\mathbf{b}\mathbf{u} - \mathbf{b}^{k+1}\|^2 + \lambda_u \sum_i \|B_i(\mathbf{L}) - D_s^k C_i^k (D_n^k)^T\|^2 + \lambda_d \|A_\mathbf{d}\mathbf{u} - \mathbf{d}^k\|^2$

$\qquad + \lambda_d \lambda_{pd} \lambda_{sd} \sum_i \|Q_i^k - (\Psi^k)^T P_i(A_\mathbf{d}\mathbf{u})\|^2 + \mu \|\mathbf{v}_{j+1}^{k+1} - \mathbf{u} + \mathbf{q}_j^{k+1}\|^2$

$\qquad$ s.t. $\mathbf{L} = \operatorname{albedo}(\mathbf{u})$

$\mathbf{q}_{j+1}^{k+1} = \mathbf{q}_j^{k+1} + \mathbf{v}_{j+1}^{k+1} - \mathbf{u}_{j+1}^{k+1}$

**End**

$\mathbf{u}^{k+1} = \mathbf{v}_J^{k+1}$

**Supplementary Table 1 Parameters of the experimental setup**

| Parameter | Explanation |
|---|---|
| $n_x, n_y, n_z$ | The number of voxels in the depth, horizontal and vertical directions |
| $p_x, p_y, p_z$ | The patch sizes of the albedo in the depth, horizontal and vertical directions |
| $q_y, q_z, q_t$ | The patch sizes of the virtual signal in the horizontal, vertical and temporal directions |
| $r$ | The number of neighboring blocks of each local albedo block |
| $w$ | Searching window size of the albedo block in the process of block matching |
| $M$ | The number of measurement pairs |
| $s$ | The patch size of the measured signal in the temporal direction |

**Supplementary Table 2 Sizes and explanations of the optimization variables**

| Variable | Size | Explanation |
|---|---|---|
| $\mathbf{L}$ | $n_x \times n_y \times n_z$ | Albedo |
| $\mathbf{n}$ | $n_x \times n_y \times n_z \times 3$ | Unit surface normal |
| $\mathbf{u}$ | $n_x \times n_y \times n_z \times 3$ | Directional albedo |
| $\tilde{\mathbf{b}}$ | $M \times n_t$ | The measured signal with arbitrary spatial pattern |
| $\mathbf{b}$ | $M \times n_t$ | The approximated signal of the ideal signal of the measurement pairs |
| $\mathbf{d}$ | $n_y \times n_z \times n_t$ | The virtual confocal signal |
| $D_s$ | $p_x p_y p_z \times p_x p_y p_z$ | The spatial dictionary learned from the albedo |
| $D_n$ | $r \times r$ | The dictionary that captures non-local correlations of the albedo |
| $C_i$ | $p_x p_y p_z \times r$ | The matrix of transform-domain coefficients of the albedo block $i$ |
| $S_i$ | $s \times 1$ | The vector of Wiener coefficients of the $i^{th}$ patch of $\mathbf{b}$ |
| $\Psi$ | $q_y q_z q_t \times q_y q_z q_t$ | The spatial dictionary learned from $\mathbf{d}$ and $A_\mathbf{d}\mathbf{u}$ |
| $Q_i$ | $q_y q_z q_t \times 1$ | The vector of transform-domain coefficients of the dictionary $\Psi$ |

**Supplementary Table 3 Parameters of the reconstruction algorithm**

| Parameter | The sub-problems involved | The sub-problem in which the parameter is fixed |
|---|---|---|
| $s_u$ | (1.2), (2.3) | (2.3) |
| $s_b$ | (1.1), (2.1) | (2.1) |
| $s_d$ | (1.5), (2.5) | (2.5) |
| $\sigma_d$ | (1.4), (2.2) | (1.4) |
| $\lambda_u$ | (2.3) | (2.3) |
| $\lambda_b$ | (1.1), (2.1) | (2.1) |
| $\lambda_d$ | (1.5), (2.3), (2.5) | (2.3) |
| $\lambda_{pu}$ | (1.3), (2.4) | (1.3) |
| $\lambda_{pb}$ | (2.1) | (2.1) |
| $\lambda_{pd}$ | (2.3), (2.5) | (2.3) |
| $\lambda_{sb}$ | (1.4), (2.1), (2.2) | (1.4) |
| $\lambda_{sd}$ | (1.6), (2.3), (2.6) | (1.6) |
| $\lambda_{fd}$ | (1.6), (2.6) | (1.6) |
| $\lambda_{bd}$ | (1.5), (2.1), (2.5) | (2.1) |

**Supplementary Table 4 Execution time of the instance of the statue with 200 randomly distributed confocal measurements and 64 × 64 virtual confocal signal**

| Sub-problem | Explanation | CPU time (s) | |
|---|---|---|---|
| Computing the forward operator of the physical model | | 0.1937 | |
| **Stage 1: Initialization** | | | |
| (1.1) | Initializing $\mathbf{b}$ | 0.0017 | |
| (1.2) | Initializing $\mathbf{u}$ | 60.9705 | |
| (1.3) | Initializing $D_s$, $D_n$ and $\mathbf{C}$ | 7.0017 | |
| (1.4) | Initializing $\mathbf{S}$ | 0.8245 | |
| (1.5) | Initializing $\mathbf{d}$ | 1.9342 | |
| (1.6) | Initializing $\Psi$ and $\mathbf{Q}$ | 10.8064 | |
| **Stage 2: Iteration** | | | |
| (2.1) | Updating $\mathbf{b}$ | 0.0267 | 0.7261 |
| (2.2) | Updating $\mathbf{S}$ | 0.`0333 | 0.0165 |
| (2.3) | Updating $\mathbf{u}$ | 82.4961 | 81.7546 |
| (2.4) | Updating $D_s$, $D_n$ and $\mathbf{C}$ | 4.8466 | 4.4856 |
| (2.5) | Updating $\mathbf{d}$ | 2.0045 | 2.0358 |
| (2.6) | Updating $\Psi$ and $\mathbf{Q}$ | 10.6700 | 10.5273 |
| **Total excution time (s)** | | | |
| 281.3559 | | | |

**Supplementary Table 5 Execution time of the instance of the statue with 200 randomly distributed confocal measurements and 32 × 32 virtual confocal signal**

| Sub-problem | Explanation | CPU time (s) | |
|---|---|---|---|
| Computing the forward operator of the physical model | | 0.1892 | |
| **Stage 1: Initialization** | | | |
| (1.1) | Initializing $\mathbf{b}$ | 0.0018 | |
| (1.2) | Initializing $\mathbf{u}$ | 63.3050 | |
| (1.3) | Initializing $D_s$, $D_n$ and $\mathbf{C}$ | 7.4353 | |
| (1.4) | Initializing $\mathbf{S}$ | 0.8742 | |
| (1.5) | Initializing $\mathbf{d}$ | 1.2871 | |
| (1.6) | Initializing $\Psi$ and $\mathbf{Q}$ | 3.5379 | |
| **Stage 2: Iteration** | | | |
| (2.1) | Updating $\mathbf{b}$ | 0.0165 | 0.8005 |
| (2.2) | Updating $\mathbf{S}$ | 0.0378 | 0.0134 |
| (2.3) | Updating $\mathbf{u}$ | 40.3780 | 44.9235 |
| (2.4) | Updating $D_s$, $D_n$ and $\mathbf{C}$ | 4.5323 | 4.6030 |
| (2.5) | Updating $\mathbf{d}$ | 1.2301 | 1.1797 |
| (2.6) | Updating $\Psi$ and $\mathbf{Q}$ | 3.3823 | 3.3888 |
| **Total excution time (s)** | | | |
| 181.1166 | | | |

**Supplementary Table 6 Execution time of the instance of the statue with 200 randomly distributed confocal measurements and 16 × 16 virtual confocal signal**

| Sub-problem | Explanation | CPU time (s) | |
|---|---|---|---|
| Computing the forward operator of the physical model | | 0.1918 | |
| **Stage 1: Initialization** | | | |
| (1.1) | Initializing $\mathbf{b}$ | 0.0015 | |
| (1.2) | Initializing $\mathbf{u}$ | 67.0203 | |
| (1.3) | Initializing $D_s$, $D_n$ and $\mathbf{C}$ | 7.4327 | |
| (1.4) | Initializing $\mathbf{S}$ | 0.8427 | |
| (1.5) | Initializing $\mathbf{d}$ | 0.8562 | |
| (1.6) | Initializing $\Psi$ and $\mathbf{Q}$ | 1.1980 | |
| **Stage 2: Iteration** | | | |
| (2.1) | Updating $\mathbf{b}$ | 0.0216 | 0.9710 |
| (2.2) | Updating $\mathbf{S}$ | 0.0361 | 0.0184 |
| (2.3) | Updating $\mathbf{u}$ | 34.5548 | 39.0236 |
| (2.4) | Updating $D_s$, $D_n$ and $\mathbf{C}$ | 4.4882 | 3.8257 |
| (2.5) | Updating $\mathbf{d}$ | 0.7508 | 0.7671 |
| (2.6) | Updating $\Psi$ and $\mathbf{Q}$ | 1.1771 | 1.1989 |
| **Total excution time (s)** | | | |
| 164.3765 | | | |

**Supplementary Table 7 Execution time of the instance of the statue with 200 randomly distributed confocal measurements and 8 × 8 virtual confocal signal**

| Sub-problem | Explanation | CPU time (s) | |
|---|---|---|---|
| | Computing the forward operator of the physical model | 0.2316 | |
| **Stage 1: Initialization** | | | |
| (1.1) | Initializing $\mathbf{b}$ | 0.0016 | |
| (1.2) | Initializing $\mathbf{u}$ | 67.0207 | |
| (1.3) | Initializing $D_s$, $D_n$ and $\mathbf{C}$ | 7.3978 | |
| (1.4) | Initializing $\mathbf{S}$ | 0.9241 | |
| (1.5) | Initializing $\mathbf{d}$ | 0.5542 | |
| (1.6) | Initializing $\Psi$ and $\mathbf{Q}$ | 0.5237 | |
| **Stage 2: Iteration** | | | |
| (2.1) | Updating $\mathbf{b}$ | 0.0068 | 0.8318 |
| (2.2) | Updating $\mathbf{S}$ | 0.0470 | 0.0149 |
| (2.3) | Updating $\mathbf{u}$ | 23.6866 | 28.4683 |
| (2.4) | Updating $D_s$, $D_n$ and $\mathbf{C}$ | 4.5322 | 4.4342 |
| (2.5) | Updating $\mathbf{d}$ | 0.5168 | 0.4928 |
| (2.6) | Updating $\Psi$ and $\mathbf{Q}$ | 0.4959 | 0.4342 |
| **Total excution time (s)** | | | |
| 140.6152 | | | |